\documentclass[11pt, letterpaper]{article}
\usepackage[margin=1in]{geometry}
\usepackage[utf8]{inputenc}

\usepackage[dvipsnames]{xcolor}
\usepackage{float}
\usepackage{wrapfig}
\usepackage{graphicx}
\graphicspath{ {./images/} }

\usepackage{subcaption}
\usepackage{enumitem}

\usepackage{braket}

\usepackage{mathtools}
\usepackage{amsmath}
\usepackage{bbm}
\usepackage{amssymb}
\usepackage{amsfonts}
\usepackage{amsthm}
\usepackage{bm}
\usepackage{mathrsfs}
\usepackage{tikz-cd}

\numberwithin{question}{section}

\numberwithin{question}{section}

\theoremstyle{remark}

\theoremstyle{definition}

\newcommand{\catname}[1]{{\normalfont\textbf{#1}}}

\newcommand{\Mod}{\catname{Mod}}
\newcommand{\Rep}{\catname{Rep}}

\renewcommand{\Vec}{\catname{Vec}}

\DeclareMathOperator{\Hom}{Hom}

\newcommand{\R}{\mathbb{R}}
\newcommand{\C}{\mathbb{C}}
\newcommand{\Z}{\mathbb{Z}}

\newcommand{\cC}{\mathcal{C}}

\newcommand{\cM}{\mathcal{M}}

\newcommand{\cZ}{\mathcal{Z}}

\usepackage{jheppub} 

\hypersetup{
    colorlinks,
    citecolor=magenta,
    filecolor=black,
    linkcolor=blue,
    urlcolor=magenta
}

\usepackage[status=draft,inline,nomargin]{fixme} 
\fxusetheme{color} 
\FXRegisterAuthor{ko}{ako}{KO} 

\usepackage[export]{adjustbox}

\DeclareMathOperator{\End}{End}
\DeclareMathOperator{\id}{id}

\renewcommand{\bar}{\overline}

\DeclareMathOperator{\Strip}{\textbf{Str}}
\newcommand{\Str}[2]{\Strip_{#1}(#2)}
\DeclareMathOperator{\Tub}{\textbf{Tube}}
\DeclareMathOperator{\symTQFT}{\textbf{SymTQFT}}

\DeclareMathOperator{\Fib}{\textbf{Fib}}
\DeclareMathOperator{\TY}{\textbf{TY}}

\title{\center{Representation Theory of Solitons}}

\abstract{Solitons in two-dimensional quantum field theory exhibit patterns of degeneracies and associated selection rules on scattering amplitudes. We develop a representation theory that captures these intriguing features of solitons. This representation theory is based on an algebra we refer to as the \emph{strip algebra}, $\Str{\cC}{\cM}$, which is defined in terms of the non-invertible symmetry, $\cC,$ a fusion category, and its action on boundary conditions encoded by a module category, $\cM$. The strip algebra is a $C^*$-weak Hopf algebra, a fact which can be elegantly deduced by quantizing the three-dimensional Drinfeld center TQFT, $\mathcal{Z}(\cC),$ on a spatial manifold with corners.  These structures imply that the representation category of the strip algebra is also a unitary fusion category which we identify with a dual category $\cC_{\cM}^{*}.$  We present a straightforward method for analyzing these representations in terms of quiver diagrams where nodes are vacua and arrows are solitons and provide examples demonstrating how the representation theory reproduces known degeneracies and selection rules of soliton scattering. Our analysis provides the general framework for analyzing non-invertible symmetry on manifolds with boundary and applies both to the case of boundaries at infinity, relevant to particle physics, and boundaries at finite distance, relevant in conformal field theory or condensed matter systems.
}

\author[*]{Clay C\'ordova,}
\author[*]{Nicholas Holfester,}
\author[**]{Kantaro Ohmori }

\affiliation[*]{Kadanoff Center for Theoretical Physics \& Enrico Fermi Institute, University of Chicago}
\affiliation[**]{Faculty of Science, University of Tokyo, Japan}

\emailAdd{clayc@uchicago.edu}
\emailAdd{nholfester@uchicago.edu}
\emailAdd{kantaro@hep-th.phys.s.u-tokyo.ac.jp}

\begin{document}

\maketitle

\section{Introduction}

Over the past several years, higher symmetry \cite{Gaiotto:2014kfa} has emerged as a new principle for organizing and understanding the structure of Quantum Field Theories (QFTs).  The most novel such symmetries are generally encoded by non-invertible topological operators which commute with the energy-momentum tensor, but in general do not act by unitary operators on the Hilbert space \cite{Frohlich:2009gb,Carqueville:2012dk, Bhardwaj:2017xup, Chang_2019, Chang:2022hud}.   These operators are algebraically richer than groups and are the frontier in a new and evolving paradigm to control the dynamics of strongly-coupled systems.  Recent applications include in particular, the action of such symmetries on local and extended operators \cite{Bartsch:2022mpm,Delcamp:2022sjf,Bartsch:2022ytj,Bartsch:2023pzl,Bhardwaj:2023wzd,Bartsch:2023wvv,Bhardwaj:2023ayw,Copetti:2024onh}, and the corresponding selection rules \cite{Chang_2019, Thorngren:2019iar, thorngren2021fusion, Lin:2023uvm, Rayhaun:2023pgc, Cordova:2023qei}, anomalies \cite{Apte:2022xtu, Kaidi:2023maf,Zhang:2023wlu,Cordova:2023bja,Antinucci:2023ezl,Thorngren:2019iar}, and implications for the structure of vacua \cite{Chang_2019,Huang:2021zvu,Inamura:2021szw,Choi:2021kmx,Kaidi:2021xfk,Choi:2022zal,Li:2023ani,Bhardwaj:2023idu,Bhardwaj:2023fca,Bhardwaj:2024qrf, Robbins:2021ibx}.

Despite these successes, there are crucial physical contexts where our understanding of higher symmetry remains in its infancy.  Most notably for this work are its applications to particle physics.  Indeed, some of the most foundational questions about a global symmetry are: what are its implications for the spectrum of particles, and what patterns of masses are allowed?  In this work we will answer this question in full generality for the case of 2d QFTs whose symmetry is algebraically dictated by fusion categories, generalizing previous work on particle spectra \cite{Cordova:2024vsq}, and the S-matrix \cite{Copetti:2024rqj}, as well as direct lattice investigations of models with non-invertible symmetries \cite{Jia:2023xar,Molnar:2022nmh,Ruiz-de-Alarcon:2022mos, OBrien:2017wmx,Aasen:2020jwb,Eck:2023gic,Cheng:2022sgb, Seiberg:2023cdc,Seiberg:2024gek, Delcamp:2022sjf,Lootens:2022avn,Jia:2024jht,Bridgeman:2022gdx} In addressing this and related questions, the key additional complication that we must face is the implication of symmetries on open spatial manifolds, where boundary conditions are essential.  This is the case for particle-like states where space is the real line, but it is also the case for boundaries at finite distance which may form physical impurities for instance in a condensed matter system.  Our algebraic characterization of the implications of symmetries in open systems applies robustly to either context, and extends the growing body of literature on the interplay of symmetry and boundary conditions \cite{Moore:2006dw, Kitaev_2012, Fuchs_2013, Thorngren:2020yht, Collier:2021ngi, Choi:2023xjw, Bhardwaj:2017xup, Huang:2021zvu}.

\subsection{The Action of Non-Invertible Symmetry on States}

To set the stage more precisely, consider a $2d$ bosonic quantum field theory $\mathcal{T}$ having finite symmetry $\cC$, a fusion category.  In the following, we denote the simple objects of $\cC$, i.e.\ the basic topological symmetry lines, by letters from the first half of the alphabet ($a,b,c,\cdots$). In particular such lines enjoy a fusion algebra:
\begin{equation}
    a\otimes b =\sum_{c}N_{ab}^{c} c, \hspace{.2in}N_{ab}^{c}\in \mathbb{Z}_{\geq0}.
\end{equation}
For a brief summary of the complete definition of a fusion category, see Appendix \ref{app:fusion_cat}. A question of basic importance is how $\cC$ acts on the states of the theory. Among other things, an understanding of this action governs both the degeneracies and selection rules enforced by the existence of the symmetry $\cC$. In practice, the description of the action of $\cC$ heavily depends on if the spatial manifold has boundary or not.

\subsubsection{The Circle and the Tube Algebra}

When $\mathcal{T}$ is placed on a spatial manifold without boundary, i.e.\ a circle, the action of $\cC$ on states is captured by the Tube algebra of $\cC$, $\Tub(\cC)$ \cite{ocneanu1993,ocneanu2001operator,Chang_2019,Tachikawa:2017gyf,Bhardwaj:2023idu,Bartsch:2023wvv,Bullivant:2019fmk}. $\Tub(\cC)$ is constructed from the topological lines and junctions contained in $\cC$. The elements of the algebra can be represented graphically as diagrams of the form shown below:
\begin{equation}\label{tubepic}
    \includegraphics[width=2cm,valign=m]{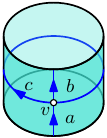}
\end{equation}
with $a,b,c \in \cC$ and $v$ a topological junction at their intersection. The multiplication in the algebra is defined by vertical stacking of elements and simplifying the stacked diagram using the associator data of $\cC$. Schematically, it is defined by the series of manipulations
\begin{equation}
    \includegraphics[width=11.5cm, valign=m]{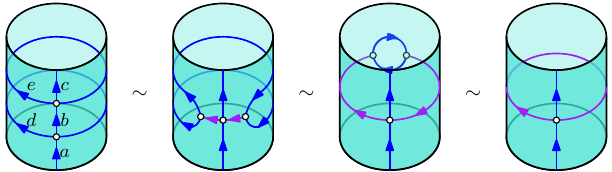}
\end{equation}
where we have used $\sim$ since we do not write the coefficients, which are constructed from $F$-symbol of $\cC$. For a complete expression for the algebra see for example \cite{Tachikawa:2017gyf,Bartsch:2023wvv}. 

 The fusion category $\cC$ naturally acts on states on $S^{1}$ which are decorated by topological lines inserted at points in space (e.g.\ the lines $a$ and $b$ in \eqref{tubepic}.) These include standard states on $S^{1}$, equivalent to local operators in a conformal theory, and twisted sectors, equivalent to operators at the end of topological lines in a conformal theory.  More explicitly, denote by $\mathcal{H}_{a}$ the Hilbert space of the theory quantized on the circle with a topological line $a$ inserted at a point in space.  Then the tube algebra acts naturally on the direct sum:     $\bigoplus_{a\in \cC} \mathcal{H}_{a}$,
 and commutes with the Hamiltonian.  The implied multiplets are given by the representations of $\Tub(\cC)$.  When $\Tub(\cC)$ acts on operators it is typically referred to as ``lassoing'' \cite{Chang_2019}.
 
The algebra $\Tub(\cC)$ is a $C^*$-weak Hopf algebra (See Appendix \ref{app:WHA} for a brief definition) and so on general grounds its representation theory is governed by the theory of unitary fusion categories. In fact, its representation theory is well known to be a familiar fusion category, the Drinfeld center of $\cC$ \cite{Tachikawa:2017gyf, evans1995ocneanu,Muger,Izumi} 
\begin{equation}\label{eq:tube_is_center}
    \Rep(\Tub(\cC)) \simeq \cZ(\cC).
\end{equation}
This fact has several immediate implications:
\begin{itemize}
    \item The representations of the algebra $\Tub(\cC)$ can be studied directly from $\cC$ without directly considering $\Tub(\cC)$.  For instance, $\cZ(\cC)$ can be constructed directly from $\cC$ using \cite{Thorngren:2019iar,levin2005}, and each object in $\cZ(\cC)$ corresponds to a representation of $\Tub(\cC)$. 
    \item The representations  $\Rep(\Tub(\cC))$ admits a topological characterization by the $3d$ Turaev-Viro TQFT constructed from $\cC$ \cite{Lin:2022dhv,Bhardwaj:2023idu,Bartsch:2023wvv}, commonly referred to as the symmetry TQFT of $\cC$, $\symTQFT(\cC)$ \cite{Fuchs:2002cm, Fuchs:2003id, Fuchs:2004dz, Fuchs:2004xi, Gaiotto:2014kfa, Gaiotto:2020iye, Apruzzi:2021nmk, Freed:2022qnc, Chatterjee:2022kxb, Inamura:2023ldn, Kaidi:2022cpf, Bhardwaj:2023bbf,  Apruzzi:2023uma, Baume:2023kkf,Heckman:2024obe,Brennan:2024fgj, Antinucci:2024zjp, Bonetti:2024cjk, Apruzzi:2024htg,Putrov:2024uor}. This enables one to study the implications of the symmetry $\cC$ on states by leveraging the techniques of TQFT.
\end{itemize}

\subsubsection{Open Spatial Manifolds and the Strip Algebra}

We now move to the case when $\mathcal{T}$ is placed on a manifold requiring spatial boundary conditions. In this setting the category $\cC$ no longer fully characterizes the symmetry of the theory. Specifically, $\cC$ does not capture how the defect lines interact with the boundary conditions. 
We denote the collection of of simple boundary conditions by letters from the second half of the alphabet ($m,n,\cdots$).  The basic interplay of the symmetry and the boundaries is a fusion rule:
\begin{equation}\label{modulefusionintro}
    a\otimes m=\sum_{n}\tilde{N}^{n}_{am} n, \hspace{.2in}\tilde{N}^{n}_{am}\in \mathbb{Z}_{\geq0}.
\end{equation}
In physical terms, this rule encodes the fact that a given symmetry operator $a$ may permute the boundary conditions.  Closely related to the fusion rule \eqref{modulefusionintro} is the fact that bulk topological lines may form boundary changing junctions when the act at fixed time.  We represent these by diagrams of the form:
\begin{equation}
    \includegraphics[width=1.5cm,valign=m]{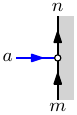}.
\end{equation}
where above, the gray shaded region is empty (so that $n,m$ are boundaries).  There is also a bulk-boundary associator, denoted by $\tilde{F},$ (see below) which permits us to rearrange compositions of junctions. Mathematically, this additional information of the interplay of the symmetry operators and boundary conditions is encoded in a $\cC$-module category $\cM$ \cite{Fuchs:2002cm,Moore:2006dw, Kitaev_2012, Fuchs_2013, Thorngren:2019iar, Choi:2023xjw, Bhardwaj:2017xup, Huang:2021zvu, Behrend_2000}. (See Appendix \ref{app:fusion_cat} for a full definition.) Together, the pair $(\cC,\cM)$ provides a complete algebraic characterization of the finite symmetry of theory in the presence of boundaries. 

An important point to emphasize is that the mathematical structure described above applies in several distinct physical contexts:
\begin{itemize}
    \item The theory $\mathcal{T}$ is quantized on an interval, $I$ of finite length.  For instance, if $\mathcal{T}$ is a conformal field theory with conformally invariant boundary conditions.   
    \item The theory $\mathcal{T}$ is quantized on non-compact spatial manifolds: $\R$ or  $[0,\infty)$.
This is because on such spatial manifolds, boundary conditions are required at spatial infinity in order to define finite energy states.  
\end{itemize}
The case of non-compact spatial manifolds is particularly natural when one describes renormalization group flows terminating in massive theories described by a spectrum of particle-like excitations.  Then, the allowed boundary conditions are the clustering vacua of theory \cite{Cordova:2024vsq}.  In this setting, the strict IR is formally described by a topological phase, and the allowed boundary conditions at infinity are determined by the module category $\mathcal{M}$ defined by the TQFT.\footnote{Depending on context, it may be natural to distinguish left and right boundary conditions.  For instance, in the case where space is semi-infinite the boundary at finite distance is naturally defined in the UV where as that at infinity is naturally defined in the IR.  For simplicity, below we work in the case where the left and right boundary conditions are defined by the same module. }  

Below, we will refer to all manifolds requiring boundary conditions as open.  The Hilbert space of the theory on open manifolds is naturally split into sectors labelled by boundary conditions as:
\begin{equation}\label{openhilbdecomp}
    \mathcal{H}\cong \bigoplus \mathcal{H}_{mn}.
\end{equation}
Our goal is to deduce the implied degeneracies and multiplets on this Hilbert space enforced by the symmetry $\cC.$  These states (and hence also the degeneracies) have distinct physical meanings depending on the contexts outlined above.  For instance:
\begin{itemize}
    \item If the theory $\mathcal{T}$ is conformal and quantized on a finite interval, then the states in $\mathcal{H}_{mn}$ are dual by the state-operator map to boundary changing operators:
    \begin{equation}
    \includegraphics[width=2.7cm, valign=m]{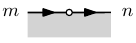}
\end{equation}
\item If the theory $\mathcal{T}$ is gapped and quantized on $\mathbb{R}$, the states in $\mathcal{H}_{mn}$ are particles or solitons depending on whether or not $m$ and $n$ are equal respectively.  For instance in a Landau-Ginzburg type model with multiple minima in the potential $V(\phi)$ for a scalar field $\phi$, the states with $m$ different from $n$ involve a field profile that interpolates between the minima: 
\begin{equation}
    \includegraphics[width=10.5cm, valign=m]{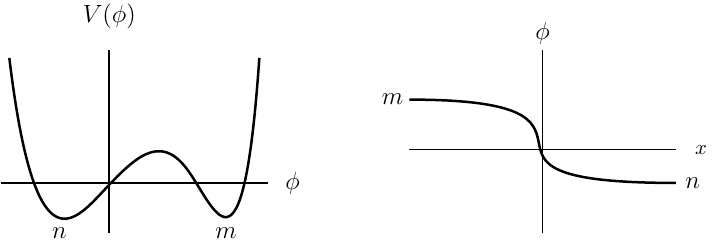}
\end{equation}
\end{itemize}
This is therefore the appropriate setting for studying how finite non-invertible symmetry both acts on particle excitations and constrains S-matrix elements.  Below we often loosely refer to states in sectors $\mathcal{H}_{mn}$ as ``solitons" regardless of whether we intend the conformal or massive application of our formalism.

In \cite{Cordova:2024vsq}, an algebra analogous to $\Tub(\cC)$ was constructed from the data of $\cC$ and $\cM$ and shown to act on the states in spatial geometries having boundary conditions. The elements of the algebra are defined by the diagrams 
\begin{equation}\label{eq:strip_alg_elt}
    \includegraphics[width=2.5cm, valign=m]{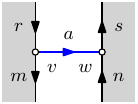}
\end{equation}
where the trivial theory is drawn in gray, $a \in \cC$, $m,n,r,s \in \cM$, and $v,w$ are topological junctions in $\cM$. The left or right boundary can be taken to represent $\pm \infty$ in the case of a non-compact spatial manifold. The composition of two elements is defined by stacking vertically and reducing the diagram using the data of the fusion category $\cC$ and the module $\cM$
\begin{equation}\label{eq:strip_mult}
    \includegraphics[width=10.5cm, valign=m]{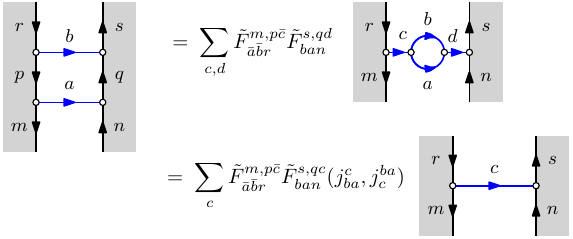}
\end{equation}
where the $\tilde{F}$-symbol and pairing $(\;,\; )$ are defined in Appendix \ref{app:fusion_cat}.\footnote{For brevity of notation our definition of $\tilde{F}$ is the inverse of the symbol used in \cite{Cordova:2024vsq}.}

This algebra is known within the mathematics literature both as the ``Ladder category'' of $\cM$ \cite{Barter_2019} and the ``annular algebra'' \cite{Bridgeman:2022gdx}.\footnote{In the prior context the algebra is instead thought of as a category having pairs of boundaries $(m,n)$ as objects and the diagrams \eqref{eq:strip_alg_elt} as morphisms.} In the physics literature it has appeared in the context of boundary conformal field theory (BCFT) \cite{Kojita:2016jwe, Konechny_2020, Petkova_2001}, as well as in Levin-Wen models, where it has been used to provide an algebraic characterization of excitations on gapped boundaries \cite{Kitaev_2012,Jia:2024jht} and to compute the fusion of topological interfaces \cite{Barter_2019}. Because of its structural resemblance to the Tube algebra, we will refer to it in this work as the ``Strip algebra'' and denote it by $\Str{\cC}{\cM}$.

The first implications of the action of $\Str{\cC}{\cM}$ for $2d$ QFTs were studied in \cite{Cordova:2024vsq}. There it was shown that $\Str{\cC}{\cM}$ acts on states quantized on open spatial manifolds and that this action can enforce novel particle degeneracies. The goal of this paper is to provide a systematic understanding of the implications of this representation theory for $\cC$-symmetric quantum field theories. 

\subsection{Summary of Results and Applications}

\subsubsection{The Structure of the Strip Algebra}
Let us summarize the main results of our analysis of the strip algebra $\Str{\cC}{\cM}.$  Since one of our main goals is to deduce the representations of $\Str{\cC}{\cM}$, our first task is to determine the general type of algebra under investigation.  As we discuss, $\Str{\cC}{\cM}$ is an example of a $C^*$-weak Hopf algebra (reviewed below).  This appropriately generalizes a group ring for invertible symmetries.  In particular, the structure of a weak Hopf algebra implies that representations of $\Str{\cC}{\cM}$ admit notions of tensor products and duals.  This is physically expected since the action of $\Str{\cC}{\cM}$ must generalize from single to multiparticle states (utilizing the tensor product of representations), and from bras to kets (utilizing the dual of a representation).

In parts of our analysis of this algebra, it is fruitful to utilize the 3d symmetry TQFT which is labelled by $\mathcal{Z}(\cC).$  The physical picture underlying this is the idea of geometrically separating the symmetry data, which is topological information, from the physical data of the theory $\mathcal{T}$, which is generally non-topological information.  Practically, this means that we view our 2d theory $\mathcal{T}$ as arising from reduction on an interval with a topological boundary, which we label by $\cC$, encoding the symmetry data on one side and a physical boundary on the other side.  In the case at hand where the theory $\mathcal{T}$ itself has a boundary $\mathcal{B}$ this picture is further enriched as shown below:
\begin{equation}
    \includegraphics[width=11.7cm,valign=b]{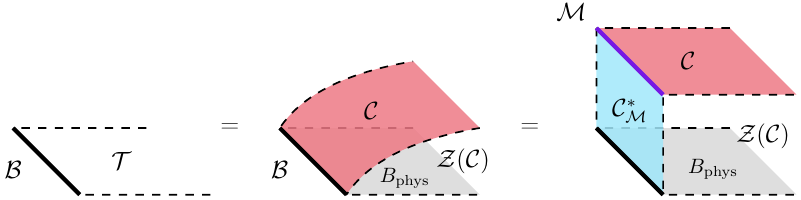}\;\;.
\end{equation}
Here, the module category $\cM$ characterizing the symmetry data of the boundary conditions appears as a corner where it forms an interface between the topological boundary $\cC$ and an associated topological boundary, labelled $\cC^{*}_{\cM}$, which is labelled by a dual fusion category canonically associated to the module $\cM.$

The perspective of quantizing (evaluating) the symmetry TQFT on a manifold with corners is particularly illuminating when considering the weak Hopf algebra structure of the strip algebra $\Str{\cC}{\cM}$ \cite{JohnsonFreydReutter:Hopf, JohnsonFreydReutter:QuantumHpty}. Following \cite{ReutterPItalk,JohnsonFreydPItalk,ReutterOxfordtalk}, here each spatial slice of the 3d theory is viewed as a square with objects in the module category controlling the data at the corners:
\begin{equation}
    \includegraphics[width=3.3cm,valign=m]{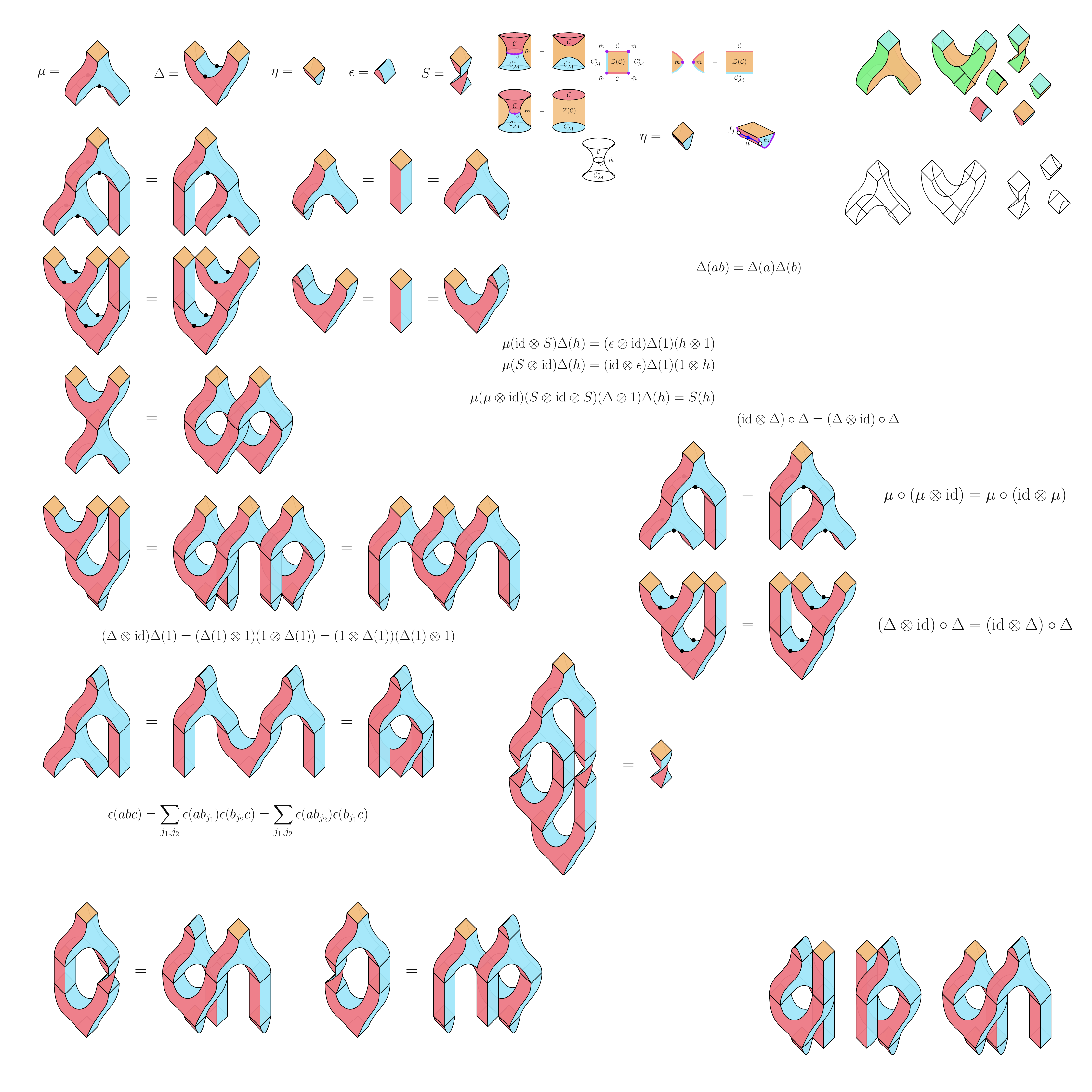}.
\end{equation}
Time evolution of this spatial slice constructs natural bordisms which may then be used to define the operations in the algebra.  For instance, the algebra has a natural multiplication, $\mu,$ and comultiplication, $\Delta$:
\begin{equation}
    \mu: \Str{\cC}{\cM}\otimes_\C \Str{\cC}{\cM} \to \Str{\cC}{\cM}, \hspace{.4in} \Delta: \Str{\cC}{\cM} \to \Str{\cC}{\cM}\otimes_\C \Str{\cC}{\cM},
\end{equation}
which are in turn visualized geometrically as (time running bottom to top):
\begin{equation}
    \includegraphics[width=6.3cm,valign=b]{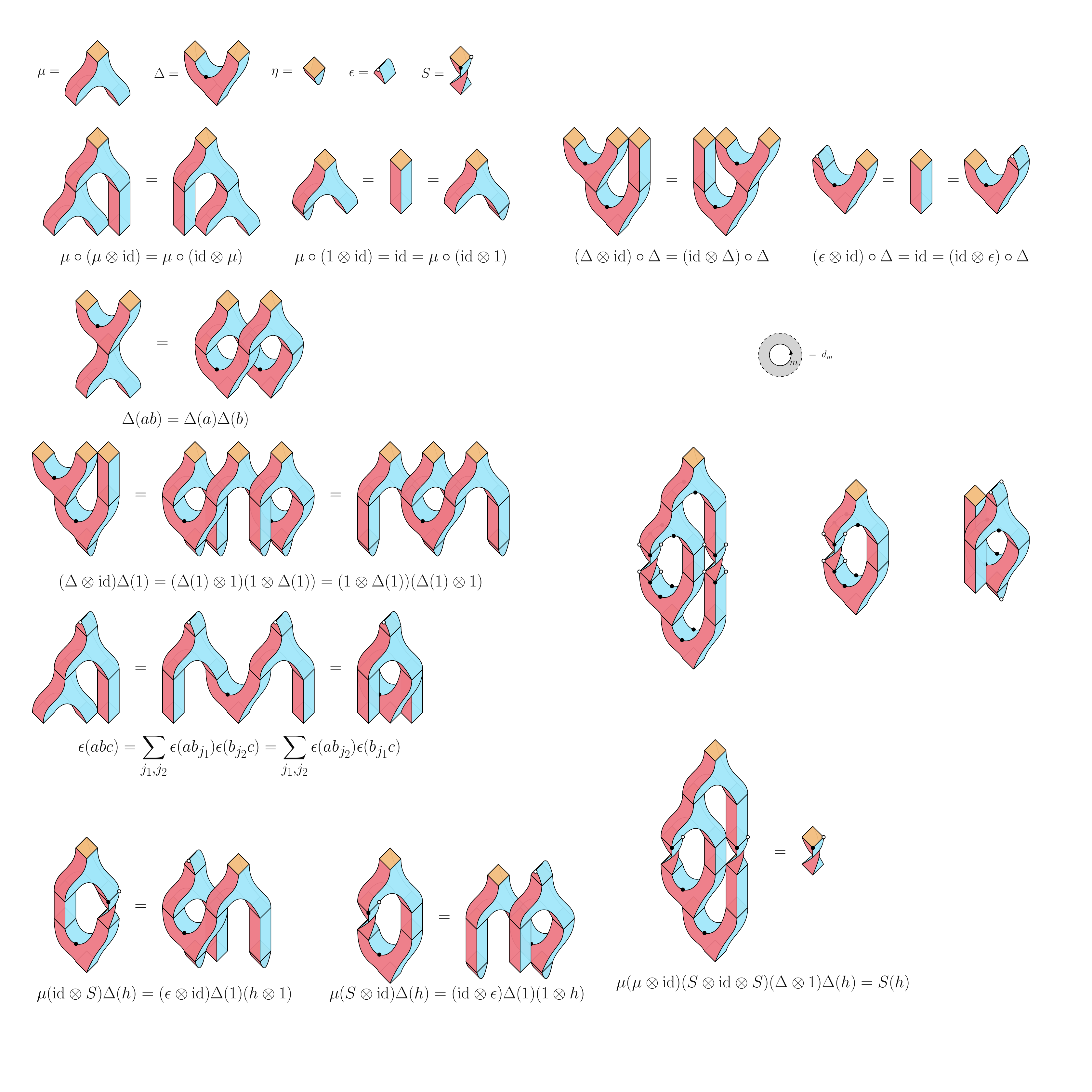}\;\;.
\end{equation}
The topological nature of this construction then allows one to directly verify algebraic axioms.  For instance associativity and coassociativity arise as follows: 
\begin{equation}
    \includegraphics[width=10.7cm,valign=b]{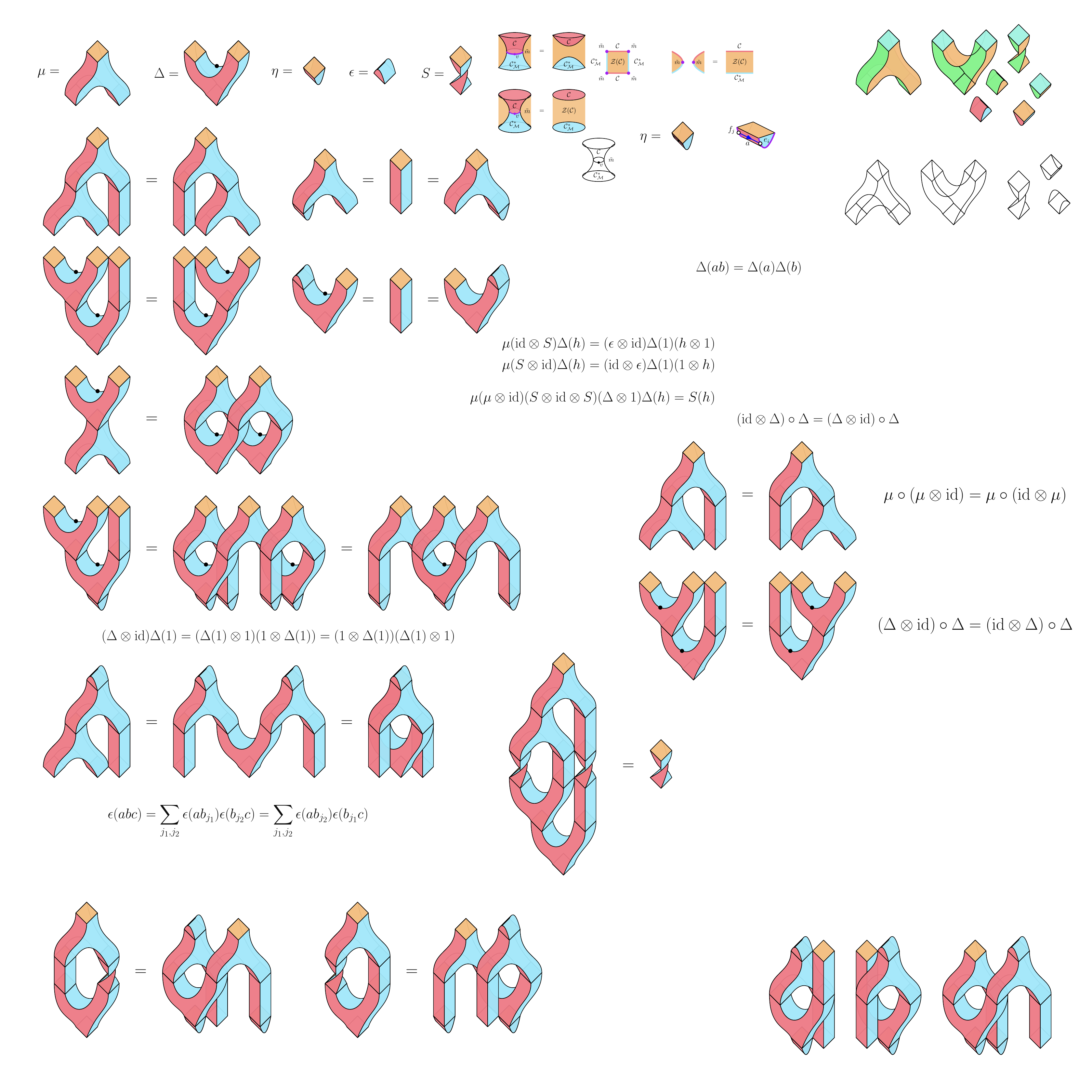}\;\;.
\end{equation}

\subsubsection{The Category of Representations}

Having characterized the properties of the strip algebra $\Str{\cC}{\cM}$, we then apply our understanding to characterize its representations.  Our goal is to deduce the complete representation category, including a list of all indecomposible representation and the fusion rules under tensor products.   A succinct summary of our result is that the category of representations is precisely the dual category $\cC^{*}_{\cM}$ which appeared above.  

To clarify this result we first recall that $\cC^{*}_{\cM}$ is the defined as the set of homomorphisms from $\mathcal{M}$ to itself which are linear over the fusion category $\cC$:
\begin{equation}
    \cC^{*}_{\cM}:=\Hom_{\cC}(\mathcal{M},\mathcal{M}).
\end{equation}
The condition of being linear over the fusion category $\cC$ means that an element $F:\cM\rightarrow \cM$ obeys the equation:
\begin{equation}
    a\otimes F(m) \cong F(a\otimes m),
\end{equation}
where $a$ is any symmetry line in the fusion category $\cC$.\footnote{More carefully, the functor $F$ is required to come with a coherent collection of natural isomorphisms \cite{ostrik2001module}.}  Graphically, this is represented as:\footnote{Here the circle labelled by $F$ denotes the action of $F$ rather than a junction in $\cM$.}
\begin{equation}
    \includegraphics[width=5.5cm,valign=b]{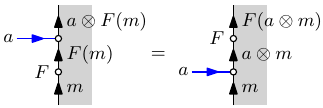}.
\end{equation}

The connection between such homomorphisms and representations of the strip algebra $\Str{\cC}{\cM}$ can be seen physically be identifying the operators that create states (solitons) in the open Hilbert space sectors $\mathcal{H}_{mn}$ defined in \eqref{openhilbdecomp}.  These are operators that reside at the end of topological lines (ray operators) in the Hilbert space sectors $\mathcal{H}_{a}$ on the circle.  For instance, in the diagram:
\begin{equation}
    \includegraphics[width=2.8cm,valign=m]{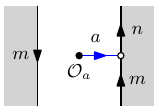},
\end{equation}
the operator $\mathcal{O}_{a}$ is one such operator which can create a soliton in $\mathcal{H}_{mn}$ out of the vacuum in sector $\mathcal{H}_{mm}$.  Notice that since $\mathcal{O}_{a}$ changes the boundary labels on one side of space it is naturally a map from $\mathcal{M}$ to $\mathcal{M}.$  Moreover we can act on $\mathcal{O}_{a}$ with the strip algebra to generate a new soliton creation operator $\tilde{\mathcal{O}}_{bab^{*}}$.  Schematically:
\begin{equation}
    \includegraphics[width=12cm,valign=b]{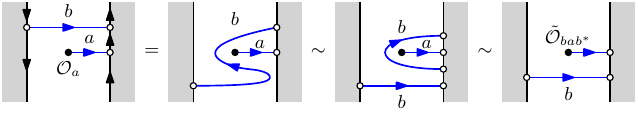}\;\;.
\end{equation}
The orbit of these soliton creation operators under the action of the strip algebra is then naturally a $\cC$-linear map from $\mathcal{M}$ to $\mathcal{M},$ i.e.\ an object in $\cC^{*}_{\cM}$, and also by construction a representation of $\Str{\cC}{\cM}$.  Thus, via the soliton creation operators, we see directly the correspondence between the dual category $\cC^{*}_{\cM}$ and the representations of the Strip algebra.

Given a general module $\cM$ describing boundary conditions, the dual category $\cC^{*}_{\cM}$ of representations is in principle computable.  However, the result is particularly simple in the case where $\cM$ is given by the so-called regular module.  Physically, this is the case where the fusion category symmetry $\cC$ is fully spontaneously broken by the boundary conditions (or vacua in the case of a massive RG flow).  Mathematically, this means that the boundary conditions in the regular module are in one to one correspondence with the elements of $\cC$ itself and that the bulk-boundary fusion is simply the bulk fusion algebra of $\cC$.  In this case it is known that the dual category of representations simply reproduces the initial fusion category.  In equations:
\begin{equation}
    \cC^{*}_{\cM}|_{\cM=\cC}\simeq \cC.
\end{equation}
Hence, when the fusion category symmetry is spontaneously broken, the resulting soliton representations are intrinsically determined by $\cC$ itself.

Our next task is pass from this abstract description of the representations of the strip algebra, to an explicit presentation of the irreducible representations and implied patterns of degeneracy.  We find that the representations can be elegantly encoded in quiver diagrams (i.e.\ graphs with directed arrows).  Concretely, given an object in $\alpha \in \cC^{*}_{\cM},$  and $n\in \mathcal{M}$ there are bulk boundary fusion rule:
\begin{equation}
    m\otimes \alpha =\sum_{m}\tilde{N}_{m\alpha}^{n}n, \hspace{.2in}\tilde{N}_{m\alpha}^{n}\in \mathbb{Z}_{\geq0}.
\end{equation}
We write this as a quiver where each node corresponds to an element in $\cM$ and there are $\tilde{N}_{m\alpha}^{n}$ arrows from $m$ to $n$.  In the associated representation, each such arrow then corresponds to a state (one-dimensional linear subspace) in the Hilbert space sector $\mathcal{H}_{mn}$.  Note also, that in the case of full spontaneous symmetry breaking, where $\cM$ is a regular module, these quiver diagrams are determined completely by the fusion rules of $\cC.$

In the context of applications to massive RG flows with elements of $\cM$ in correspondence to the clustering vacua these quiver diagrams may be interpreted in terms of particle-like excitations:
\begin{itemize}
    \item Arrows connecting distinct nodes are solitons interpolating between distinct vacua.
    \item Arrows connecting a node to itself are particle excitations above a single vacuum state.
\end{itemize}
In general, one finds that the resulting quivers contain both particles and solitons and hence imply novel mass degeneracies. 

As a simple example, consider the Fibonacci fusion category with one non-trivial simple object $W$ obeying the fusion rule:
\begin{equation}
    W\otimes W= 1+W.
\end{equation}
The unique indecomposible module of the Fibonacci fusion category symmetry is regular and hence the category of representations in this case is isomorphic again to the Fibonacci fusion category.  Correspondingly, there are two irreducible representations represented by the two quiver diagrams: 
\begin{itemize}
    \item One representation with two degenerate particle excitations:
    \begin{equation}
    \includegraphics[width=3cm,valign=m]{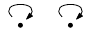}
\end{equation}
    \item One representation with three degenerate states: a particle, and two solitons:
    \begin{equation}
    \includegraphics[width=2.5cm,valign=m]{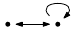}
\end{equation}
\end{itemize}
Analogous diagrams illustrating the soliton representations may be drawn for any fusion category $\cC$ and pattern $\cM$ of boundary symmetry action.  Several explicit constructions are carried out in Section \ref{sec:multiplets}.  It is worth noting that the connection between weak Hopf algebras and soliton representations was anticipated in the mathematics literature in the foundational work \cite{bohm1999weak, bohm2000weak}.  Furthermore, vacuum-soliton quiver diagrams have also been extensively studied in the context of supersymmetric (2,2) Landau Ginzburg models \cite{Cecotti:1992rm}, perhaps relating to the extensive categorical structure of solitons in these models \cite{Gaiotto:2015zna, Gaiotto:2015aoa}.  

Finally, in Section \ref{sec:select}, we carry out an analysis of the selection rules imposed on the S-matrix by the representation theory of the strip algebra. Here the key tools are: the existence of the tensor product of representations which exists because, as remarked above, $\Str{\cC}{\cM}$ is a weak Hopf algebra, as well as the $C^{*}$-structure.  The latter in particular implies that the strip algebra enjoys a version of Schur's lemma.  Hence a non-zero S-matrix element must involve a multiparticle in-state and multiparticle out-state which each contain a common representation of $\Str{\cC}{\cM}$.  Conversely, when no such common representation exists, the corresponding S-matrix element vanishes yielding a selection rule.

\section{Properties of the Strip Algebra}\label{sec:strip_alg_properties}

In this section we describe the algebraic properties of $\Str{\cC}{\cM}$ and explicitly analyze a collection of examples. We work in the setting of unitary QFTs and so both $\cC$ and $\cM$ are taken to be unitary throughout.

\subsection{Semisimplicity and \texorpdfstring{$C^*$}{C star}-Weak Hopf Structure}\label{sec:weak_hopf_defs}

The strip algebra $\Str{\cC}{\cM}$ is naturally both a $C^*$-algebra and a weak Hopf algebra. For discussion of both structures as well as their mutual compatibility see \cite{nikshych2000finite,szlachanyi2000finite}. A concise collection of definitions can also be found in Appendix \ref{app:WHA}. Briefly, a $C^*$-algebra is a complex algebra equipped with a map that is meant to mimic the properties of the adjoint in an operator algebra on a Hilbert space. This map is an involutive anti-linear, anti-homorphism of the algebra that is required to be compatible with the norm of the algebra. $C^*$-structures appear naturally in unitary theories. For $\Str{\cC}{\cM}$, the operation is defined by applying the so called $\dagger$-structure in $\cM$ to each boundary junction along with a weighting by quantum dimensions.\footnote{The $\dagger$-structure in a $\cC$-module category $\cM$ defines the action of orientation reversal on junctions. See Appendix \ref{app:fusion_cat} and references therein for further discussion. See this appendix for the definition of boundary quantum dimensions as well.} This produces the element
\begin{equation}
    \includegraphics[width=8.7cm,valign=m]{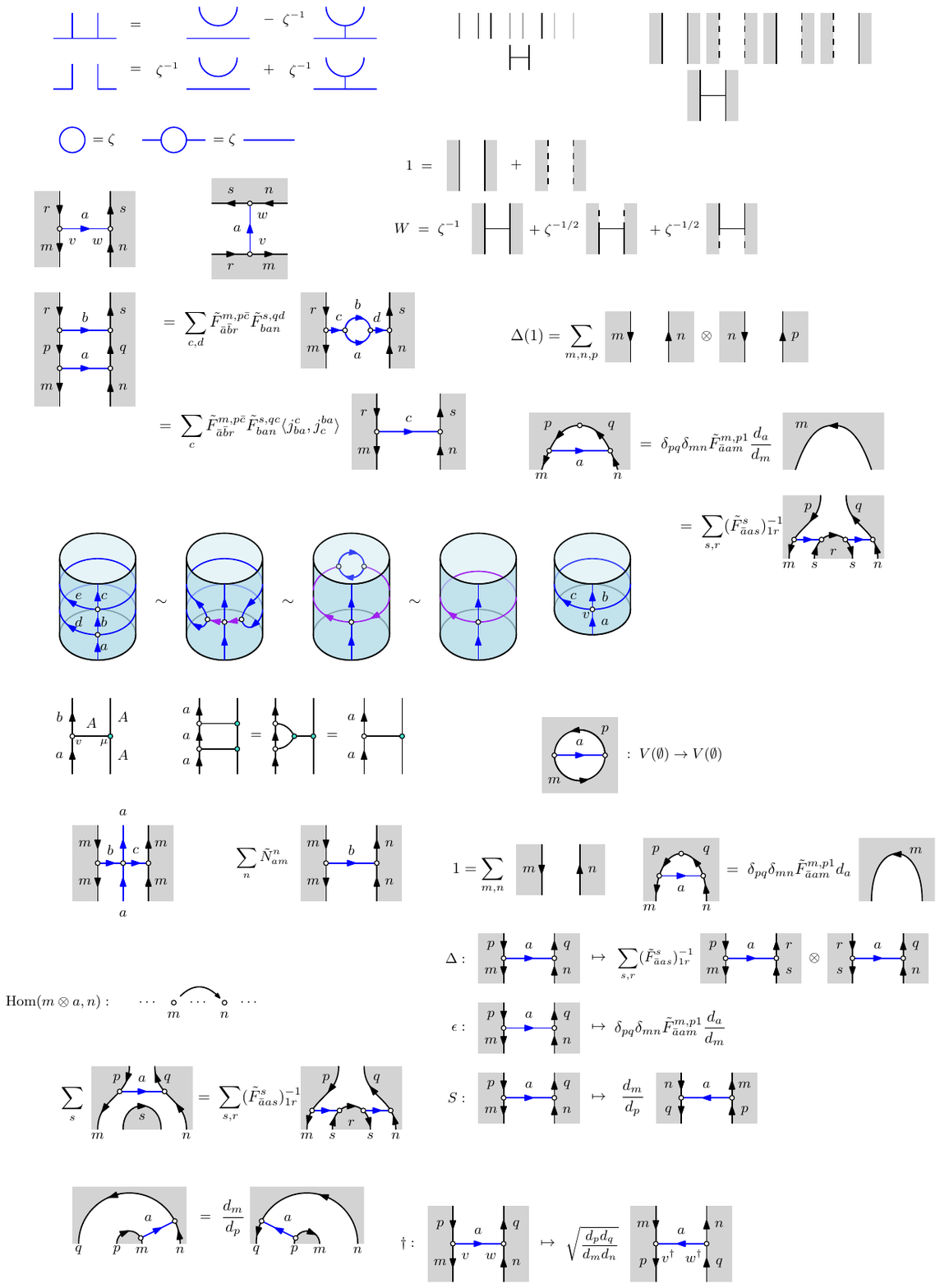}.
\end{equation}
The specific prefactor is chosen so that the consistency conditions reviewed in Appendix \ref{app:WHA} will be satisfied for the weak Hopf algebra maps that we now define.

$C^*$-algebras are semisimple and so their representation theory is well behaved. Among other things, their representations are completely reducible and as algebras they decompose into matrix algebras over their irreducible representations
\begin{equation}
    \Str{\cC}{\cM} \cong \bigoplus_i \End(W_i), \quad W_i \; \textrm{irreducible}.
\end{equation}
These are of course familiar properties in the context of finite group theory. In fact, we will see later that the group algebra of a finite group $\C[G]$ can be realized as a strip algebra. Such properties make both $\Str{\cC}{\cM}$ and its representation theory reasonably tractable to study directly by hand.

The structure of a weak Hopf algebra \cite{bohm1999weak} may be less familiar to the reader. Here we provide a gentle introduction with more details provided in Appendix \ref{app:WHA}.
Weak Hopf algebras provide the most general context for obtaining representations which have notions of tensor products and duals \cite{hayashi1999canonical,ostrik2001module}. Physically, these properties correspond to admitting a notion of action on multi-particle states and having charge conjugates. Therefore, they provide the most general setting for the study of selection rules in 2 dimensions.
To define these notions for a complex algebra, $H$, three additional linear maps are needed\footnote{Since there will be many instances of the symbol $\otimes$ being used throughout this paper, we will more carefully use $\otimes_\C$ to denote the tensor product of complex vector spaces.}
\begin{equation}
    \Delta : H \to H \otimes_\C H,  \qquad \epsilon : H \to \C, \qquad S : H \to H,
\end{equation}
which are called the coproduct, counit, and antipode respectively. To ensure the proper behavior of representations, these maps are required to satisfy a collection of compatibility conditions which we review in Appendix \ref{app:WHA}.
Algebraically, the behavior of tensor product and dual representations are governed by the following two properties: $\Delta$ is required to be a (not necessarily unit preserving) algebra homomorphism
\begin{equation}
    \Delta(ab) = \Delta(a)\Delta(b),
\end{equation}
which implies that $\Delta(1)$ is an idempote, and $S$ is required to be an anti-algebra homomorphism
\begin{equation}
    S(ab) = S(b)S(a) \qquad S(1) = 1.
\end{equation}

Given two representations, $R_1$ and $R_2$, of $H$, their tensor product is defined as follows. Using $\Delta$, there is a natural action of $H$ on $R_1\otimes_\C R_2$. However, since $\Delta(1)$ is not necessarily the identity in $H \otimes_\C H$, this may not be a representation of $H$, as we require the unit $1\in H$ acts on a representation as an identity. This failure though is easily remedied. Since the element $\Delta(1)$ is idempotent, it defines a projection operator onto the subspace of $R_1 \otimes_\C R_2$ where $\Delta(1)$ acts as the identity. It is \textit{this} subspace that is defined as the tensor product of the representations
\begin{equation}\label{eq:tensor_product_def}
    R_1 \otimes R_2 := \Delta(1)(R_1 \otimes_\C R_2).
\end{equation}
The reader may rightfully complain that the motivation for this construction appears rather opaque. However, it can be  naturally understood in the context of $\Str{\cC}{\cM}$, where, as will be shown momentarily, it reflects the physics of solitons.

The definition of duals is much more clear: taking $h \in H$ and $\phi \in \Hom_\C(R,\C) =: R^\vee$, $H$ acts on $\phi$ as
\begin{equation}
    (h\phi)(v) = \phi(S(h)v), \quad v \in R.
\end{equation}

We have already defined the multiplication in $\Str{\cC}{\cM}$. The coproduct, counit, and antipode can be easily motivated by thinking of the strip algebra as acting in the $2d$ TQFT which has $\cM$ as its category of topological boundary conditions. Another approach for constructing $\Str{\cC}{\cM}$ as a weak Hopf algebra using a $3d$ TQFT will be demonstrated in Section \ref{sec:WHA_from_symtft}. Using the $2d$ approach, the coproduct is defined by the following operation\footnote{Generally the boundary junction vector spaces may be more than one dimensional, meaning the boundary $\tilde{F}$-symbol should have additional indices after a basis choice. To simplify notation we suppress such indices.}
\begin{equation}\label{eq:strip_coprod}
    \includegraphics[width=9.5cm, valign=m]{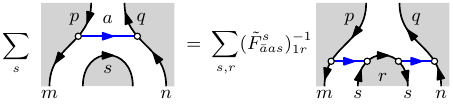},
\end{equation}
where a boundary crossing is applied to join the line to the center. The counit is defined as the coefficient obtained from contracting the line on a cap and weighting by a boundary quantum dimension $d_m$,\footnote{See Appendix \ref{app:fusion_cat}.}
\begin{equation}
    \includegraphics[width=9cm, valign=m]{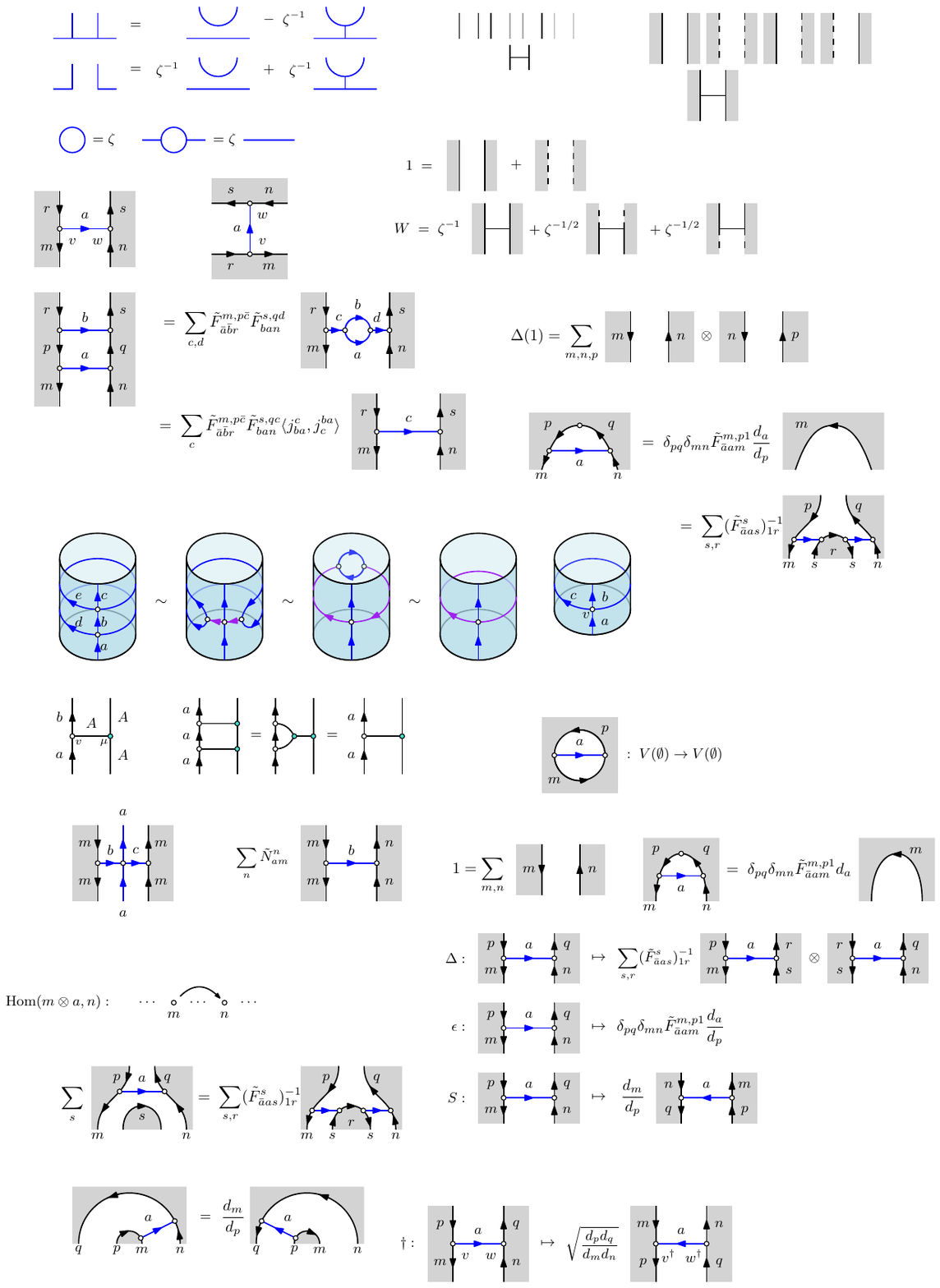},
\end{equation}
where $d_a$ is the quantum dimension of $a$. This formula uses the canonically normalized topological junction of a simple bulk line defect as a basis element. Finally, the antipode it defined as the manipulation\footnote{The present authors thank the authors of \cite{AliAhmad:2025bnd} for noting a missing factor in this formula as given in the initial version of this paper.}
\begin{equation}
    \includegraphics[width=7.2cm, valign=m]{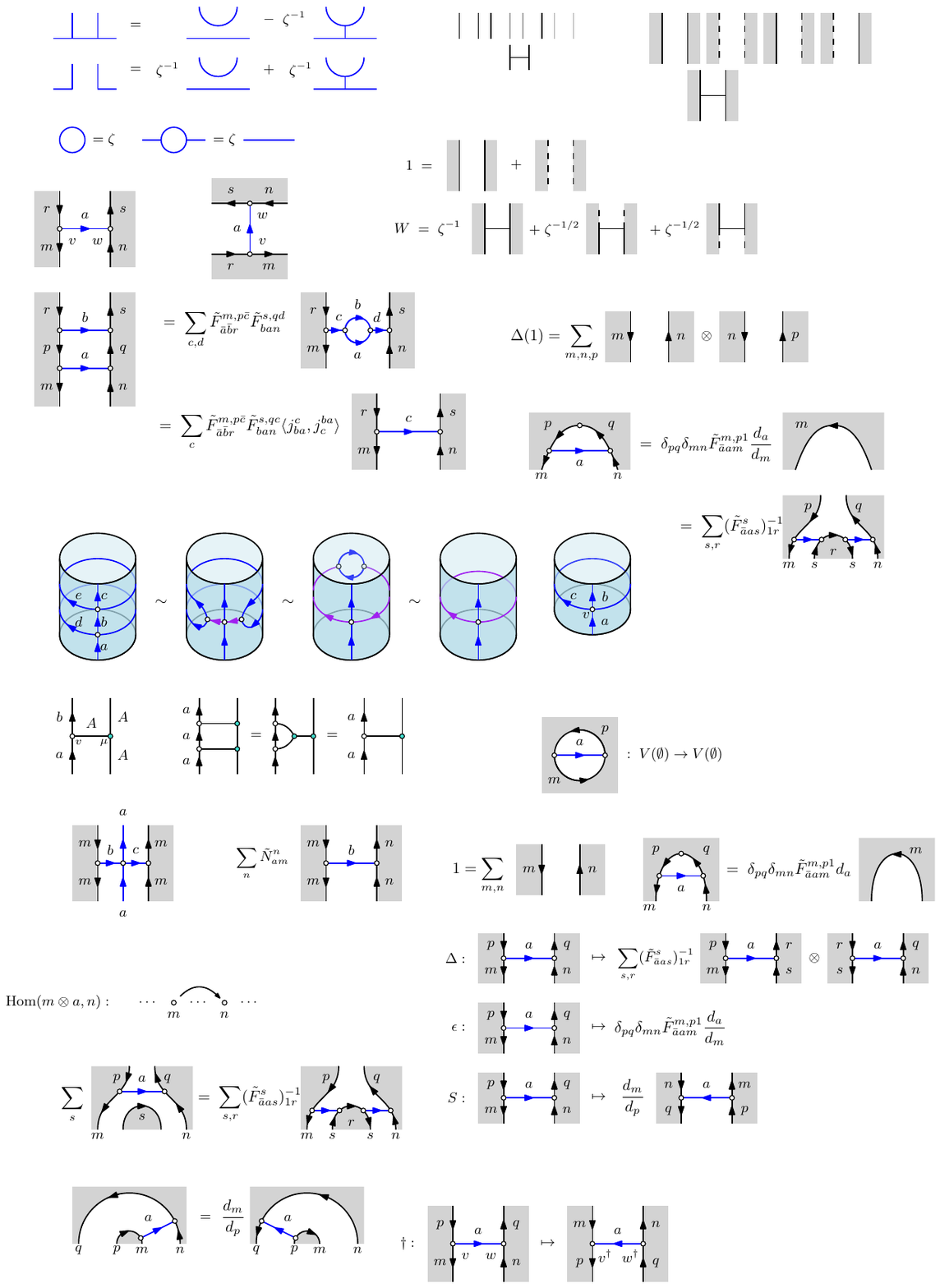}
\end{equation}
which naturally exchanges an action on a vector space for an action on its dual. In summary, the weak Hopf algebra structure of $\Str{\cC}{\cM}$ is defined by the collection of maps
\begin{equation}\label{eq:weak_hopf_operations}
    \includegraphics[width=12.5cm,valign=m]{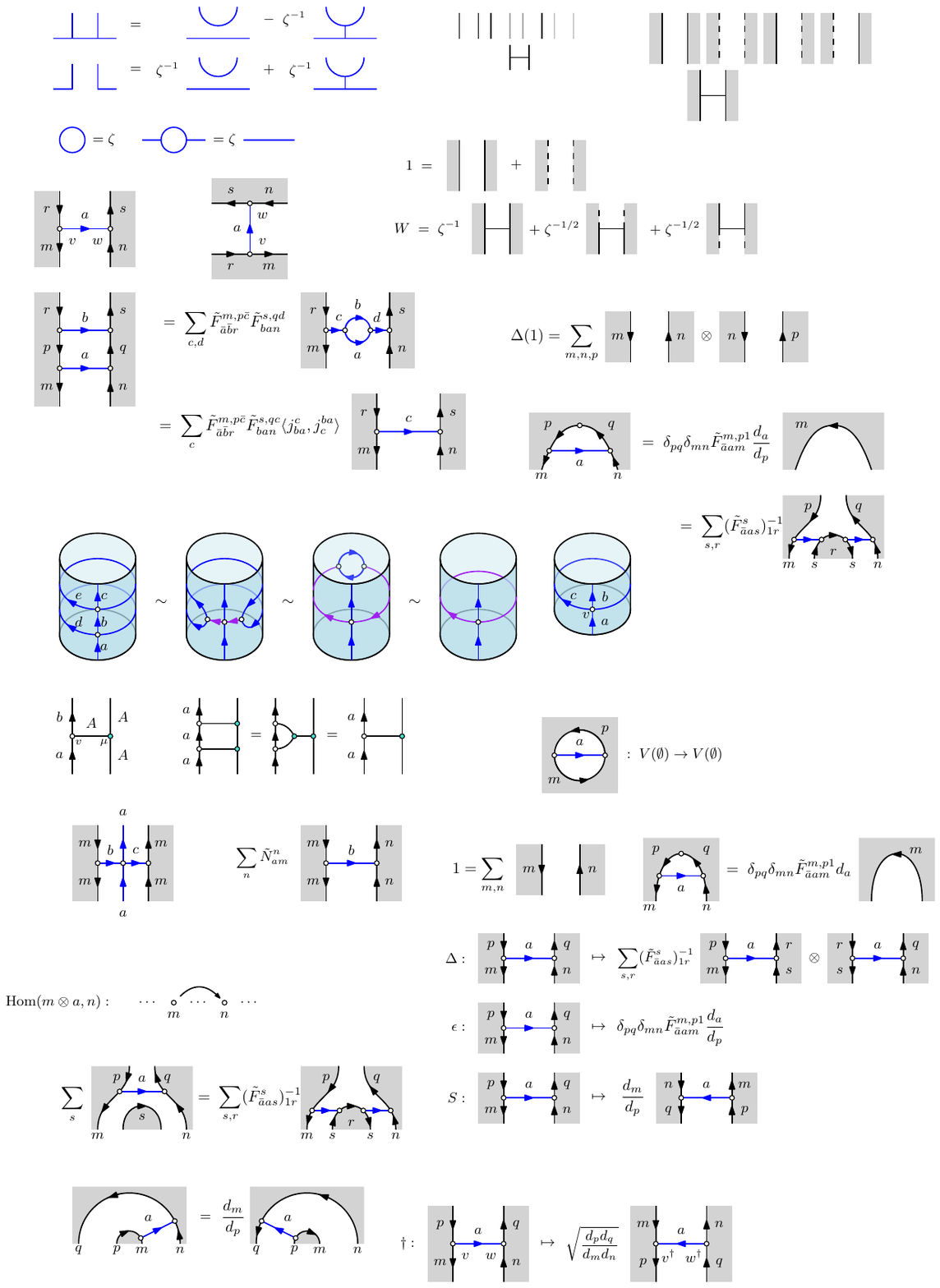}.
\end{equation}
The fact that these maps satisfy the compatibility conditions required for a weak Hopf algebra can be checked directly either symbolically or using topological manipulations. We delay their verification to Section \ref{sec:WHA_from_symtft} where they are more easily verified from the the $3d$ perspective. This collection of maps defines the same weak Hopf algebra as in \cite{Kitaev_2012}.

With the above coproduct, the tensor products of representations of $\Str{\cC}{\cM}$ is easily understood. The unit in $\Str{\cC}{\cM}$ is the element
\begin{equation}
    \includegraphics[width=3.7cm, valign=m]{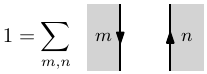}.
\end{equation}
From \eqref{eq:weak_hopf_operations}, we see that
\begin{equation}
    \includegraphics[width=7.5cm,valign=m]{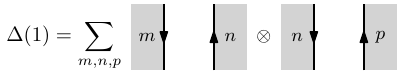},
\end{equation}
which is the diagonal component of $1 \otimes 1$ in $\Str{\cC}{\cM} \otimes_\C \Str{\cC}{\cM}$. The projection defined by $\Delta(1)$ therefore plays the role of requiring the two adjacent boundaries of a tensor product to be equal. The physical interpretation of this is clear: on $\R$, a multiparticle state must have the same vacuum interpolating between localized excitations. From this point of view, the tensor product structure of a weak Hopf algebra's representations is entirely natural.

The representation theory of the strip algebra captures how topological defect lines organize the states of a quantum field theory having boundary conditions. Because of its $C^*$-weak Hopf structure, it will enforce degeneracies and impose selection rules. The remainder of this paper aims to both demonstrate how to efficiently analyze the implications of the existence non-invertible symmetries as well as to illustrate interesting and representative examples of novel behavior non-invertible symmetries can enforce. 

\subsection{Relation to the Tube Algebra}
To make contact with the existing literature, it is useful to note that $\Tub(\cC)$ is closely related to a strip algebra. Geometrically this follows from the folding trick. Folding the cylinder across a central plane gives a doubled theory with symmetry $\cC \boxtimes \cC^{op}$ now living on a strip\footnote{$\cC^{op}$ is an orientation reversal of $\cC$. It is the same as $\cC$ but with opposite order of fusion: $a \otimes^{op} b = b \otimes a$.} 
\begin{equation}
    \includegraphics[width=9cm,valign=m]{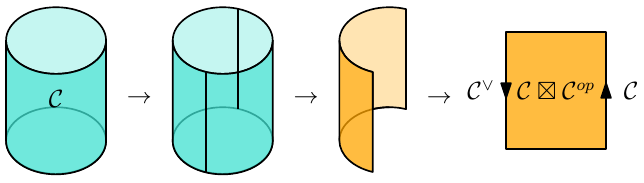}.
\end{equation}

Placing two junctions labelling elements of $\Tub(\cC)$ along the intersection of this plane and the cylinder and then folding, a diagram is obtained with a bulk line valued in $\cC \boxtimes \cC^{op}$, boundaries labelled by $\cC$, and junctions labelled by the action of $\cC\boxtimes \cC^{op}$ on $\cC$, where the action of $\cC^{op}$ on $\cC$ is by fusion on the right
\begin{equation}\label{eq:strip_from_tube_morph}
    \includegraphics[width=6.3cm,valign=m]{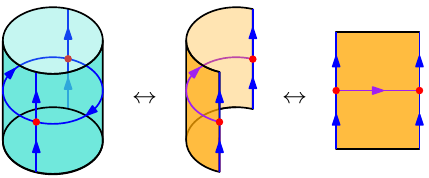}.
\end{equation}
The diagram on the right-hand side is precisely an element of $\Str{\cC\boxtimes \cC^{op}}{\cC}$. The diagram on the left-hand side is an element of $\Tub(\cC)$ only after reducing it to have a single vertical topological line. Since the strip algebra treats such configurations as distinct, we see that $\Tub(\cC)$ is naturally a quotient of $\Str{\cC\boxtimes \cC^{op}}{\cC}$
\begin{equation}
    \Str{\cC\boxtimes \cC^{op}}{\cC} \to \Tub(\cC).
\end{equation}
Conversely, $\Tub(\cC)$ admits two natural inclusions into the strip algebra defined by taking either vertical line in \eqref{eq:strip_from_tube_morph} to be the identity line in $\cC$. Therefore, $\Tub(\cC)$ is also (non-canonically) a subalgebra of $\Str{\cC\boxtimes \cC^{op}}{\cC}$
\begin{equation}
    \Tub(\cC) \to \Str{\cC\boxtimes \cC^{op}}{\cC}.
\end{equation}
Generally neither subalgebra closes under the comultiplication however. Later we will see another manifestation of these relationships at the level of the representations of each algebra.

\subsection{Examples}
In this section we explicitly study $\Str{\cC}{\cM}$ for a few representative classes of examples.
These include both invertible cases and non-invertible cases. For invertible (group) symmetry, we start from the case when the symmetry is unbroken in the infrared, obtaining the corresponding group ring $\C [G]$. The spontaneously broken case and anomalous case follow. For non-invertible symmetries, we study the cases of the Fibonacci category and the Tambara-Yamagami category.  The last part of this section contains a general method of analysis at its end.

\subsubsection{\texorpdfstring{$G$}{G} Symmetry: Unbroken}\label{sec:strip_grp_unbroken}
Let's begin by considering the structure of $\Str{\cC}{\cM}$ for group symmetry. Taking $G$ a finite group, this means $\cC = \Vec_{G}$, the category of $G$-graded vector spaces. A simple object of the category is labeled by an element $g$ in $G$, representing the topological defect implementing the symmetry action by $g$. This category describes a non-anomalous $G$ symmetry. The symmetry-preserving phase of the symmetry is specified by the module category $\Vec$, the category of (finite dimensional) vector spaces.\footnote{
    In general, when a fusion category $\cC$ admits $\Vec$ as its module category, the functor $\cC \to \Vec$, $a \mapsto a \otimes \C$ is monoidal and called a fiber functor. $\cC$ admitting a fiber functor means that $\cC$ can act on the trivial vacuum, i.e.\ it is non-anomalous \cite{Thorngren:2019iar,Thorngren:2020yht,Bhardwaj:2023idu,Zhang:2023wlu}}
 This has a few interpretations \cite{Choi_2023}. Most literally, it means that $G$ acts trivially on the boundary condition and so preserves the corresponding vacuum on $\R$. 

We can chose junctions so that
\begin{equation}\label{eq:grp_unbroken_loops}
    \includegraphics[width=4cm, valign=m]{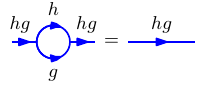}.
\end{equation}
The boundary junction normalizations can also be fixed so that the boundary $\tilde{F}$-symbols are trivial
\begin{equation}\label{eq:grp_unbroken_associator}
    \includegraphics[width=4.5cm]{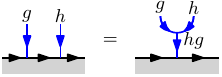}.
\end{equation}
Note that the boundary has no label in this case because there is only one vacuum.
The algebra $\Str{\Vec_G}{\Vec}$ has a basis of elements
\begin{equation}
    \includegraphics[width=1.7cm, valign=m]{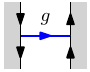}, \quad g \in G.
\end{equation}
Using \eqref{eq:grp_unbroken_loops} and \eqref{eq:grp_unbroken_associator}, their multiplication is
\begin{equation}
    \includegraphics[width=7cm]{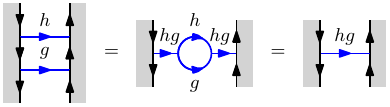}.
\end{equation}

Therefore we find that
\begin{equation}
    \Str{\Vec_G}{\Vec} = \C[G].
\end{equation}
The coproduct of $\Str{\Vec_G}{\Vec}$ is simply the diagonal map and the antipode is inversion
\begin{equation}
    \Delta(g) = g \otimes g \qquad S(g) = g^{-1}.
\end{equation}
In particular, in this case the coproduct and the unit satisfies
\begin{equation}
    \Delta(1) = 1 \otimes 1.
\end{equation}
A weak Hopf algebra satisfying this condition is called Hopf algebra.\footnote{Historically, weak Hopf algebras were introduced as a generalization of Hopf algebras by relaxing this axiom. This additional condition to be a Hopf algebra is satisfied whenever the symmetry is not spontaneously broken.} 
Therefore, $\Str{\Vec_G}{\Vec}$ is $\C[G]$ equipped with its canonical Hopf algebra structure. 

The formula for the $C^*$-structure on $\C[G]$ is given by
\begin{equation}
    g^* = g^{-1},
\end{equation}
together with the norm
\begin{equation}
    \lVert  g \rVert  = 1.
\end{equation}
Note that the $*$-structure is antilinear, i.e.\ for a complex number $c$, it acts on $cg \in \C[G]$ as
\begin{equation}
    (c g)^* = c^* g^{-1}.
\end{equation}

In particular, the associated representation theory is simply $\Rep(G)$: the standard (unitary) representations of a finite group. This is the expected result and recovers the familiar action of a finite group in a non-spontaneously broken phase. 

\subsubsection{\texorpdfstring{$G$}{G} Symmetry: Spontaneously Broken}\label{sec:strip_G_SSB}
Similarly, we can consider a spontaneously broken phase of $G$. The fully spontaneously broken phase corresponds to choosing the regular module category of $\Vec_G$. This is because each boundary, and so vacuum on $\R$, is labelled by an element of $G$ and acted upon by the action of $G$ on itself. 

The only thing that differs from the unbroken case is that we must now keep track of boundary labels. We can use the same normalization for junctions as \eqref{eq:grp_unbroken_loops} and normalize boundary junctions so that the $\tilde{F}$-symbol is trivial
\begin{equation}
    \includegraphics[width=4.3cm]{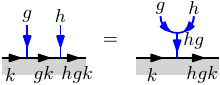}.
\end{equation}
The algebra $\Str{\Vec_G}{\Vec_G}$ has a basis of elements
\begin{equation}
    \includegraphics[width=1.7cm, valign=m]{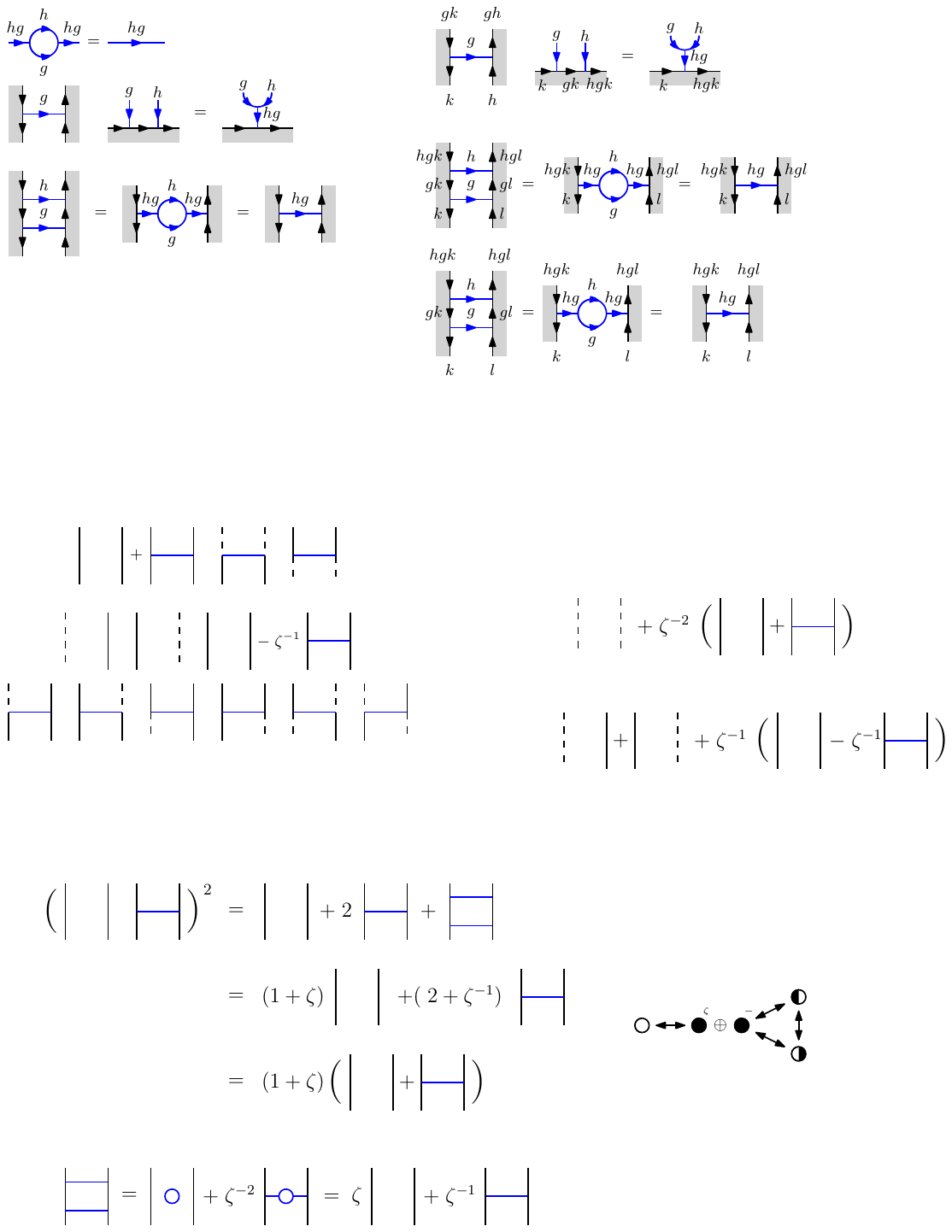},
\end{equation}
whose non-zero multiplications are
\begin{equation}
    \includegraphics[width=7cm]{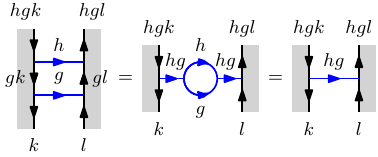}.
\end{equation}

This algebra is more complicated than the unbroken case. However its structure becomes clear by observing that elements of the form
\begin{equation}
    \includegraphics[width=1.7cm, valign=m]{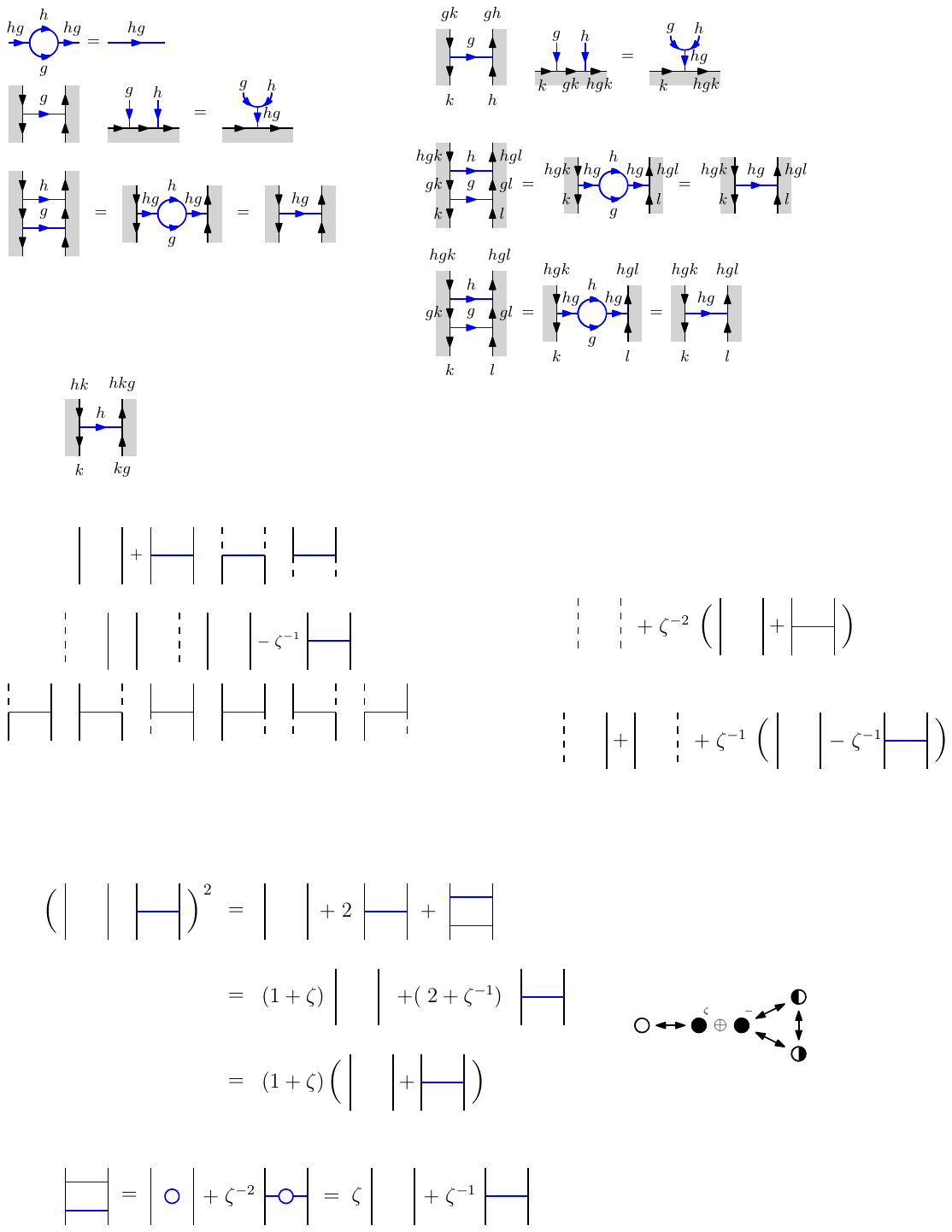}, \quad k, h \in G.
\end{equation}
for fixed $g\in G$ form subalgebras $A_g \subset \Str{\Vec_G}{\Vec_G}$. Because $G$ acts both freely and transitively on itself, these algebras are simple and provide a decomposition of the algebra
\begin{equation}
    \Str{\Vec_G}{\Vec_G} = \bigoplus_{g\in G} A_g.
\end{equation}
The algebras $A_g$ are examples of groupoid algebras, the generalization of the group algebra of a group to groupoids \cite{nikshych2000finite}. 

Since each $A_g$ is simple, it is a matrix algebra and one can check that $|A_g| = |G|^2$. It follows from the general theory of finite dimensional semisimple algebras that the irreducible representations of $\Str{\Vec_G}{\Vec_G}$ have dimension $G$ and are labelled by elements $g \in G$. The coproduct in $\Str{\Vec_G}{\Vec_G}$ defines the tensor product of irreducible representations to be multiplicative over the $g$ index
\begin{equation}
    V_g \otimes V_h = V_{gh}.
\end{equation}
This is all as it should be: physically, it is simply the familiar statement that in a fully spontaneously broken phase, the action of $G$ permutes all vacua and moreover that all vacua are identical. It is also consistent with the usual thinking of solitons as labelled by group elements with their bound states being labelled by the group product.

Note that, while the representation $V_1$ is the unit in the tensor product:
\begin{equation}
    V_1 \otimes V_g = V_{g},
\end{equation}
$V_1$ is also $|G|$-dimensional. This is the point where the representation theory of weak Hopf algebras crucially differs from that of Hopf algebras, and in particular that of finite groups. For the latter, the unit in the tensor product is the trivial, i.e.\ one-dimensional, representation on which the algebra or group acts trivially.

\subsubsection{Anomalous \texorpdfstring{$G$}{G} Symmetry}\label{sec:strip_anomalous_G}
It is natural to ask how anomalies present themselves in the structure $\Str{\cC}{\cM}$. For example, in the study of the tube algebra it is known that the anomaly of a $G$-symmetry can be manifested as a projective representation in twisted sectors of $G$ \cite{Bartsch:2023wvv}. We will see that the analogous question for the strip algebra is subtle. 

Recall that a symmetry $G$ with anomaly $[\omega] \in H^3(G,U(1))$ is described by the category $\Vec_G^\omega$, which has objects and fusions labeled by $G$ and $F$-symbol give by a representative cocycle $\omega$. If $[\omega]$ is non-zero, then it is known that the theory does not admit a symmetry preserving boundary condition \cite{Thorngren:2020yht,Bhardwaj:2017xup}. However, the fully spontaneously broken phase always exists and we will consider this here. The new feature of the analysis due to the anomaly is that there are now non-trivial crossings 
\begin{equation}
    \includegraphics[width=6.2cm, valign=m]{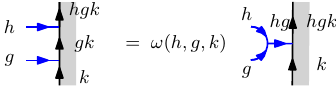}.
\end{equation}
The presence of the anomaly obstructs us from being able to choose a basis of junctions satisfying \eqref{eq:grp_unbroken_loops}. Instead, the given basis of junctions will satisfy
\begin{equation}
    \includegraphics[width=5.7cm, valign=m]{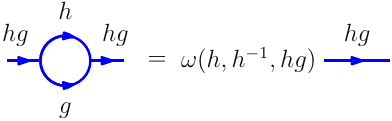}~.
\end{equation}
As a result, the multiplication in $\Str{\Vec_G^\omega}{\Vec_G^\omega}$ is modified by an overall phase
\begin{equation}
    \includegraphics[width=10.5cm, valign=m]{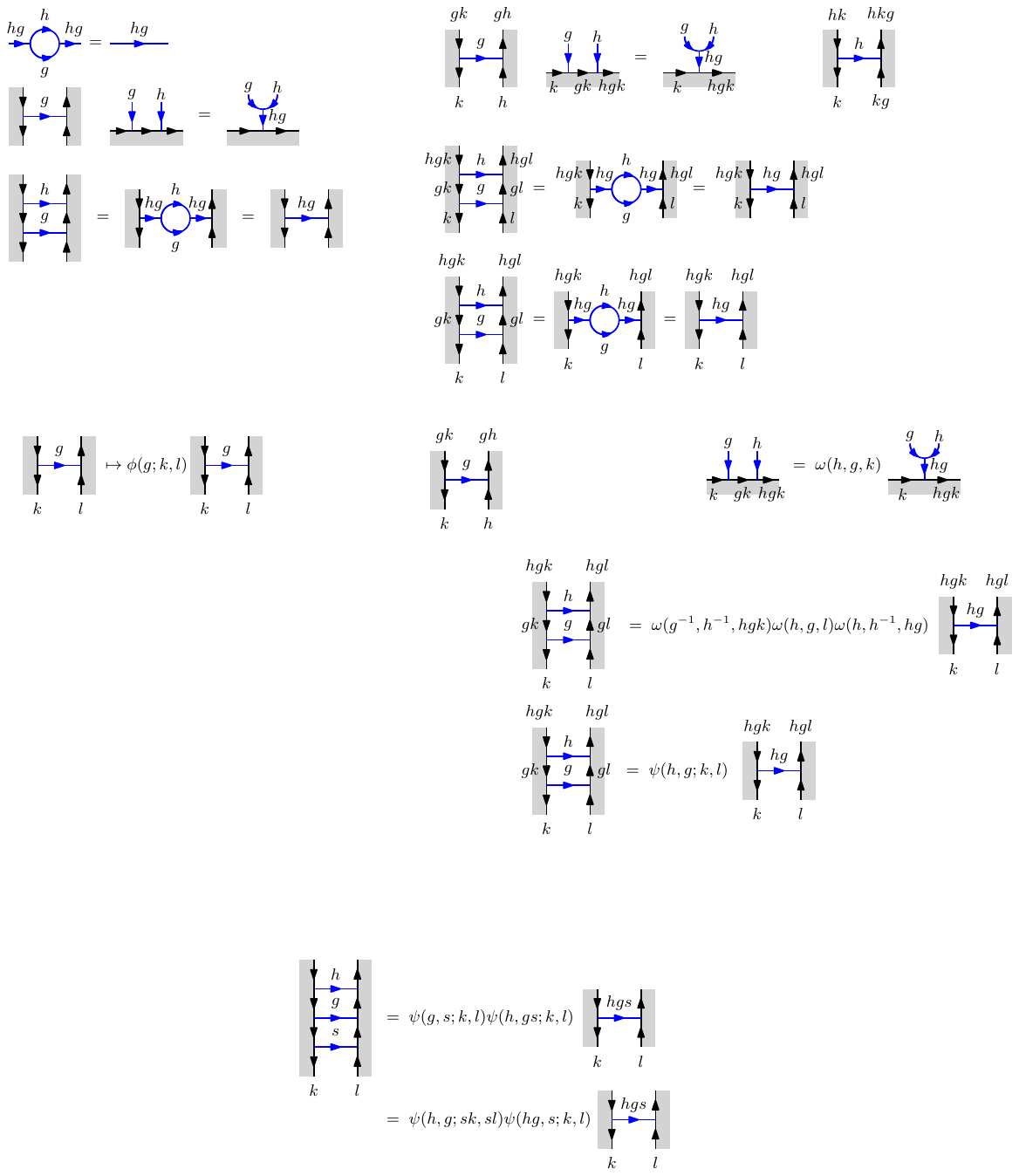}.
\end{equation}
It is useful to  consider what kind of structure this phase is. If we consider a general phase
\begin{equation}\label{eq:groupoid_cohom_phase}
    \includegraphics[width=6.5cm,valign=m]{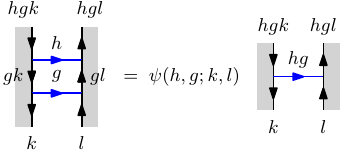}
\end{equation}
then associativity requires
\begin{equation}
    \includegraphics[width=8.5cm,valign=m]{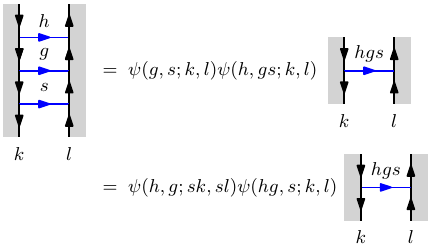},
\end{equation}
or equivalently that
\begin{equation}\label{eq:bdry_2cocycle_condition}
    \frac{\psi(g,s;k,l)\psi(h,gs;k,l) }{\psi(hg,s;k,l)\psi(h,g;sk,sl)} = 1.
\end{equation}

This condition is a mild generalization of the cocycle condition for group cohomology, differing due to the change of the boundary label. Since we can multiply basis elements by phases, the phase \eqref{eq:groupoid_cohom_phase} is defined up to multiplication by a term 
\begin{equation}
    \psi(h,g;k,l) \mapsto \psi(h,g;k,l)\frac{\phi(g;k,l)\phi(h;gk,gl)}{\phi(hg;k,l)}.
\end{equation}
This factor resembles a generalization of a coboundary
\begin{equation}
    \delta \phi(h,g;k,l) = \frac{\phi(g;k,l)\phi(h;gk,gl)}{\phi(hg;k,l)},
\end{equation}
though again differs by a change of boundary label. These equations suggest that $\psi$ is an object in a cohomology theory. This is indeed correct, $\psi$ is a groupoid cohomology class. We will not delve into this subject because, as we will see in a moment, the theory trivializes the case at hand. The curious reader, however, may consult \cite{Bartsch:2023wvv} for a discussion in the context of the tube algebra.

The phase factors, $\psi$, can always be trivialized in $\Str{\Vec_G^\omega}{\Vec_G^\omega}$. To see this, note that using the cocycle condition \eqref{eq:bdry_2cocycle_condition}, all phases can be fixed in terms of a single choice of element 
\begin{equation}
     \psi(h,g;s,sl) = \frac{\psi(g,s;1,l)\psi(h,gs;1,l) }{\psi(hg,s;1,l)}.
\end{equation}
This follows from the transitivity of the action of $G$ on itself. Because $G$ acts freely on itself, the values of $\psi(h,g;1,l)$ for different $l$ are not related by the cocycle condition. Therefore every allowed collection of phases is specified by a choice of $\psi(h,g;1,l)$ for every $l\in G$. In terms of these phases, the condition for $\psi$ to be trivializable is then
\begin{equation}
    \frac{\psi(h,s;1,l)\psi(g,hs;1,l) }{\psi(gh,s;1,l)}\frac{\phi(g;s,sl)\phi(h;gs,gsl)}{\phi(hg;s,sl)} = 1
\end{equation}
which has solution
\begin{equation}
    \phi(h;s,sl) = \psi^{-1}(h,s;1,l).
\end{equation}

This demonstrates that, as algebras
\begin{equation}
    \Str{\Vec_G^\omega}{\Vec_G^\omega} \cong \Str{\Vec_G}{\Vec_G}.
\end{equation}
Importantly, however, both algebras do differ in their weak Hopf structure. Specifically, the coproduct of algebras for different $[\omega]$ are inequivalent under an algebra isomorphism, which manifests itself as a difference in the tensor products of both representations. We will see later that this distinction is mathematically the same as the distinction between the categories $\Vec_G$ and $\Vec_G^\omega$ themselves. In particular, when one considers the entire weak Hopf structure, the strip algebra is sensitive to the anomaly of $G$. A similar analysis may be performed for boundary conditions preserving a non-anomalous subgroup $H$ of $G$ with analogous conclusions holding.

\subsubsection{Fibonacci}\label{sec:strip_fib}
We now shift to studying examples of $\Str{\cC}{\cM}$ for non-invertible symmetry. The simplest such example is the Fibonacci fusion category, $\Fib$, which has two simple objects, $\{1,W\}$, with fusion
\[W^2 = 1 + W.\]
This symmetry is realized in the Golden Chain and the Tricritical Ising Model \cite{Chang_2019}. To simplify diagrammatics we will represent $W$ with an unlabelled line and $1$ by the absence of a line. A basis of junction vectors can be fixed by the normalization conventions
\begin{equation}\label{eq:fib_loop_norm}
    \includegraphics[width=7 cm, valign=m]{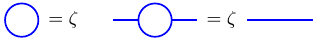}
\end{equation}
where 
\begin{equation}
    \zeta = \frac{1+\sqrt{5}}{2}, \quad \zeta^2 = \zeta + 1.
\end{equation}
In this basis, the non-trivial $F$-symbols are\footnote{``Non-trivial'' meaning that all other $F$-symbols that can be non-zero are $1$.} \cite{Huang:2021zvu}
\begin{equation}
    F_{WWW}^W = \begin{pmatrix} \zeta^{-1} & \zeta^{-1} \\ 1 & -\zeta^{-1} \end{pmatrix}
\end{equation}
so that diagrammatically the crossings 
\begin{equation}\label{eq:fib_associator}
    \includegraphics[width=9cm, valign=m]{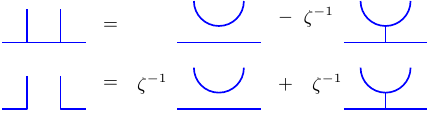}
\end{equation}
hold.

$\Fib$ is known to have a single indecomposable module category corresponding to its regular module category, $\Fib$. Physically, this is the statement that $\Fib$ is always spontaneously broken. The strip algebra $\Str{\Fib}{\Fib}$ has a basis of $13$ elements
\begin{equation}
    \includegraphics[width=8.5cm]{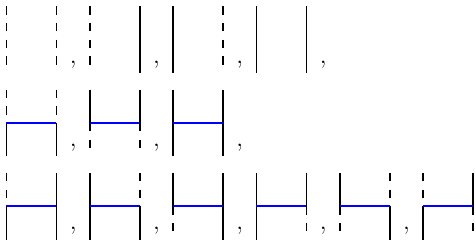}
\end{equation}
where, for notational simplicity, we now suppress the grayed region as the boundary is clear.

Since $\Str{\Fib}{\Fib}$ is semisimple, it will be the easiest to understand by identifying its decomposition into matrix algebras. This can be done directly in this example. The only factorization of $13$ into a sum of squares is $13 = 4 + 9$.\footnote{In principle we should worry about factors of $1$, but we will see explicitly at the end that this is ruled out since there are not enough invertible elements in this algebra.} Therefore $\Str{\Fib}{\Fib}$ decomposes into two simple subalgebras
\begin{equation}
    \Str{\Fib}{\Fib} \cong A_1 \oplus A_2, \quad \dim(A_1) = 4, \; \dim(A_2) = 9.
\end{equation}
Note that from this we can deduce that $\Str{\Fib}{\Fib}$ has two irreducible representations of dimensions $2$ and $3$. We will later recover this fact by a more indirect approach.

Our goal then is to find two collections of maps which are mutually zero when multiplied and form closed algebras between themselves. This is not too bad here. Without doing any calculation it is clear that
\begin{equation}
    \includegraphics[width=11.5cm]{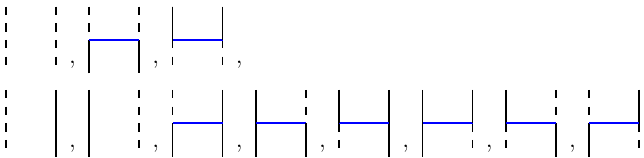}
\end{equation}
are two collections which mutually multiply to zero. The subalgebras generated by these collections are
\begin{equation}
    A_1 : \quad \includegraphics[width=7cm, valign=b]{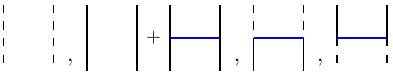}
\end{equation}
and
\begin{equation}
    A_2 : \quad \includegraphics[width=8.5cm, valign=b]{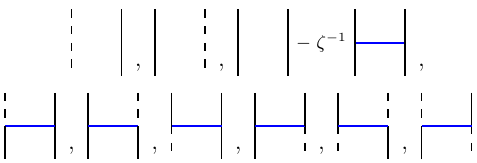}\;.
\end{equation}
This explicit expression for the subalgebras shows what boundary sectors assemble into a multiplet under the action of $\Str{\Fib}{\Fib}$. In particular, it reproduces the mixed particle and soliton multiplet discussed in \cite{Cordova:2024vsq}.

It is interesting to note that $\Str{\Fib}{\Fib}$ has a non-trivial subalgebra which fixes a pair of boundary conditions. Using \eqref{eq:fib_loop_norm} and $\eqref{eq:fib_associator}$, one can compute
\begin{equation}\label{eq:fib_sub_alg_WW}
    \includegraphics[width=9.5cm]{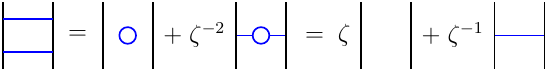}\;.
\end{equation}
This algebra resembles the fusion in $\Fib$, but with spurious factors of the quantum dimension $\zeta$ appearing from the boundary $\tilde{F}$-symbol. Physically, this means that the vacuum labelled by $W$ supports a modified version of the $\Fib$ fusion as a symmetry. The appearance of such modified fusion algebras is a generic feature when the symmetry of the theory contains a non-invertible symmetry.

The identity in each subalgebra is 
\begin{equation}
    1_{A_1} = \includegraphics[width=5.5cm,valign=m]{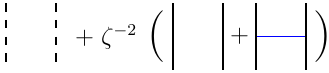}\;,
\end{equation}
\begin{equation}
    1_{A_2} = \includegraphics[width=7.3cm,valign=m]{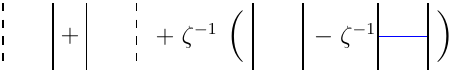}\;,
\end{equation}
which also define the projector into each subalgebra. This again makes explicit the non-trivial mixing of boundary sectors under the action of $\Str{\Fib}{\Fib}$. This provides a complete understanding of $\Str{\Fib}{\Fib}$ as an algebra. 

\subsubsection{Fibonacci: Reducible Case}
We can also consider reducible module categories of $\Fib$. Such a module category corresponds to a theory having collections of boundary conditions which are not connected by the action of the bulk $\Fib$ symmetry of the theory. A semisimple module category splits as a sum of indecomposable module categories. Therefore, for $\Fib$, all semisimple module categories are of the form
\begin{equation}
    \cM = \bigoplus^n_{i=1} \Fib
\end{equation}
for some $n$. This adds an additional index to the sectors of the theory and has the sole effect of dressing the above analysis by an index on the left and right boundaries. Therefore the strip algebra is simply
\begin{equation}
    \Str{\Fib}{\cM} = \bigoplus_{i,j=1}^n \Str{\Fib}{\Fib}.
\end{equation}

An irreducible representation can be conveniently written as $(R)_{i,j}$ where $R$ is an irreducible representation of $\Str{\Fib}{\Fib}$, and $i,j$ specifies the sectors at each boundary. 
The tensor product is
\begin{equation}
    (R_1)_{i,j} \otimes (R_2)_{k,l} = \delta_{j,k}\;(R_1\otimes R_2)_{i,l}.
\end{equation}
Note that this representation category is multi-fusion, meaning that the tensor unit $1$ is non-simple:
\begin{equation}
    1 = \bigoplus_{i} (1)_{i,i}.
\end{equation}

\subsubsection{\texorpdfstring{$\Z_2$}{Z2} Tambara-Yamagami}\label{sec:strip_TYZ2}
The $\Z_2$ Tambara-Yamagami fusion category, $\TY(\Z_2)$, have $3$ simple objects  $\{1,\eta, D\}$ with fusions
\begin{equation}
    \eta^2 = 1, \quad \eta D = D \eta = D, \quad D^2 = 1 + \eta.
\end{equation}
The line $D$ is called the duality line and is non-invertible. $\TY(\Z_2)$ is well known to be the symmetry of the $2d$ Ising CFT and naturally appears whenever a theory is self dual under $\Z_2$ gauging \cite{thorngren2021fusion, Chang_2019,Fr_hlich_2004}. 

To simplify notation, in diagrams we will denote lines by different colors: $\eta$ will be orange and $D$ will be blue. Fixing a basis of junctions by the normalization choices
\begin{equation}
    \includegraphics[width=7cm]{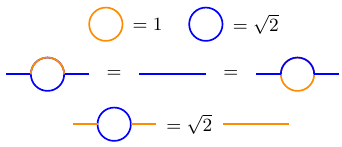}
\end{equation}
the non-trivial $F$-symbols are \cite{Moore:1988qv,Chang_2019}\footnote{There are two $\TY(\Z_2)$ fusion categories with positive quantum dimensions which differ by a choice of sign in their $F$-symbol, corresponding to a choice of Frobenius-Schur indicator. In this example we take a trivial Frobenius-Schur indicator \cite{Chang_2019,Huang:2021zvu}. }
\begin{equation}
    \includegraphics[width=8cm]{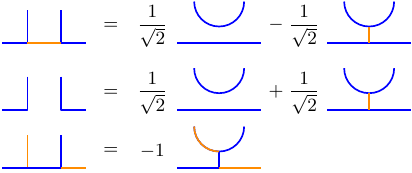}.
\end{equation}

Like $\Fib$, $\TY(\Z_2)$ has a single indecomposable module category, regular module category $\TY(\Z_2)$. The corresponding strip algebra $\Str{\TY(\Z_2)}{\TY(\Z_2)}$ has a basis of $34$ elements\footnote{In this example for convenience we will break our implicit notational convention and label boundaries and lines in the same way.}
\begin{equation}\label{eq:strip_ty_elements}
    \includegraphics[width=12cm]{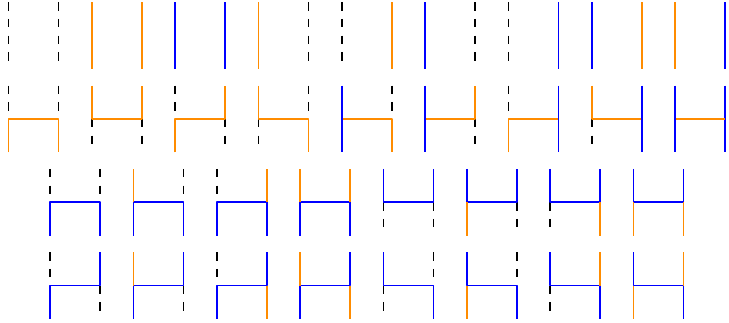}.
\end{equation}
Already this algebra is quite large. We will proceed with a more systematic analysis that demonstrates the general techniques which can be used to study strip algebras. 

Because $\Str{\TY(\Z_2)}{\TY(\Z_2)}$ is semisimple, its study can be entirely phrased as a problem in computing idempotes. Recall that an idempote in an element of an algebra which squares to itself
\begin{equation}
    e_i^2 = e_i \neq 0.
\end{equation}
A collection of idempotes are called orthogonal if
\begin{equation}
    e_i e_j = \delta_{ij}e_i.
\end{equation}
and complete if 
\begin{equation}
    1 = e_1 + \cdots + e_n.
\end{equation}
A complete collection of idempotes is called minimal if the identity cannot be further decomposed into a larger sum of idempotes. An individual idempote is called indecomposable if it cannot be written as the sum of two orthogonal idempotes.

Let $A$ denote a generic finite dimensional semisimple algebra and $\{e_i\}$ a complete collection of orthogonal idempotes. Every orbit $Ae_i \subset A$ is a subrepresentation and by completeness the regular representation has a decomposition
\begin{equation}
    A = \bigoplus_i A e_i.
\end{equation}
The representation $A e_i$ is reducible if and only if $e_i$ can be written as a sum of orthogonal idempotes. Therefore constructing a minimal complete collection of orthogonal idempotes solves the problem of constructing the irreducible representations of $A$. 

Two representations $Ae_i$ and $Ae_j$ are equivalent precisely when there is an element $a \in A$ such that $Ae_i = Ae_j a$. Summing over the idempotes of equivalent representations, one obtains a central idempote (an idempote that commutes with all elements of $A$). Central idempotes therefore specify isomorphism classes of representations of $A$. The central idempotes which can't be written as sums of orthogonal central idempotes also label the simple subalgebras that $A$ decomposes into. We will continue to analyze $\Str{\TY(\Z_2)}{\TY(\Z_2)}$ by identifying such idempotes. 

The strip algebra always has an obvious complete collection of idempotes corresponding to diagrams without a horizontal line inserted. For $\Str{\TY(\Z_2)}{\TY(\Z_2)}$ these are
\begin{equation}
    \includegraphics[width=13.5cm]{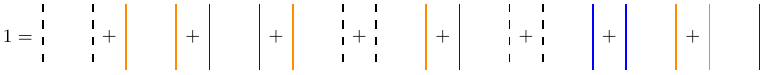}.
\end{equation}
Let's ask if any of these summands can admit a further decomposition. This requires finding an element in $\Str{\TY(\Z_2)}{\TY(\Z_2)}$ which is orthogonal to all the summands besides the one we are trying to decompose. The condition of being orthogonal to an above summand is equivalent to requiring the element not have the boundary conditions labelled by that summand on its top or bottom. Therefore, an idempote which decomposes one of the summands must have the same top and bottom boundary conditions as that summand. 

This significantly narrows down the number of elements to consider. In $\Str{\TY(\Z_2)}{\TY(\Z_2)}$ there is only one possibility: \includegraphics[width=0.45cm, valign=b]{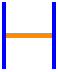}. Using that
\begin{equation}
    \includegraphics[width=4.5cm]{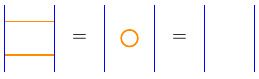},
\end{equation}
by direct calculation one can see that
\begin{equation}\label{eq:TY_non_triv_idempotes}
    \includegraphics[width=7cm]{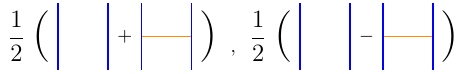}
\end{equation}
are orthogonal idempotes decomposing \includegraphics[width=0.45cm,valign=b]{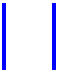}. This exhausts all candidates and so we have found a minimal complete collection of idempotent elements in $\Str{\TY(\Z_2)}{\TY(\Z_2)}$. 

All that remains to do is to analyze which indecomposable idempotes are equivalent. This is equivalent to asking if there are two elements $a,b$ in the algebra such that 
\begin{equation}\label{eq:idempote_equivalence_condition}
   e_i = be_j a,
\end{equation}
for $e_i,e_j$ two idempotes. The structure of the junctions again drastically restricts the number of elements we need to consider. For example, beginning with \includegraphics[width=0.45cm,valign=b]{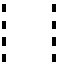}, the only possible non-zero junctions are
\begin{equation}
    \includegraphics[width=5cm]{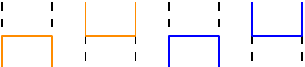},
\end{equation}
which yield
\begin{equation}
    \includegraphics[width=8.5cm,valign=m]{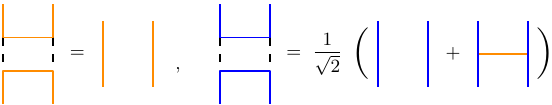}.
\end{equation}
Picking another idempote and continuing, one finds the three equivalence classes
\begin{equation}\label{eq:TY_minimal_idempotes}
    \includegraphics[width=5.5cm, valign=b]{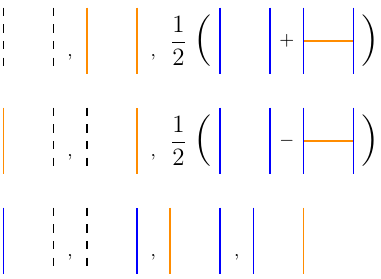}.
\end{equation}
From this, we deduce that $\Str{\TY(\Z_2)}{\TY(\Z_2)}$ has three irreducible representations, two of dimension $3$, and one of dimensions $4$. The four dimensional representation reproduces the soliton degeneracies studied in \cite{Cordova:2024vsq}. 

Let's consider the structure of these representations in more detail. By construction, the Hilbert space of the theory decomposes into a sum over boundary sectors
\begin{equation}
    \mathcal{H} = \bigoplus_{m,n}\mathcal{H}_{mn}, \quad m,n \in \{1,\eta,D\}.
\end{equation}
This decomposition can be decomposed further by using the minimal complete collection of idempotes. For example, we saw that the $\mathcal{H}_{DD}$ sector identity is decomposable and splits into two idempotes. The physical interpretation is clear: unlike the vacua labelled by $1$ and $\eta$, the vacuum corresponding to $D$ supports an unbroken $\Z_2$ symmetry. The $\mathcal{H}_{DD}$ subspace therefore has a natural further decomposition into charged and uncharged sectors
\begin{equation}
    \mathcal{H}_{DD} = \mathcal{H}_{DD}^+ \oplus \mathcal{H}_{DD}^-.
\end{equation}
The idempotes we have calculated are the associated characters. The fact that the different vacua of the same theory can realize symmetry differently may seem curious to reader primarily familiar with grouplike symmetry. This is a simple demonstration of the fact that the vacua resulting from spontaneous breaking of non-invertible symmetry can be physically distinguished. From \eqref{eq:TY_minimal_idempotes} we also see that this $\Z_2$ decomposition governs which multiplet a state in $\mathcal{H}_{DD}$ belongs to. Specifically, a particle in $\mathcal{H}_{DD}$ \textit{without} charge mixes with other particles while a particle \textit{with} charge mixes with solitons and that the $\Z_2$ charge of this particle becomes the $\Z_2$ topological charge of the solitons. This demonstrates a rich interplay between particle and solitonic excitations enforced by non-invertible symmetry!

Finally, it is fruitful to more explicitly consider the significance of the equivalence relation between idempotes \eqref{eq:idempote_equivalence_condition}. Using the refined decomposition of $\mathcal{H}$, by definition the minimal idempotes act as the identity on each summand. The condition for two indecomposable idempotes to be equivalent is therefore that there exist two \textit{invertible} maps between the associated sectors furnished by $\Str{\cC}{\cM}$. This makes transparent that every summand of a representation of $\Str{\cC}{\cM}$ is equivalent to every other. For the case of $\Str{\TY(\Z_2)}{\TY(\Z_2)}$ one can explicitly check that every map between sectors admits an inverse in this sense.

All of this discussion can be organized very concisely using a graph:
\begin{equation}
    \includegraphics[width=4.5cm,valign=m]{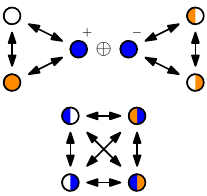}
\end{equation}
where the coloring of the dots denotes the boundary conditions and we have suppressed the identity map on each sector which would be an arrow from a node to itself. This diagram is best thought of as a category, where the composition of arrows is the composition in $\Str{\TY(\Z_2)}{\TY(\Z_2)}$. Each connected component labels an irreducible representation. The statement that equivalent minimal idempotes must admit inverses between their associated sectors is the statement that one pair of arrows between each node must be isomorphisms. The fact that there happens to be one such pair here means that this graph is really a groupoid!

We could also have produced an analgous graph for $\Str{\Fib}{\Fib}$
\begin{equation}
    \includegraphics[width=4.3cm,valign=m]{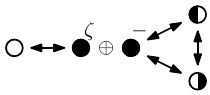}
\end{equation}
which is again a groupoid. Here the labels denote the charges of $\mathcal{H}_{WW}$ under the subalgebra \eqref{eq:fib_sub_alg_WW} of $\Str{\Fib}{\Fib}$.

\subsection{A Systematic Approach to Study}
The final example demonstrated a systematic approach to analyzing $\Str{\cC}{\cM}$ based on the general analysis of finite dimensional semisimple algebras. To summarize. For a general fusion category $\cC$ and module category $\cM$, the algebra $\Str{\cC}{\cM}$ has a natural set of complete idempotes corresponding to the identity line insertions between simple boundaries
\begin{equation}
    1 = \sum_{m,n} e_{m,n}.
\end{equation}
It is possible for an idempote $e_{m,n}$ to decompose only if there is a non-trivial subalgebra which fixes the sector labelled by $(m,n)$. This splits the problem into studying the idempotes, or equivalently the representations, of the subalgebra defined for each sector. Such subalgebras will be related to the fusion rules of $\cC$, but deformed by the associator of $\cM$. Having computed all idempotes, one finally must determine which are equivalent. This requires producing two elements in $\Str{\cC}{\cM}$ which intertwine the pair of idempotes. 

In principle these are steps that one can follow to study both $\Str{\cC}{\cM}$ and its representations. While we have been able to carry this out fully in the prior examples, for a general $\cC$ and $\cM$, this can be a formidable task. In the next section we turn to a more indirect approach for studying the representation of $\Str{\cC}{\cM}$. The utility of this second approach will become apparent when we later reproduce the pages of above analysis in single lines of work.

\section{Representation Theory and the Symmetry TQFT}

Depending on the desired application, the algebra $\Str{\cC}{\cM}$ may be cumbersome to work with directly. For example, it requires knowledge of the explicit forms of bulk and boundary $F$-symbols which can be complicated to both compute and work with for categories of even moderate size. Fortunately, questions about degeneracies or selection rules enforced by $\cC$, which are governed by the representation theory of $\Str{\cC}{\cM}$, can be studied more directly in terms of $\cC$ and $\cM$. 

\subsection{The Dual Category}

Associated to any fusion category $\cC$ and $\cC$-module category $\cM$, there is another category denoted $\cC^*_\cM$, called the dual category of $\cC$ with respect to $\cM$. Algebraically this is defined as the category of endomorphisms of $\cM$,\footnote{Here $\text{op}$ denotes the opposite monoidal structure on $\End_\cC(\cM)$. Its inclusion in this definition is a choice of convention to have $\cC^*_\cM$ act on $\cM$ from the right.}
\begin{equation}
    \cC^*_\cM := \End_\cC(\cM)^{\text{op}}.
\end{equation}
Physically, this category naturally arises as the dual symmetry after gauging the non-invertible symmetry associated to $\cM$ \cite{Bhardwaj:2017xup, thorngren2021fusion,Fuchs:2002cm}. For our purposes we are interested in this category because it is known to encode the representation theory of $\Str{\cC}{\cM}$ (see for example \cite{Douglas_2019,Barter_2019})
\begin{equation}\label{eq:strip_rep_is_dual_cat}
    \Rep(\Str{\cC}{\cM}) \simeq \cC^*_\cM.
\end{equation}
This equivalence is an analogue of the relation
\begin{equation}\tag{\ref{eq:tube_is_center}}
    \Rep(\Tub(\cC)) \simeq \cZ(\cC),
\end{equation}
which from the chain of equivalences
\begin{equation}
    \Rep(\Str{\cC\boxtimes \cC^{op}}{\cC}) \simeq (\cC \boxtimes \cC^{op})^*_\cC \simeq \cZ(\cC) \simeq \Rep(\Tub(\cC)),
\end{equation}
shows that $\Tub(\cC)$ and $\Str{\cC \boxtimes \cC^{op}}{\cC}$ are Morita equivalent algebras. 

It will become clear in the following sections that the equivalence between the representations of $\Str{\cC}{\cM}$ and the dual category $\cC_\cM^*$ is both conceptually and practically very useful, providing a powerful organizing tool for thinking about and working with the representations of $\Str{\cC}{\cM}$. Moreover, depending on the application, it can be substantially easier to work with than studying the algebra $\Str{\cC}{\cM}$ directly. This will likewise become apparent in the following sections. 

Equivalence \eqref{eq:strip_rep_is_dual_cat} has a very natural interpretation in terms of the symmetry topological field theory, $\symTQFT(\cC)$ \cite{Fuchs:2002cm, Fuchs:2003id, Fuchs:2004dz, Fuchs:2004xi, Gaiotto:2014kfa, Gaiotto:2020iye, Apruzzi:2021nmk, Freed:2022qnc, Chatterjee:2022kxb, Inamura:2023ldn, Kaidi:2022cpf, Bhardwaj:2023bbf, Brennan:2024fgj, Antinucci:2024zjp, Bonetti:2024cjk, Apruzzi:2024htg, Apruzzi:2023uma, DelZotto:2024tae}. For a fusion category $\cC$, the symmetry TQFT is the Turaev-Viro theory whose topological lines form the Drinfeld center $\cZ(\cC)$. It is well known that the bulk of a $2d$ quantum field theory $\mathcal{T}$ having symmetry $\cC$ can be realized from a reduction of $\symTQFT(\cC)$ on an interval, $I$. This means that $\mathcal{T}$ placed on a manifold without boundary, $M$, is equivalent to $\symTQFT(\cC)$ placed on $M \times I$, for a particular choice of boundary conditions. On one boundary, we put a non-topological ``physical" boundary where all the dynamical degrees of freedom in $\mathcal{T}$ resides. On the other, we put the topological ``symmetry boundary" which contains all of the lines in $\cC$,
\begin{equation}
    \includegraphics[width=7.5cm,valign=b]{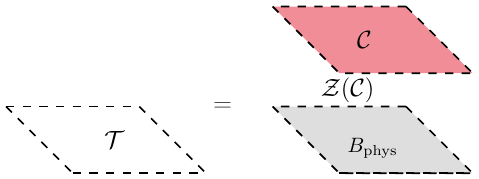}\;\;.
\end{equation}
In this construction, a local operators of $\mathcal{T}$ factors into a triple of data, consisting of an operator on the physical boundary, a topological line in the bulk TQFT, and a topological operator on the symmetry boundary,
\begin{equation}
    \includegraphics[width=7.5cm,valign=b]{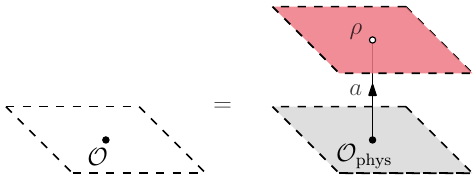}\;,
\end{equation}
where here $a \in \cZ(\cC)$.

It is natural in our context to consider how this construction is modified when the manifold $M$ has boundary (or more generally requires boundary conditions). Let $\mathcal{B}$ denote a boundary condition of $\mathcal{T}$. Rewriting the bulk of $\mathcal{T}$ using the symmetry TQFT, a corner is realized as the boundary is approached:
\begin{equation}\label{eq:symtft_with_bdry}
    \includegraphics[width=6.5cm,valign=b]{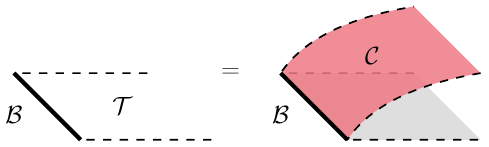}\;\;.
\end{equation}

The collection of all boundary conditions of $\mathcal{T}$ forms a module category, $\cM$, under the action of $\cC$. Its simple objects, $m \in \mathcal{M}$, specify indecomposable boundary conditions, i.e. boundary conditions without a topological local operator confined to the boundary \cite{Cordova:2024vsq, Bhardwaj:2017xup, Huang:2021zvu, Choi_2023}. A symmetry TQFT for a manifold with boundary should capture the symmetry action of $\cC$ on the boundary conditions of $\mathcal{T}$ on its topological symmetry boundary. It should therefore be relative to the symmetry TQFT of $\mathcal{T}$ \cite{freed2014relativequantumfieldtheory}, and so realized by a topological boundary of $\symTQFT(\cC)$. Additionally, the topological interfaces of this boundary with the topological boundary labelled in $\cC$ must form the module category $\cM$. In the Turaev-Viro theory it is known that there always exists one such topological boundary: the boundary having lines forming $\cC^*_\cM$ \footnote{The data governing boundary conditions and interfaces between them can be read off from the 3-category of fusion categories \cite{johnson2017op,douglas2020dualizable}. An alternative approach is the orbifold completion of closed TQFT; see e.g.\ \cite{Kitaev_2012,Fuchs_2013, Carqueville:2016kdq,Carqueville:2023aak}.} 

A natural guess for the symmetry TQFT resolving the boundary conditions of $\mathcal{T}$ is therefore
\begin{equation}
    \includegraphics[width=7.5cm,valign=b]{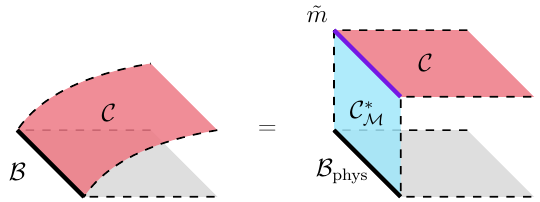}\;,
\end{equation}
where at the corner is an element $\tilde{m} \in \mathcal{M}$ that we leave unspecified for the moment. This boundary does indeed end up being the correct choice, which can be seen by a construction using gauging \cite{Diatlyk:2023fwf, Cordova:2023jip}.\footnote{This follows from a mild generalization of the construction of the physical boundary presented in \cite{Lin:2022dhv}. In this construction, the physical boundary is constructed by stacking the original $2d$ theory with a topological boundary of the $3d$ symmetry TQFT and performing a diagonal gauging. When the $2d$ theory has a boundary, in order to compatibly gauge at the boundary, an additional topological boundary of the 3d theory must be fixed. Moreover, its interface with the symmetry boundary must furnish the module category $\cM$. Since there is only one such boundary there is no ambiguity and one can proceed as in the reference.} Paralleling the standard nomenclature, we will refer to the corner between the physical boundary and the topological boundary as the ``physical corner'' and the corner between the two topological boundaries as the ``symmetry corner''.

To specify the symmetry corner, which is one-dimensional, we need to specify an object $\tilde{m}$ in the module category $\mathcal{M}$. Different choices of object will produce different boundary conditions in the full theory. When studying irreducible boundary conditions of $\mathcal{T}$, $\tilde{m}$ will be a simple object. Another particularly natural choice for $\tilde{m}$ is the sum over the simple elements:
\begin{equation}\label{eq:sum_over_simples}
    \tilde{m} = \sum_{m:\text{simple}} m \in \mathcal{M}.
\end{equation}
This choice corresponds to considering all of the indecomposable boundary conditions contained in $\mathcal{B}$.\footnote{Having introduced both $\cC^*_\cM$ and $\tilde{m}$, the expert may note that the algebra $\Str{\cC}{\cM}$ admits a concise definition as $\End_{\cC^*_\cM}(\underline{\Hom}_\cM(\tilde{m},\tilde{m}))$, where $\underline{\Hom}$ denotes the interal-hom of $\cM$ viewed as a $\cC^*_\cM$-module category.}

In the relative symmetry TQFT, a local boundary operator $\mathcal{O}$ in $\mathcal{B}$ is realized by a triple consisting of a local operator in the physical corner, a topological line in the $\cC^*_\cM$ boundary, and a topological local operator in the symmetry corner: 
\begin{equation}
    \includegraphics[width=8cm, valign=b]{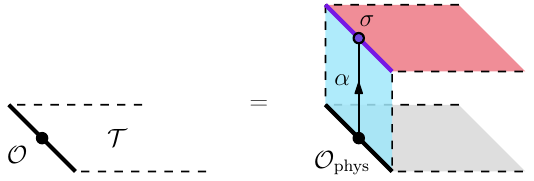}\;,
\end{equation}
where $\alpha$ is a line in $\cC_{\cM}^*$ and $\sigma$ is a morphism in $\cM$, $\sigma \in \Hom_{\cM}(\tilde{m} \otimes \alpha, \tilde{m})$. Here $\tilde{m}\otimes \alpha \coloneqq \alpha(\tilde{m})$ is the action of $\alpha\in\cC_{\cM}^* = \End_{\cM}(\cM)$ on $\tilde{m}$. More generally, boundary changing operators between two boundaries, $\mathcal{B}_i$ and $\mathcal{B}_j$, of $\mathcal{T}$ admit an analogous factorization, now with different topological interfaces in the symmetry corner. The action of $\Str{\cC}{\cM}$ on a boundary changing operator $\mathcal{O}$ is then realized by the action of $\Str{\cC}{\cM}$ on the topological junction $\sigma$ at the symmetry corner: 
\begin{equation}
\includegraphics[width=11cm,valign=b]{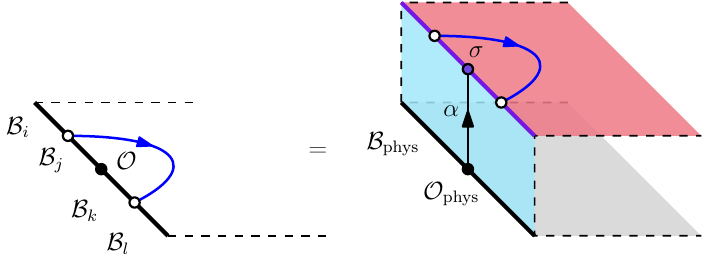}.
\end{equation}
Moreover, if $\alpha$ is simple, the action on topological junctions is transitive. Therefore, for a fixed choice of physical corner local operator, $\alpha$ is a complete invariant of the action of $\Str{\cC}{\cM}$. This defines the equivalence,
\begin{equation}
    \cC^*_\cM \xrightarrow{\sim} \Rep(\Str{\cC}{\cM}), \hspace{.2in}\alpha \mapsto \Hom_{\cM}(\tilde{m}\otimes \alpha,\tilde{m}),
\end{equation}
which reproduces \eqref{eq:strip_rep_is_dual_cat} with $\tilde{m}$ as in \eqref{eq:sum_over_simples}.

This construction directly parallels previous analyses of the representations of $\Tub(\cC)$ using the symmetry TQFT \cite{Lin:2022dhv,Bhardwaj:2023idu,Bartsch:2023wvv}, which can be related to our present considerations by folding. Folding $\symTQFT(\cC)$ on itself, the symmetry boundary becomes $\cC \boxtimes \cC^{op}$ and the new topological boundary remains $\cZ(\cC)$. The corner interface between these boundaries is labelled by $\cC$ thought of as a bimodule category of $\cZ(\cC)$ and $\cC \boxtimes \cC^{op}$. Using again that $\cZ(\cC) \simeq (\cC \boxtimes \cC^{op})^*_\cC$, this is seen to be a special case of the analysis using the relative symmetry TQFT. 

The practical utility of using \eqref{eq:strip_rep_is_dual_cat} to study representation of $\Str{\cC}{\cM}$ rests on how computable the dual category $\cC^*_\cM$ is. In general, this depends sensitively on the choices of $\cC$ and $\cM$. Fortunately, however, there is a general class of interesting examples where this category is known a priori. Consider taking $\cM = \cC$. In the relative symmetry TQFT this means that the topological corner should be decorated elements of $\cC$. However, topological lines in $\cC$ are interfaces from the boundary labelled by $\cC$ to itself $\cC$, which implies that \cite{etingof_tensor_2015} (example 7.12.3)
\begin{equation}
    \cC^{op} \simeq \cC_\cC^*.
\end{equation}
Therefore, when $\cC$ is fully spontaneously broken, we have that
\begin{equation}\label{eq:reg_rep_is_category}
    \Rep(\Str{\cC}{\cC}) \simeq \cC.
\end{equation}
This makes studying the representations that appear in fully spontaneously broken phases very easy if the symmetry category is well understood. We will see a few such physically interesting examples in Section \ref{sec:multiplets}.

\subsection{Weak Hopf Algebras from the Symmetry TQFT}\label{sec:WHA_from_symtft}
It turns out that the symmetry TQFT not only naturally contains the representation theory of the $\Str{\cC}{\cM}$, but can actually reconstruct the entire weak Hopf algebra itself. Such a construction originally appeared in \cite{Freed_2022,Severa_2002} and was rediscovered in \cite{JohnsonFreydReutter:Hopf, JohnsonFreydReutter:QuantumHpty} (see also the talks \cite{ReutterPItalk,JohnsonFreydPItalk,ReutterOxfordtalk}). Here we review this result, following \cite{ReutterPItalk,JohnsonFreydPItalk,ReutterOxfordtalk}. The ingredients of the construction are the following. 

The vector space of elements of the algebra is defined by evaluating the symmetry TQFT on a $2d$ square. This requires a choice of data on each boundary as well as each corner. For the symmetry TQFT of a fusion category $\cC$, topological boundary conditions are known to be in bijection with $\cC$-module categories \cite{johnson2017op,douglas2020dualizable,Thorngren:2019iar,Fuchs_2013}. In this correspondence, an object of the module category $\cM$ labels the corner while the dual category $\cC_\cM^*$ labels the adjacent boundary. Fixing $\cC$ and $\cM$, the data on the square is
\begin{equation}\label{eq:square}
    \includegraphics[width=3.3cm]{images/square_quantization.pdf},
\end{equation}
where at each corner we put the sum over simple objects $\tilde{m}$, \eqref{eq:sum_over_simples} of $\cM$. With this, the state space of the TQFT on the square is specified. This vector space has a natural basis given by the states obtained by attaching the following decorations to the square from below:
\begin{equation}\label{eq:elt_in_strip_symtft}
    \includegraphics[width=2.5cm,valign=m]{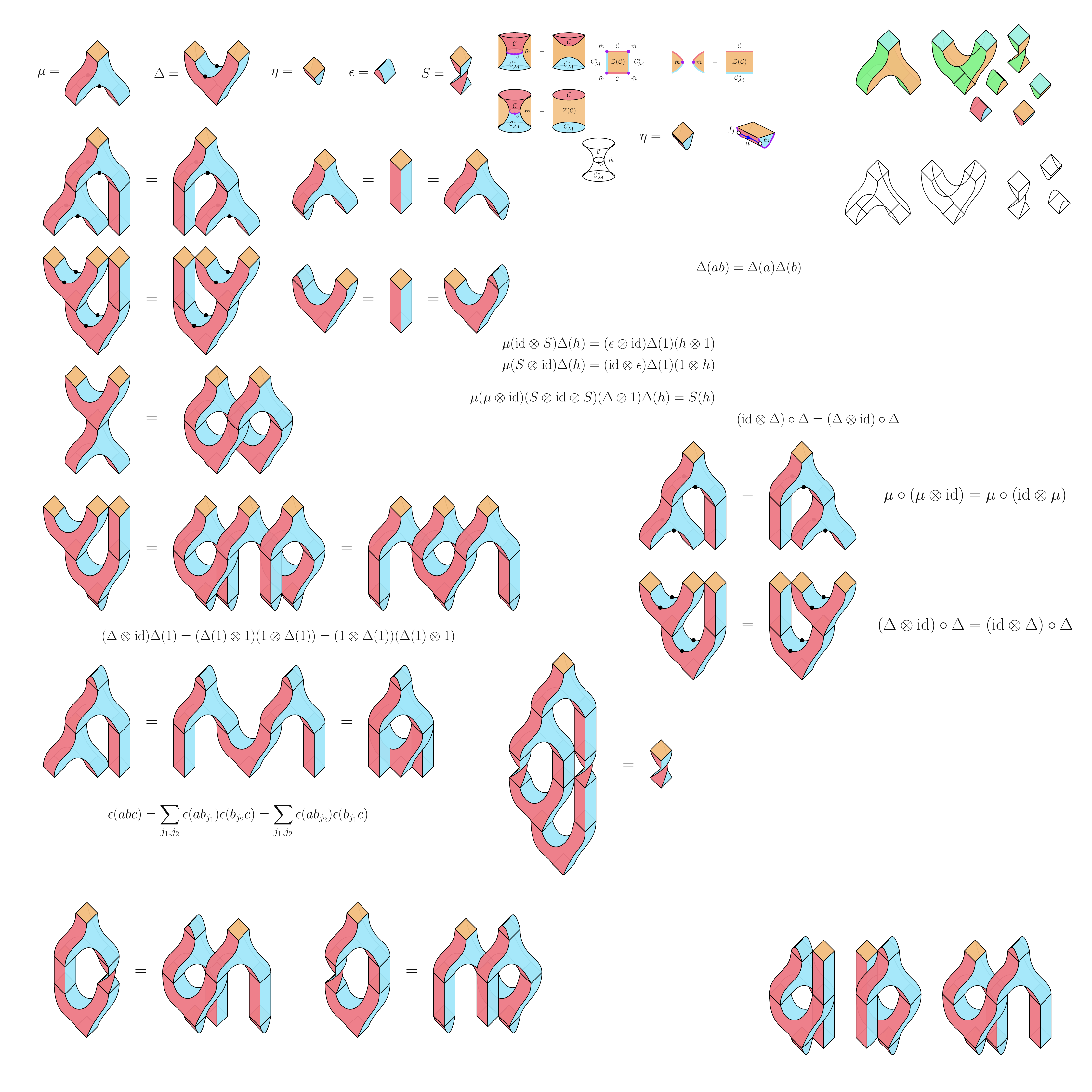},
\end{equation}
where $a$ is a topological defect in the $\cC$-boundary and $e_i$ and $f_j$ are topological junctions that below to a choice of basis on each boundary. This is a state-operator correspondence for this state-space.\footnote{Dually, one could have also constructed a different basis by using the other pair of parallel boundaries.} These decorated bordisms define states when viewed as bordisms from the empty manifold to the square by specifying a linear map
\begin{equation}
    \ket{a; i,j}: \C \to \mathcal{H}(I \times I).
\end{equation}

In the following it will be important to know that the interface, $\tilde{m}$, has an important property. Namely, when it is decorated by a specific local operator, it can be ``retracted''. This means the following. Consider the symmetry TQFT with a solid cylinder excised. On the newly introduced cylindrical boundary, place the boundary condition labeled by $\cC$ on one half, and the boundary condition labeled by $\cC^*_\cM$ on the other. Join the two by the topological interface $\tilde{m}$ along their common cycle. If the topological operator
\begin{equation}
    v = \sum_{m \in \cM} d_m \id_m,
\end{equation}
where each $m$ is simple and $d_m$ is the quantum dimension of the element $m$ (see Appendix \ref{app:fusion_cat}), is placed on the interface then it can be shrank to nothing (i.e.\ ``retracted'')
\begin{equation}\label{eq:retr_bdry}
    \includegraphics[width=5cm, valign=m]{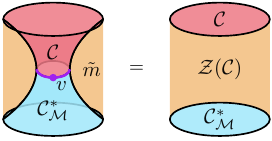}~.
\end{equation}
One way to see that the junction $v$ has this property is the following. Consider the configuration on the right hand side of \eqref{eq:retr_bdry}. Its reduction on the interval produces the $2d$ $\cC$-symmetric TQFT with topological boundary conditions described by $\cM$ \cite{Huang:2021zvu}. In this TQFT, it is known that the simple boundary conditions $\cM$ correspond to idempotent local operators which sum to the identity.\footnote{
This is a general fact about 2d TQFT whose algebra of local operators is semisimple. See for example \cite{Moore:2006dw}.
} Specifically, the idempote corresponding to the simple boundary condition $m$ is given by the boundary $m$ placed on a circle and dressed by a local operator giving it weight $d_m$. Summing over these dressed boundaries gives the operator $v$. Therefore, a puncture assigned $\tilde{m}$ as its boundary conditions and dressed by $v$ can be shrank to nothing. 

We will now naturally equip the state space assigned to the square with the structure of a weak Hopf algebra. This will be done using decorated bordisms. There exist many choice of decorations that can be used, with their differences corresponding to a choice of basis. To most directly match our notion in Section \ref{sec:strip_alg_properties}, we will choose to work with the following decorated bordisms
\begin{equation}\label{eq:weak_hopf_bordisms}
    \includegraphics[width=12.5cm,valign=m]{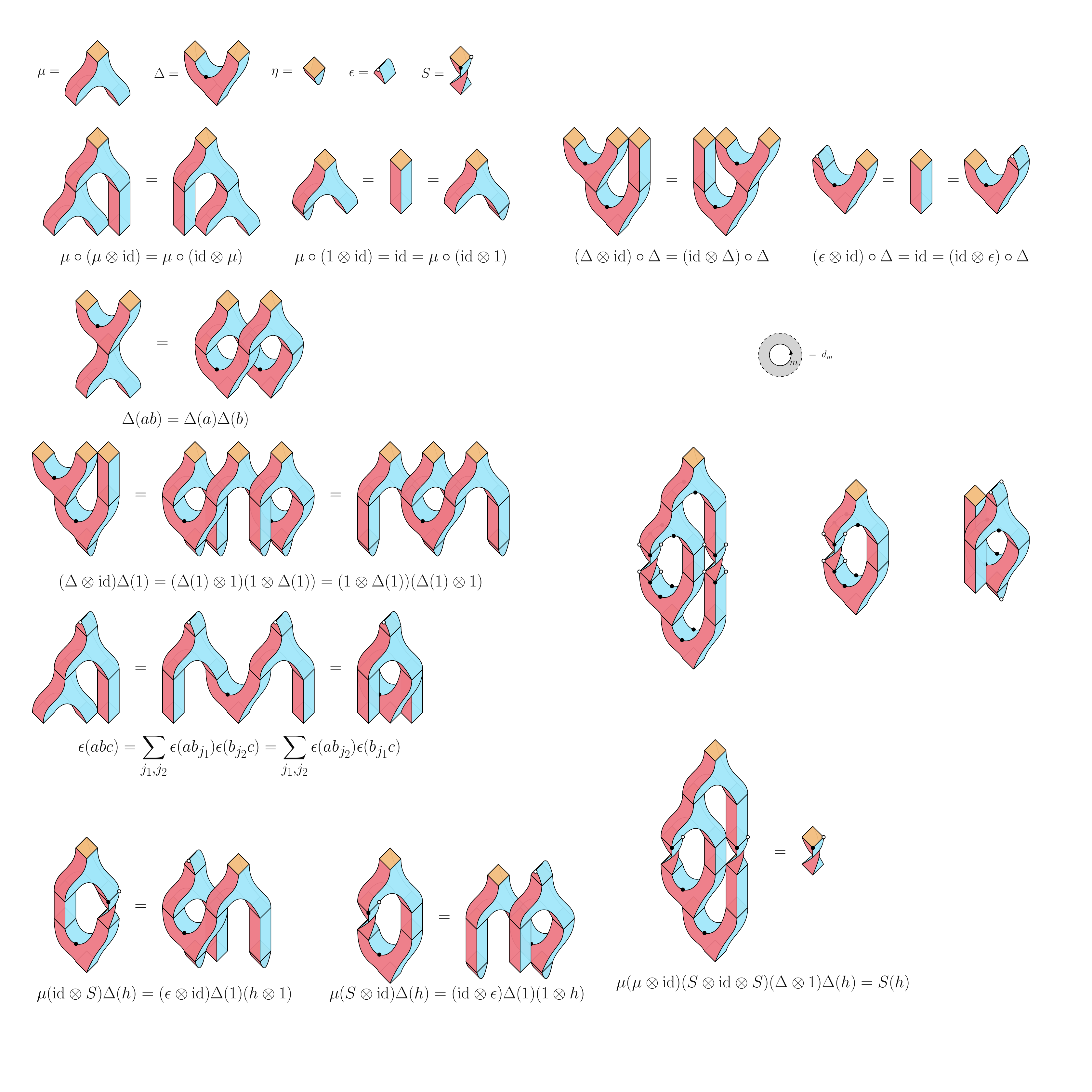}.
\end{equation}
One could alternatively chose to work with more symmetrically decorated bordisms. The bordisms are decorated by local operators on corners. These operators are
\begin{equation}
    \bullet = \sum_{m \in \cM} d_m \id_m, \quad \circ = \sum_{m \in \cM} d_m^{-1} \id_m,
\end{equation}
where $m$ are simple boundary conditions. That is,
\begin{equation}
    \bullet = v, \quad \bullet \circ = \circ \bullet = \id,
\end{equation}
with $v$ as above. Other, more symmetric choices of decoration could be made, but we will proceed with this choice for convenience. 

Our first task is to compare the maps defined by these bordisms to those used in Section \ref{sec:strip_alg_properties}. The multiplication and unit maps are clearly seen to be the same. On two basis elements, the multiplication bordism defines the map
\begin{equation}
    \includegraphics[width=7.5cm,valign=m]{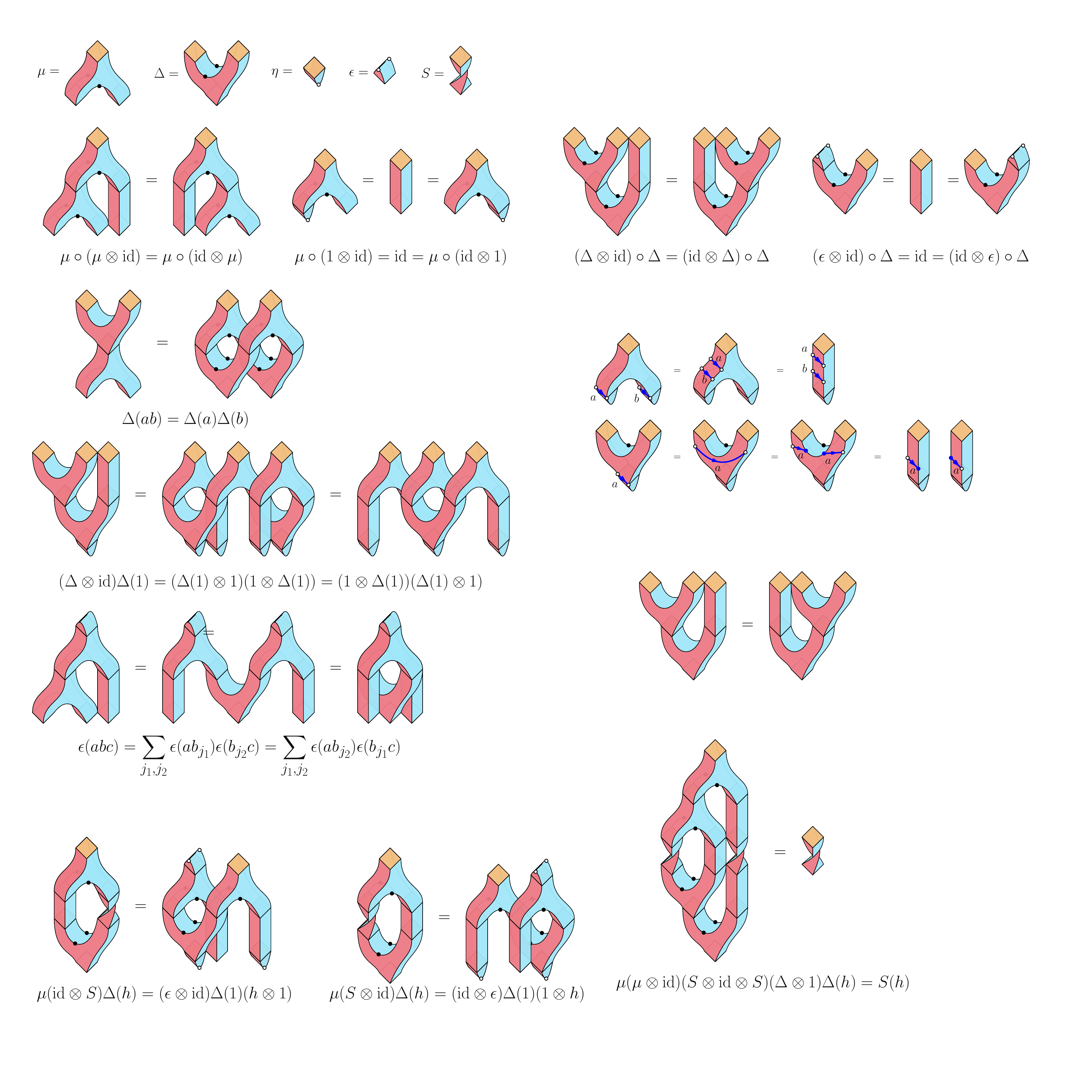}~,
\end{equation}
which is the same as the multiplication in \eqref{eq:strip_mult}. The comultiplication bordism is computed on basis elements to be
\begin{equation}\label{eq:bordism_coprod_cut}
    \includegraphics[width=11.5cm,valign=m]{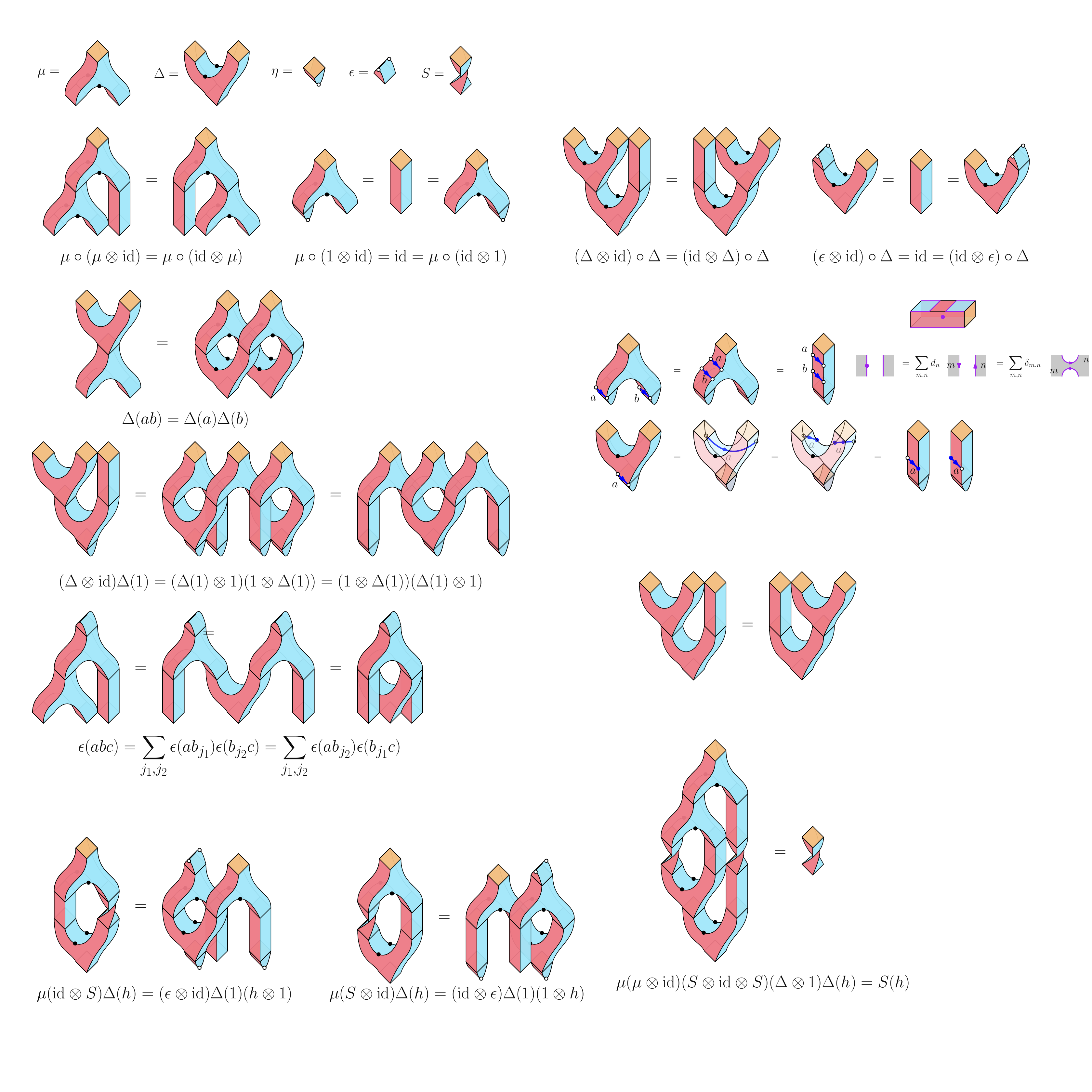}~,
\end{equation}
where we have suppressed boundary crossings and boundary and junction labels for schematic ease. For the second and third manipulation, we've made the diagrams more transparent so to more easily see the line moved onto the backside. This produces the same map as \eqref{eq:strip_coprod}. The final equality requires a bit of work to demonstrate but can be shown using arguments in $2d$ TQFT. Specially, forgetting about the line defect for now, a unit bordism and comultiplication bordism topologically are
\begin{equation}
    \includegraphics[width=3cm,valign=m]{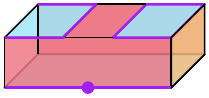}~.
\end{equation}
Now, by bringing the red boundary of the unit, which is now the red square in the top component, to the boundary of the coproduct, which is the bottom boundary, we obtain a $2d$ patch that is a $\cC$-symmetric $2d$ TQFT whose topological boundary conditions are described by $\cM$. We can then perform the $2d$ computation \cite{Huang:2021zvu}
\begin{equation}
    \includegraphics[width=10.5cm,valign=m]{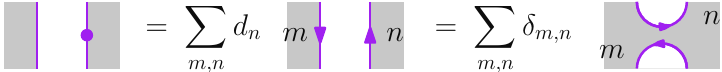}~,
\end{equation}
to cut this patch. This then produces the right hand side of \eqref{eq:bordism_coprod_cut}. The other maps are easily checked and can be seen to produce the pre-factors appearing in \eqref{eq:weak_hopf_operations}.

The verification that the bordisms \eqref{eq:weak_hopf_bordisms} do indeed define the structure of a weak Hopf algebra on the state space is now a simple exercise in topological manipulations. Going down the list of axioms review in Appendix \ref{app:WHA}, most are manifest. For example, the conditions for the algebra and coalgebra structures are immediate
\begin{equation}
    \includegraphics[width=11.5cm,valign=m]{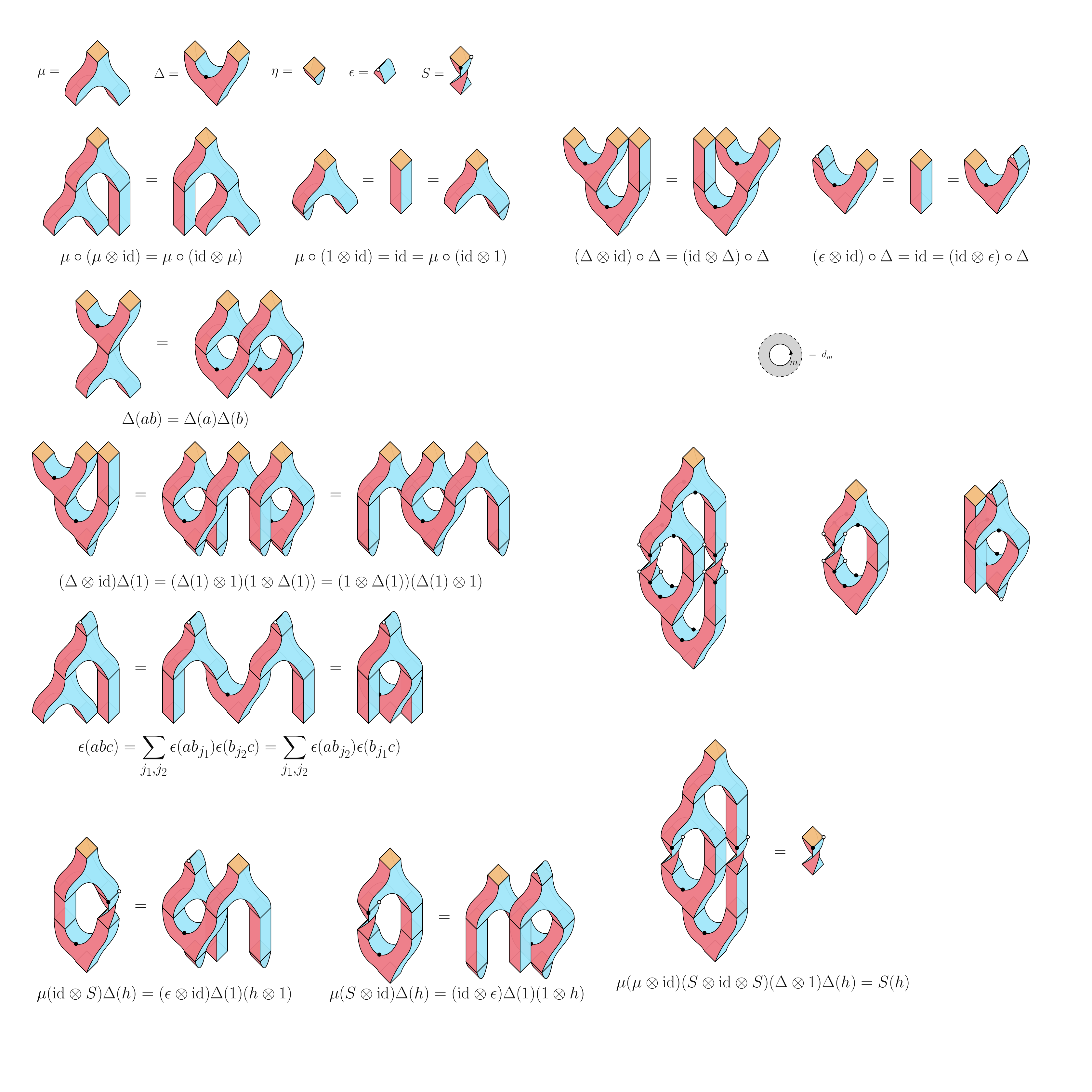}\;\;,
\end{equation}
\begin{equation}
    \includegraphics[width=11.5cm,valign=m]{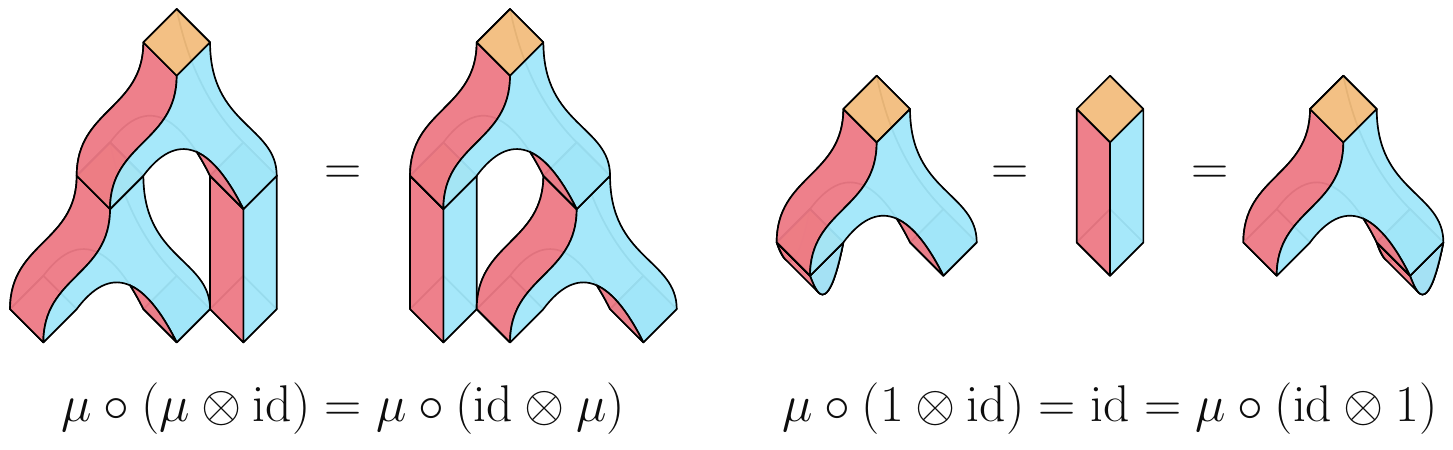}\;\;.
\end{equation}

Similarly, with a little work the reader can convince themself that the following diagrams are topological equivalent
\begin{equation}
    \includegraphics[width=11cm,valign=m]{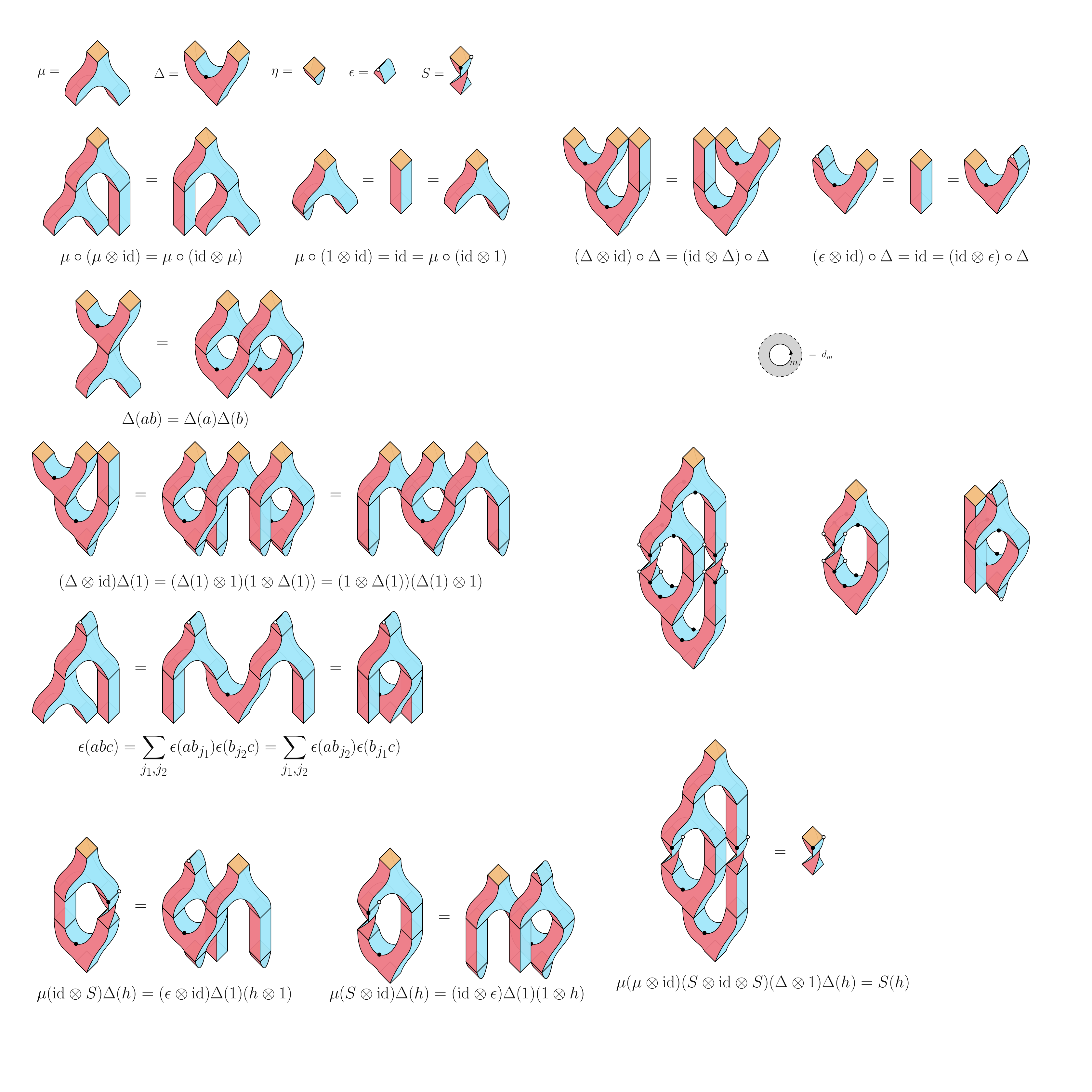}\;\;,
\end{equation}
\begin{equation}
    \includegraphics[width=9cm,valign=m]{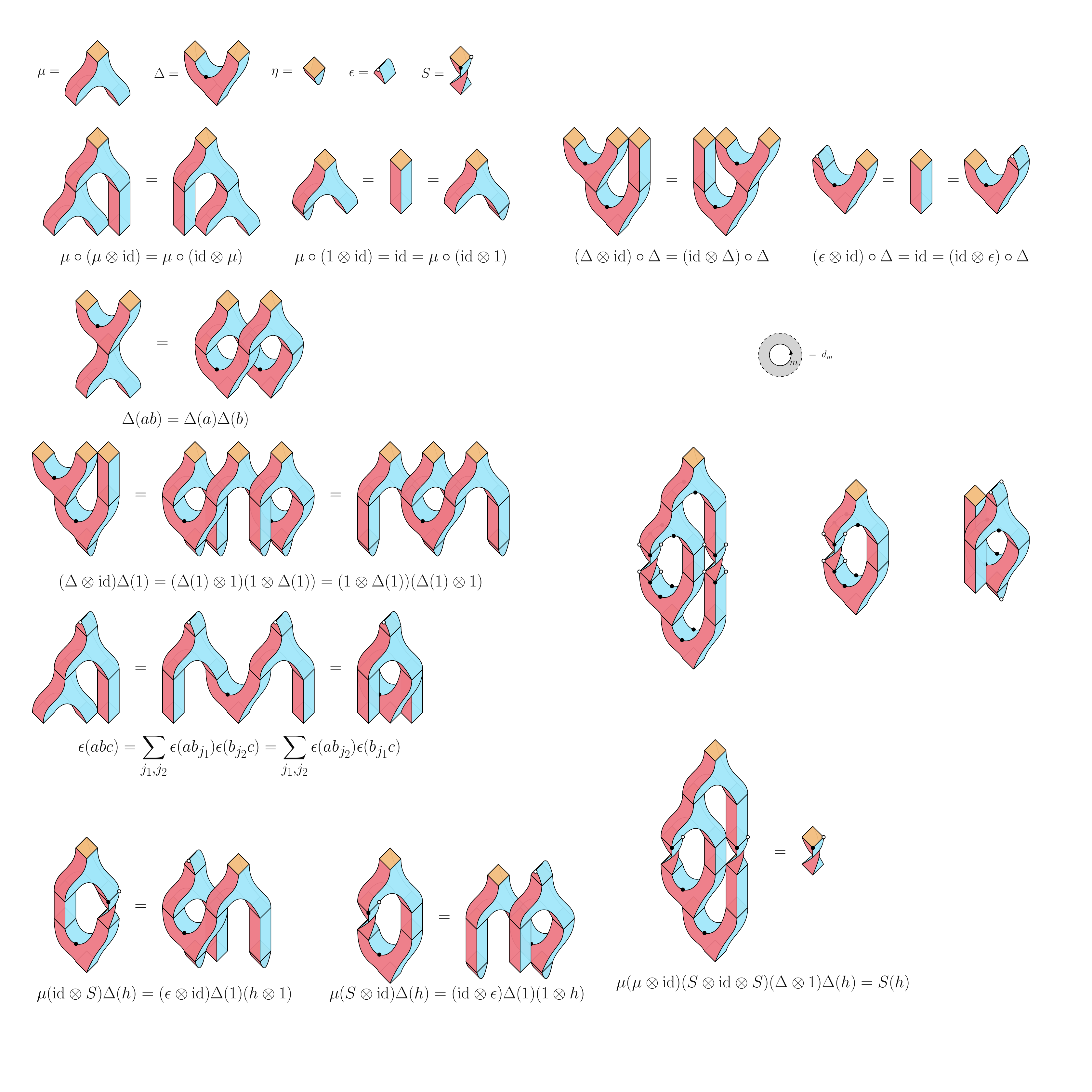}\;\;,
\end{equation}
and so the unit-coproduct and counit-product compatibility conditions hold.\footnote{In the right hand diagrams there are additional maps hidden from view by the coproduct.} In both diagrams we've added extra identity insertions so to better clarify the geometry of diagrams.

The remaining axioms require a bit more care. For example, the compatibility between the product and coproduct requires that the following maps are equal
\begin{equation}
    \includegraphics[width=5.5cm, valign=m]{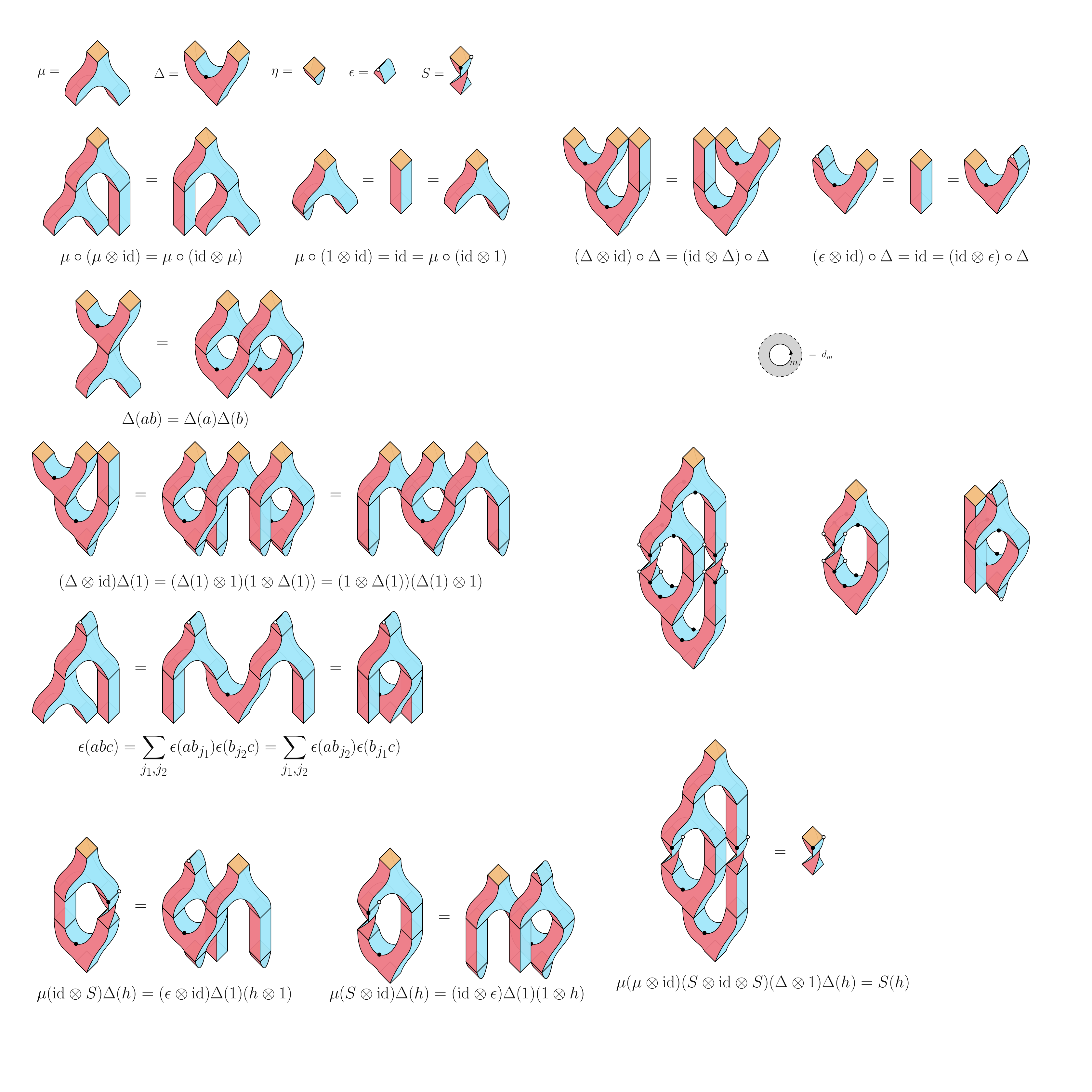}~.
\end{equation}
These bordisms are not topologically equivalent: the second figure has as non-trivial $1$-cycle while the first does not. However, the operation they define is in fact equal. This is because the four local operators along the internal corner in the second figure compose to the local operator $v$ and then using \eqref{eq:retr_bdry}, the $1$-cycle traced by that corner can be shrank to nothing and removed. After this operation the hole is removed an the spaces are equivalent.

Similar considerations must be made when verifying the antipode axioms. The first two conditions require
\begin{equation}
    \includegraphics[width=6cm,valign=m]{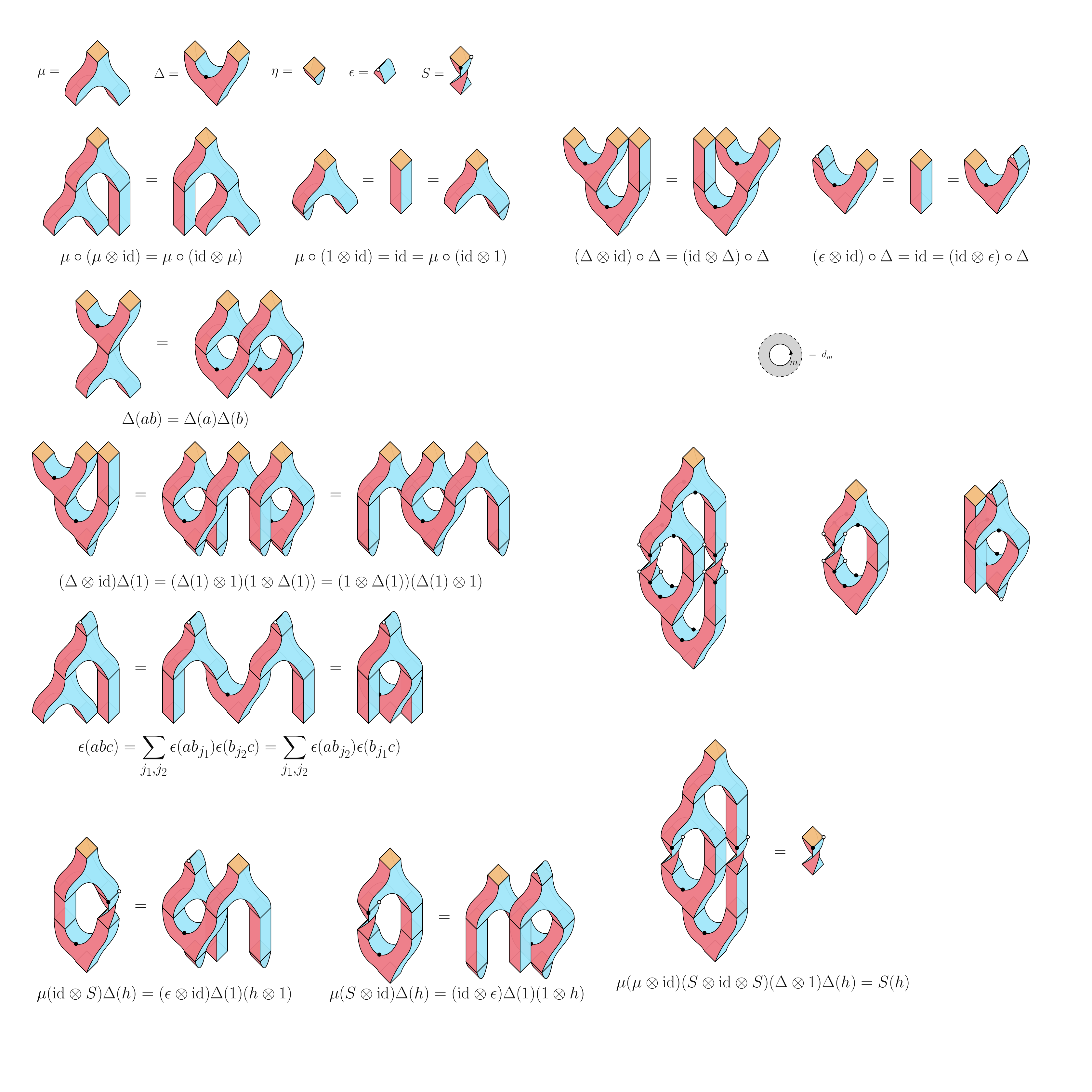}\;\;,
\end{equation}
\begin{equation}
    \includegraphics[width=6cm,valign=m]{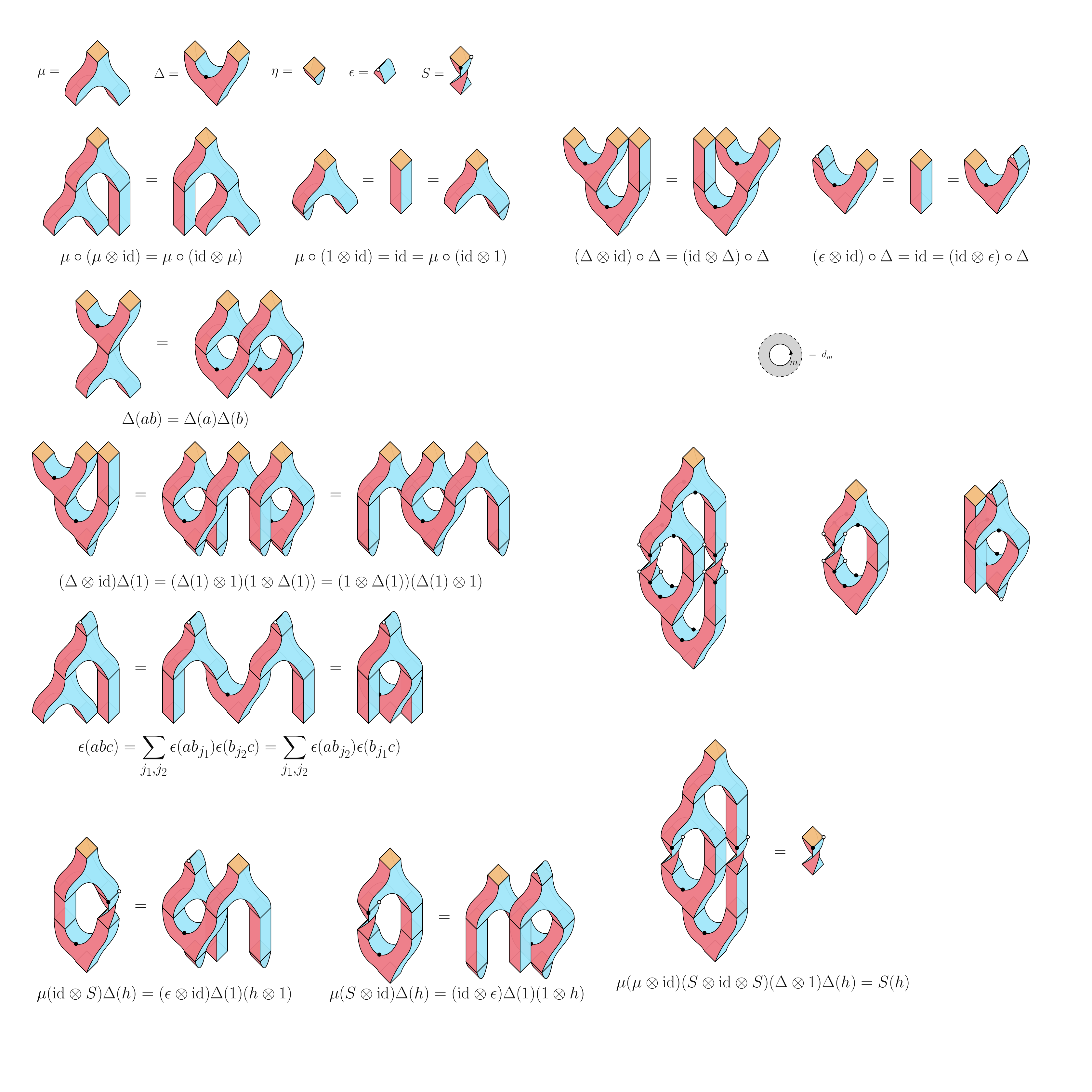}\;\;.
\end{equation}
In each case the left hand side contains a non-trivial cycle which the right hand side does not and so the figures are topological in-equivalent. However, the maps are again seen to be equal by contracting the corner along this cycle to nothing, with the operator $v$ having contributions from the topological local operators dressing the product, coproduct, and antipode. This is again the case for the remaining antipode compatibility condition
\begin{equation}
    \includegraphics[width=7.2cm,valign=m]{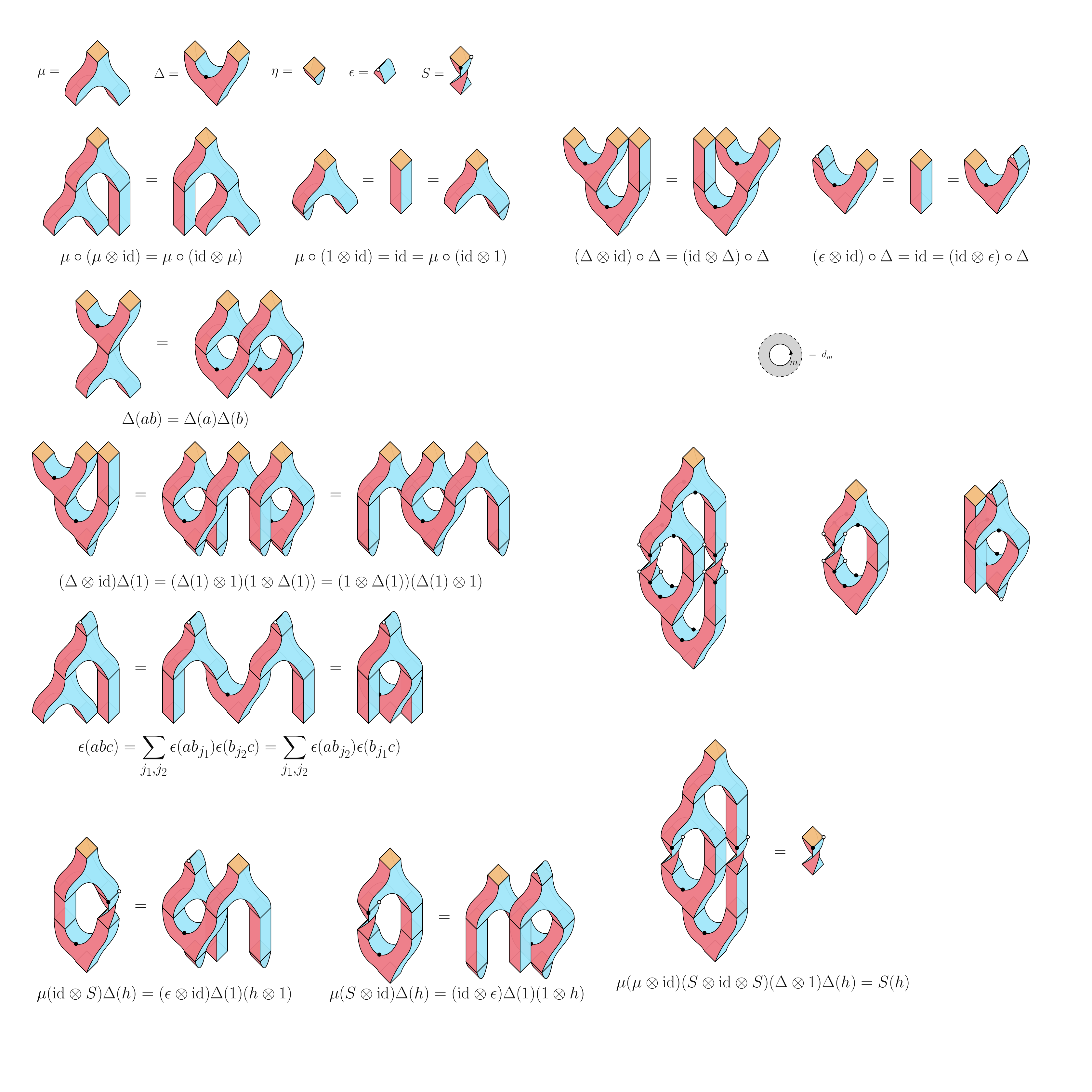},
\end{equation}
where now two $1$-cycles are reduced. These final three identities are trickier to visualize than the prior identities. A systematic way to see them is to study their corners (including the dressed operator) in order to show that they are the same between the left and right hand diagrams. For example, in the final diagram, both the left and right hand diagrams have four external corners (corners which begin and end on ingoing/outing squares) which all twist by $\pi$. The internal corners of the left hand side are precisely those that are retracted to nothing and so removed and so we see that the left hand side does recover the right hand side.

From these considerations, we've shown that the state space does indeed furnish the structure of a weak Hopf algebra. This construction of the algebra provides another tool with which to study it and makes some of its properties more manifest. For example, the axioms of a weak Hopf algebra are self dual, meaning that swapping the underlying vector space and all the structure maps with their duals gives another weak Hopf algebra. 
The above presentation of the strip algebra makes it obvious that the dual of $\Str{\cC}{\cM}$ is actually the strip algebra of $\cC^*_\cM$:
\begin{equation}
    \Str{\cC}{\cM}^\vee = \Str{\cC^*_\cM}{\cM}.
\end{equation}
This implies that for any choice of module category, the strip algebra retains all of the information about the symmetry category that defines it. Concretely, the symmetry category can always be reconstructed as 
\begin{equation}
    \Rep(\Str{\cC}{\cM}^\vee) \simeq \cC.
\end{equation}

The above construction can also provide a basis-free definition of the strip algebra. For the case of $\cM = \cC$, \eqref{eq:square} is just a disk with $\cC$-boundary with four insertion of $\tilde{m} = \sum_{\text{$a$:simple}}a \in \cC$, and the vector space associated to it is $\Hom_{\mathcal{C}}(\tilde{m}^4,1)$.

\section{Multiplets}\label{sec:multiplets}
\subsection{Quivers from Representations}

To summarize, for a theory having symmetry $\cC$ and boundary conditions $\cM$, the dual category $\cC^*_\cM$ and its action on $\cM$ fully determine the representational data of $\Str{\cC}{\cM}$. Concretely, irreducible representation of $\Str{\cC}{\cM}$ are in bijection with simple objects $\alpha \in \cC_\cM^*$ and the representation associated to $\alpha$ is built from the junction vector spaces, 
\begin{equation}\label{representations_from_dual}
    R_\alpha = \bigoplus_{m,n}\Hom(m \otimes \alpha, n),
\end{equation}
of the topological corner of the relative symmetry TQFT. From this perspective, the number, dimension, and boundary sectors of irreducible representations of the strip algebra is made manifest. Note that this information is entirely determined by the dimensions of $\Hom(m \otimes \alpha, n)$, or equivalently, the module category fusion coefficients $\tilde{N}_{m\alpha}^n$.

These coefficients can be conveniently presented using quivers, which allows for a graphical presentation of the representations of the strip algebra. To each simple object $\alpha \in \cC^*_\cM$ construct a quiver in the following way
\begin{enumerate}
    \item For every simple boundary $m\in \cM$, add one node.
    \item For any two nodes $m,n$, add $\tilde{N}_{m\alpha}^n$ arrows from $m$ to $n$.
\end{enumerate}
For example, the quiver associated to the line $\alpha$ will have a contribution of  
\begin{equation}
    \includegraphics[width=3.5cm, valign=m]{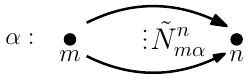}
\end{equation}
from $\Hom(m \otimes \alpha,n)$.

When the theory is quantized on $\R$, these graphs have a direct physical interpretation. Recall that in this geometry, the simple objects of $\cM$ label the clustering vacua of the theory. Each arrow between distinct nodes therefore can be thought of as labelling an excitation that interpolates between distinct vacua, i.e. a soliton. Likewise, an arrow from a node to itself can be thought of as an excitation over the vacuum labelling the node, i.e. a particle. 

Note that the ingredients used to defined the strip algebra are topological, meaning that its action on the states commutes with the Hamiltonian. 
In particular, if a single-particle state is in an representation, other states in the same representation must exist as single-particle states. In this way the representation theory of the strip algebra governs the degeneracies of particles and solitons in the theory \cite{Cordova:2024vsq}.

We emphasize that while the quiver associated to a representation in the above way contains information about dimensions and sectors contained in a representation, it does not contain \textit{all} of the information about the representation. In particular, when multiple sectors are contained in different representations, it does not specify which subspaces of these sectors are contained in each representation. If one wishes to know this, an explicit understanding of the action of the strip algebra is needed. In this case, the techniques used in Section \ref{sec:strip_alg_properties} can be employed to supplement the data contained in the quiver. We now consider some representative examples to both demonstrate how these techniques are used, as well as to provide a sense for the interesting phenomenon that can occur as a result of non-invertible symmetry.

\subsection{Invertible Symmetry}
\subsubsection{\texorpdfstring{$G$}{G} Symmetry: Unbroken}
The unbroken phase of a non-anomalous $G$ symmetry has $\cM = \Vec$ as its module category. The dual category is known to be 
\begin{equation}
    {\Vec^*_G}_{\Vec} \simeq \Rep(G).
\end{equation}
This has a simple interpretation via gauging \cite{Tachikawa:2017gyf}. A $\Vec$ module category is naturally associated to a gauging of the full, non-anomalous $G$ symmetry. The dual symmetry under this gauging is exhausted by the Wilson lines, which are described by $\Rep(G)$. 

From this, it follows that 
\begin{equation}
    \Rep(\Str{\Vec_G}{\Vec}) \simeq \Rep(G),
\end{equation}
which recovers the direct analysis in Section \ref{sec:strip_grp_unbroken}. 

For a representation $\rho \in \Rep(G)$, $\Hom(1\otimes \rho,1)$ is simply the representation $\rho$ itself. In particular, 
\begin{equation}
    \tilde{N}^{1}_{1\rho} = \dim(\rho).
\end{equation}
The quiver associated to a simple element of $\Rep(G)$ is therefore
\begin{equation}
    \includegraphics[width=3.3cm, valign=m]{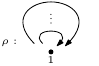},
\end{equation}
where $\dim(\rho)$ arrows are included. This simply expresses that the representation labelled by $\rho$ furnishes a multiplet of $\dim(\rho)$ particle like excitations over a single vacuum.

\subsubsection{\texorpdfstring{$G$}{G} Symmetry: Spontaneously Broken}
In the fully spontaneously broken phase of a $G$ symmetry, $\cM = \Vec_G$ and so
\begin{equation}
    \Rep(\Str{\Vec_G}{\Vec_G}) \simeq \Vec_{G}.
\end{equation}
Therefore, the irreducible representations of the strip algebra are labelled by elements $g \in G$ and their tensor product is governed by the group structure. This is the result we obtained in Section \ref{sec:strip_G_SSB}. 

To illustrate an example, consider $G = \Z_2$. $\Z_2$ has two irreducible representations. The representations labelled by $0,1 \in \Z_2$ are 
\begin{equation}
    R_0 = \Hom(0,0) \oplus \Hom(1,1) \qquad R_1 = \Hom(0,1) \oplus \Hom(1,0)
\end{equation}
and can be expressed by the two quivers
\begin{equation}
    \includegraphics[width=3.5cm, valign=1]{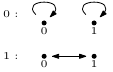}.
\end{equation}
Physically this tells us that there are two possible multiplets in the $\Z_2$ spontaneously broken phase: a multiplet of particle like excitations in each vacuum and a multiplet of soliton like excitations interpolating between each vacuum, the expected behavior.

Proceeding in this same way, the quivers corresponding to irreducible representations of the spontaneously broken phase of other groups can be easily constructed. For example, the quivers for the spontaneously broken phases of $\Z_3$ and $\Z_4$ are
\begin{equation}
    \Z_3: \includegraphics[width=4cm, valign=m]{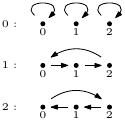}, \qquad \Z_4: \includegraphics[width=5cm, valign=m]{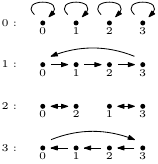}.
\end{equation}
As these examples demonstrate, the use of quivers to encode the representations of the strip algebra has the express benefit of making the excitation type contained in each multiplet manifest.

\subsubsection{Anomalous \texorpdfstring{$G$}{G} Symmetry}
When $G$ is anomalous, its fully spontaneously broken representations from the category
\begin{equation}
    \Rep(\Str{\Vec_G^\omega}{\Vec_G^\omega}).
\end{equation}
Both the elements and fusions of $\Vec_G^\omega$ and $\Vec_G$ are the same and as a result the quivers labelling the representations of each algebra are the same. Therefore, the multiplet structure of a spontaneously broken $G$-symmetry does not sense the anomaly in the symmetry. Rather, the anomaly only appears when one considers decompositions of tensor products of representations. This physically means that the anomaly can appear only in calculations involving decompositions of multiparticle states. This is a manifestation of the observation in Section \ref{sec:strip_anomalous_G} that $\Str{\Vec_G^\omega}{\Vec_G^\omega} = \Str{\Vec_G}{\Vec_G}$ as algebras but differ as weak Hopf algebras.

\subsubsection{No Symmetry}
As another check that the results we are obtaining are sensible, lets consider what this formalism tells us about a theory without symmetry. A fancy name for no symmetry is $\cC = \Vec$. A general decomposable $\Vec$-module category is of the form 
\begin{equation}
    \cM = \bigoplus^n_{i=1} \Vec
\end{equation}
for $n$ a positive integer. The representations of the associated strip algebra are therefore described by 
\begin{equation}
    \Vec^*_{\oplus \Vec} := \End\left(\bigoplus^n_{i=1} \Vec, \bigoplus^n_{j=1} \Vec\right).
\end{equation}
The simple objects in this category are the maps between summands. For concreteness, take $n =2$. Then this category has $4$ simple objects and the quivers they label are
\begin{equation}
    \includegraphics[width=9.5cm,valign=b]{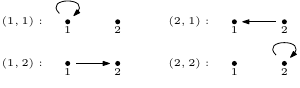}.
\end{equation}
Here, the tensor product of the representations is by composing arrows. If the arrows are not connectable, the tensor product is understood to be zero-dimensional, which can occur because of the projection involved in the definition \eqref{eq:tensor_product_def} of the tensor product.\footnote{As an aside, 
if this dual multi-fusion category were to arise as a symmetry category for a 1+1-dimensional system, it would describe the invertible symmetry $\mathbb{Z}_n^{(1)}\rtimes \mathbb{Z}_n$, i.e. a 0-form $\mathbb{Z}_n$ symmetry permuting 1-form $\mathbb{Z}_n$ symmetry. An object of the form $(i,i)$ represents a one-form symmetry operator smeared over a line, while a zero-form symmetry operator is $\oplus_i (i,i+j \mod n)$ for some $j$.} 

Physically this states that the (lack of) symmetry enforces four irreducible representations, each 1-dimensional, with all possible particle or soliton configurations allowed. That is to say, the lack of symmetry imposes no constraints of the structure of the states of theory. This is clearly as it should be, so we see that the analysis of symmetries using the dual category does indeed behave correctly for this edge case. The analysis for general $n$ is immediately clear.

\subsection{Fibonacci}\label{sec:dualcat_fib}
Recall that the single indecomposable module category of $\Fib$ is its regular module category. Therefore
\begin{equation}
    \Rep(\Str{\Fib}{\Fib}) \simeq \Fib.
\end{equation}
This implies that $\Str{\Fib}{\Fib}$ has two irreducible representations. From the fusion rules in $\Fib$, the associated quivers are 
\begin{equation}
    \includegraphics[width=3.5cm, valign=b]{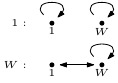}.
\end{equation}
This is immediately seen to reproduce the results we obtained with more effort in Section \ref{sec:strip_fib}. In particular, we immediately see that one representation of $\Str{\Fib}{\Fib}$ is three dimensional and exchanges soliton and particle excitations. 

To avoid the potential for confusion, we emphasize again what one can and cannot deduce from the quiver. The dimensions and sectors contained in the irreducible representations of the strip algebra are immediate from the quiver. However, more nuanced questions such as, when does a particle in the $\mathcal{H}_{WW}$ sector belong to the two dimensional or three dimensional representation, requires explicit knowledge of the action of the strip algebra. Such a question can be answered using the analysis of idempotes in Section \ref{sec:strip_fib}.

\subsubsection{Application: Tricritical Ising \texorpdfstring{$+ \;\phi_{2,1}$}{plus phi21}}
The Tricritical Ising model is a conformal field theory that is the third in an infinite family of unitary diagonal minimal models.\footnote{For details, see \cite{DiFrancesco:1997nk}.} The deformation of this CFT by the subleading $\Z_2$-odd operator, denoted $\phi_{2,1}$, is well studied \cite{Lassig:1990xy, Zamolodchikov:1990xc, Colomo_1992_1, Colomo_1992_2}. The deformed theory is integrable and gapped with two vacua. The spectrum is known to consist of a soliton, anti-soliton, and single massive particle. Moreover, the masses of these excitations are known to be degenerate. 

The symmetry of the deformed theory is known to be $\Fib$ \cite{Chang_2019,Cordova:2024vsq,Copetti:2024rqj}. As explained in \cite{Cordova:2024vsq}, this mass degeneracy should be understood as enforced by the $\Fib$ symmetry: the excitations furnish the three dimensional representation of $\Str{\Fib}{\Fib}$. Beyond single excitations, the entire spectrum of the theory will be organized by $\Str{\Fib}{\Fib}$ and the Hilbert space will decompose into irreducible representations labelled by $\Fib$.

\subsection{\texorpdfstring{$\Z_2$}{Z2} Tambara-Yamagami}\label{sec:dualcat_TY}
The single indecomposable module category of $\TY(\Z_2)$ is also its regular module category and so
\begin{equation}
    \Rep(\Str{\TY(\Z_2)}{\TY(\Z_2)}) \simeq \TY(\Z_2).
\end{equation}
This tells us that $\Str{\TY(\Z_2)}{\TY(\Z_2)}$ has three irreducible representations having quivers
\begin{equation}\label{eq:TYZ2_quiver}
    \includegraphics[width=4.7cm,valign=b]{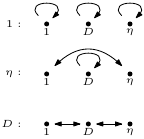}.
\end{equation}
We therefore easily see that the action of $\Str{\TY(\Z_2)}{\TY(\Z_2)}$ can furnish degeneracies between collections of particles, particles and solitons, and collections of solitons. It should not be lost on the reader how much quicker this analysis was in comparison to the direct study of the strip algebra in Section \ref{sec:strip_TYZ2}. The reader should compare the information contained in this quiver and the one presented in that section.

\subsubsection{Application: Tricritical Ising \texorpdfstring{$- \;\phi_{1,3}$}{minus phi13}}
Another well studied deformation of the Tricrtical Ising model is by its least-relevant relevant operator, denoted $\phi_{1,3}$. This deformations is known to be integrable and for the negative sign flows to a gapped phase having three vacua \cite{Zamolodchikov:1991vh}. The spectrum consists of two sets of fundamental kink-anti-kink pairs interpolating between distinct vacua which have exactly degenerate masses.

The symmetry of the deformed theory is known to be $\TY(\Z_2)$ \cite{Chang_2019,Cordova:2024vsq,Copetti:2024rqj}. Considering the possible representations of $\Str{\TY(\Z_2)}{\TY(\Z_2)}$, this degeneracy is understood to be a result of the solitons furnishing the four dimensional irreducible representation \cite{Cordova:2024vsq}. That is, the degeneracy is symmetry enforced by the non-invertible symmetry of the model and corresponds to the bottom quiver in \eqref{eq:TYZ2_quiver}. More still can be deduced from the representations. For example, the lack of stable solitons between the $1$ and $\eta$ vacua in the model imply the non-existence of a stable, $\Z_2$-charged particle over the $D$ vacuum, the content of the second quiver in \eqref{eq:TYZ2_quiver}. Note that this charge condition is not contained as data of the quiver, but rather requires more information about $\Str{\TY(\Z_2)}{\TY(\Z_2)}$; for example, it follows from our computation of idempotes in Section \ref{sec:strip_TYZ2}. We therefore see that a complete knowledge of the representations of the strip algebra provides a new tool box to analyze familiar theories.

\subsection{\texorpdfstring{$su(2)_k$}{su2k}}
While the prior examples were seen to be easily understood using the dual category, we were also able to understand them by directly analysing the strip algebra. The relationship between the dual category and representations of the strip algebra becomes especially useful as the categories $\cC$ and $\cM$ become larger and $\Str{\cC}{\cM}$ becomes more difficult to study directly. For example, let us consider $\cC = \cM = su(2)_k$, the category of lines in $SU(2)$ Chern-Simons at level $k$. This category has $k+1$ simple lines
\begin{equation}
    \{1,2,\dots, k, k+1\} \in su(2)_k
\end{equation}
and its fusion rules have simple closed forms.\footnote{See for example \cite{DiFrancesco:1997nk}.} 

In the fully spontaneously broken phase,
\begin{equation}
    \Rep(\Str{su(2)_k}{su(2)_k}) \simeq su(2)_k
\end{equation}
and so there are $k+1$ irreducible representations of the strip algebra. For varying level, the structure of these representations generically changes. However, there does exist a class of representation that persists for any $k$. For example, the quiver associated to the line $2$ is always an $A_{k+1}$ Dynkin diagram
\begin{equation}
    \includegraphics[width=7.5cm, valign=b]{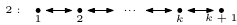}.
\end{equation}
Therefore, for any level, $su(2)_k$ furnishes a representation of solitons interpolating between a complete sequence of vacua with no additional particle excitations.

\subsubsection{Application: \texorpdfstring{$WZW_{su(2)_k}  -  J\bar{J}$}{SU(2)k WZW minus JbarJ} }
The deformation of the $su(2)_k-$WZW model by its current bilinear $\sum_a J^a(z)\bar{J}^a(\bar{z})$ has been studied in detail \cite{Ahn:1990gn}. The deforming operator is marginally relevant and for a negative coupling flows to a gapped phase. The resulting theory is integrable and its spectrum is known: there are $k+1$ vacua and $k$ soliton-anti-soliton pairs interpolating between the vacua in a complete sequence. That is, the spectrum may be represented by the $A_{k+1}$ Dynkin diagram. The masses of all of the solitons are known to be degenerate. 

This deformation preserves the $su(2)_k$ symmetry of the ultraviolet WZW model which is known to be realized in a spontaneously broken phase \cite{Ahn:1990gn}. Therefore, this particle content admits an explanation in terms of the non-invertible symmetry of the model: the solitons furnish the irreducible representation $\mathbf{2}$ of the $su(2)_k$ non-invertible symmetry.

\section{Selection Rules}\label{sec:select}

One of the primary applications of the existence of unitarily acting invertible symmetry in quantum field theories is to the study of selection rules. Given two states in a Hilbert space $\ket{\psi},\ket{\phi} \in \mathcal{H}$ and a unitary action of a symmetry group $G$ on $\mathcal{H}$, if $\ket{\psi}$ and $\ket{\phi}$ live in inequivalent irreducible representations of $G$, then it is a basic result in representation theory that their transition amplitude must vanish
\begin{equation}
    \braket{\psi|\phi} = 0.
\end{equation}
This feature applies more generally to actions of $C^*$-algebras on Hilbert spaces when they are $\dagger$-representations.\footnote{Read: ``dagger-representation''.} The notion of a $\dagger$-representation generalizes that of a unitary representation of a group to algebras which may not admit a basis of invertible elements. Specifically, it requires that the $*$-structure of the $C^*$-algebra be compatible with the inner product of $\mathcal{H}$ in the sense that
\begin{equation}
    L^* = L^\dagger,
\end{equation}
for $L$ an element of the algebra. It follows from this compatibility conditions that if two states are in inequivalent irreducible representations of the algebra, their transition amplitude must vanish. Therefore, $C^*$-algebras provide a more general setting for the study of selection rules.

The natural $C^*$-structure on the strip algebra means that the non-invertible symmetry of a quantum field theory can enforce novel selection rules on transition amplitudes of states. This is especially interesting in light of the fact that the representations of $\Str{\cC}{\cM}$ admit tensor products, meaning that it has a natural action on multiparticle states. Combining these observations, this means that the non-invertible symmetry of the theory, acting through the strip algebra $\Str{\cC}{\cM}$, enforces selection rules on structures such as the $S$-matrix of the theory (see \cite{Copetti:2024rqj} for early examples).  Given the prevalence of non-invertible symmetry in two dimensional theories, this is especially exciting, offering the possibility of learning new insights about the scattering of long studied models. 

By combining the techniques considered in Sections \ref{sec:strip_alg_properties} and \ref{sec:multiplets}, the selection rules enforced by $\Str{\cC}{\cM}$ can be studied systematically. In the spirit of demonstration, here we will discuss two simple examples which are relevant to understanding the scattering in the massive quantum field theories mentioned in Sections \ref{sec:dualcat_fib} and \ref{sec:dualcat_TY}.

\subsection{Selection Rules from Fibonacci}
Let $R_1$ and $R_W$ denote the two irreducible representations of $\Str{\Fib}{\Fib}$. An example of a selection rule on $2 \to 2$ scattering enforced by the $\Fib$ symmetry is the following. Consider the scattering of an excitation of charge $R_1$ by a excitation of charge $R_W$. From our prior analyses we know that
\begin{equation}\label{eq:fib_rep_constraint_1}
    R_1 \otimes R_W \cong R_W \ncong R_1 \otimes R_1,
\end{equation}
which implies that final configurations with two excitations of charge $R_1$ are not permitted. For the scattering of a particle by a soliton, this is the statement that at least one outgoing state must be a soliton which is physically clear. The selection rule is more interesting when considering the particle by particle scattering in this tensor product. 

We saw earlier that the subspace $\mathcal{H}_{WW}$ contains two possible particle types distinguished by their transformation under $\Str{\Fib}{\Fib}$. Under the action of the $W$ line, the particle in the $R_1$ multiplet has charge $\zeta$ while the particle in the $R_W$ multiplet has charge $-1$. The representational statement \eqref{eq:fib_rep_constraint_1} then imposes the following selection rule: a particle of charge $\zeta$ and particle of charge $-1$ cannot scatter into a pair of particles of individual charge $\zeta$. This illustrates a kind of non-invertible charge conservation. 

\subsection{Selection Rules from \texorpdfstring{$\Z_2$}{Z2} Tambara-Yamagami}
Let $R_1$, $R_\eta$, and $R_D$ denote the three irreducible representations of $\Str{\TY(\Z_2)}{\TY(\Z_2)}$. An example of a selection rule imposed on the $2 \to 2$ scattering is the following. Recall from earlier that the $\mathcal{H}_{\eta\eta}$ sector of the theory can support two types of particle excitation differing by their $\Z_2$ charge. The uncharged particle is in the $R_1$ multiplet while the charged particle is in the $R_\eta$ multiplet. The representational statements
\begin{equation}
    R_1 \otimes R_1 \cong R_1 \cong R_\eta \otimes R_\eta \qquad R_1 \otimes R_\eta \cong R_\eta \cong R_\eta \otimes R_1
\end{equation}
enforce the selection rule that the total $\Z_2$ charge be conserved. The interesting feature of this selection rule is that the action of the strip algebra tells us that it is equivalent to the selection rule that a particle pair or soliton pair cannot scatter into a particle-soliton pair, which is physically obvious. Therefore the non-invertible symmetry makes the existence of this $\Z_2$ selection rule especially transparent.

\section*{Acknowledgements}
We thank Clement Delcamp and Kansei Inamura for useful discussions. CC, NH acknowledge support from the US Department of Energy Grant 5-29073, and the Sloan Foundation. 
KO is supported by JSPS KAKENHI Grant-in-Aid No.22K13969 and No.24K00522.
CC, NH and KO also acknowledge support from the Simons Collaboration on Global Categorical Symmetries.

\appendix

\section{Fusion Categories and their Module Categories} \label{app:fusion_cat}
In this appendix we review the necessary category theory to understand this paper. For further reading see for instance \cite{Bhardwaj:2017xup,Chang_2019,etingof_tensor_2015}.

\subsection{Fusion Categories}
The symmetry of a $2d$ quantum field theory $\mathcal{T}$ consists of its topological local operators, its topological line defects, and how they interact in correlation functions. For finite symmetry, this data is known to admit an algebraic characterization in terms of tensor categories. When $\mathcal{T}$ is unitary and bosonic this is specialized to the theory of unitary multi-fusion categories. Finally, when $\mathcal{T}$ has only a single topological local operator,\footnote{Equivalently, $\mathcal{T}$ has a unique vacuum on all compact manifolds} this further reduces to the theory of unitary fusion categories. Physically a unitary fusion category encodes the topological defect lines, their topological junctions, and the relations they satisfy in correlation functions in $\mathcal{T}$. Here we provide a brief review of the background mathematics. 

First, each object $a$ in a fusion category $\cC$ represents a topological line defect in a 2d theory $\mathcal{T}$. We use the words defect and object interchangeably. Among the defects, we include the trivial one, which as no effect in correlation functions, and denote it by $1\in\cC$. Two defects $a,b$ can be summed in correlation functions to produce a new defect, $a+b$.\footnote{
    In general the coefficients of this sum must be non-negative integers so that the result has a well defined Hilbert space associated to it. A complex number instead corresponds to inserting a local operator on the defect. 
} This is a defect in a superposition. If a defect $a$ cannot be split into a sum, it is called simple. We assume that $\cC$ is semisimple, meaning that every defect is isomorphic to a sum of simple defects. We also assume that there are finitely many simple defects. Thus we are analyzing \emph{finite} symmetry in $\mathcal{T}$ (generalizing finite symmetry groups).

Two defects $a,b$ can also be placed in parallel close to each other, and the new defect created in this way is written as $a\otimes b$.
This object should also be expanded as a sum of simple objects:
\begin{equation}
    \includegraphics[width=3.5cm,valign=m]{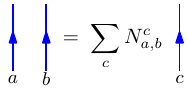}.
\end{equation}
Here, the coefficient $N_{a,b}^c \in \mathbb{Z}$ of the expansion is called the fusion coefficient.
The trivial defect $1$ should be the identity under the fusion, i.e.\ $N_{1,b}^c = N_{b,1}^c = \delta_b^c$.

Given two defects $a,b$ in $\cC$, a morphism from $a$ to $b$ is a topological defect-changing operator between the two defects. Such operators form a vector space denoted by $\Hom(a,b)$. The composition of morphisms is given by the product of operators. For simple $a,b$, we have $\Hom(a,b) = \mathbb{C}^{\delta_{a,b}}$.
Morphisms of particular importance are junctions, connecting three defects $a,b,c$. The space of junctions is denoted by $\Hom(a\otimes b,c)$. The fusion coefficients $N_{a,b}^c$ can be understood as the dimensions of the junction spaces.
We fix a basis of the junction space $\Hom(a \otimes b,c)$ for each triple of simples $a,b,c$, and denote the basis elements by $j_{ab}^c$. In a diagram, we represent a basis junction by a white dot connecting three lines:
\begin{equation}
    \includegraphics[width=2cm,valign=m]{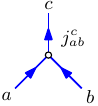}.
\end{equation}
We can form a loop by connecting a junction $j_{ab}^c$ and a junction $j_{c}^{ab}$ as below, and write the resulting coefficient as $( j^{ab}_c,j^{ab}_c )$,
\begin{equation}
    \includegraphics[width=4.2cm,valign=m]{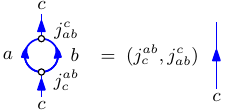}.
\end{equation}

Given a defect $a$, we let $a^\vee$ denote the dual, or orientation reversal, of $a$. The double dual $(a^\vee)^\vee$ is isomorphic to $a$. 
There are particular junctions, evaluation $\mathrm{ev}_a \in \Hom(a\otimes a^\vee,1)$ and coevaluation $\mathrm{coev}_a \in \Hom(1,a^\vee\otimes a)$ for each $a$,\footnote{ To be precise, there are two (left and right) evaluations and coevaluations for each $a$, and there are axioms for them. We omit the details here.}
\begin{equation}
    \includegraphics[width=4.2cm, valign=m]{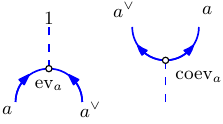}
\end{equation}
The evaluation and coevaluation satisfies the zig-zag axiom 
\begin{equation}
    \includegraphics[width=6cm,valign=m]{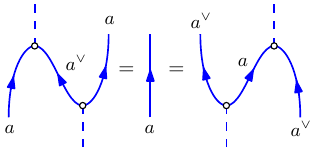}.
\end{equation}
Furthermore, we can form a loop by connecting the evaluation and coevaluation, resulting in a number $d_a$ called quantum dimension of $a$,
\begin{equation}
    \includegraphics[width=3.4cm,valign=m]{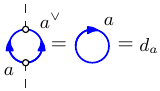}.
\end{equation}

Given three defects $a,b,c$ running in parallel, we can fuse then into a single defect $d$ in two ways: $(a\otimes b)\otimes c$ and $a\otimes (b\otimes c)$. 
Each of the two process defines a basis in $\Hom(a\otimes b\otimes c,d)$, and the two basis are related by $F_{a,b,c}^d$, called the F-symbol:
\begin{equation}
    \includegraphics[width=7.5cm,valign=m]{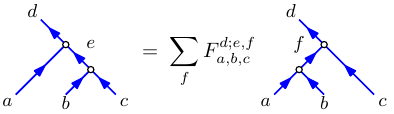}.
\end{equation}
When $N_{a,b}^c\ge 2$ for some $a,b,c$, the F-symbol has more indices to specify basis elements, but we suppress them for simplicity.

A category $\cC$ with structure as outlined above is called a multi-fusion category. When the trivial defect $1$ is simple, the category is called a fusion category.
Multi-fusion categories occurs when the theory has a non-trivial topological local (point) operator, and the simple objects contained in $1$ are topological local operators smeared over a line.

Finally, we need the unitary structure on $\cC$. 
This assigns, for each morphism $f\in \Hom(a,b)$, its dagger $f^\dagger \in \Hom(b,a)$. It further satisfies compatibility conditions with fusion, (co)evaluation, and a positivity condition. See, for example, \cite{Bhardwaj:2017xup} for details.
We note that, while we assume the semisimplicity above, it actually follows from the unitary structure \cite[Exercise 3.2.6]{PenneysLecture}. 
Sometimes it is useful to consider the category $\cC^\text{op}$ with the same objects and morphisms as $\cC$ but with the opposite fusion $\otimes$.\footnote{
    In a general category theory, $\cC^\text{op}$ often denotes the opposite category, where the direction of morphisms are reversed. However here we follow \cite{etingof_tensor_2015} and use the symbol to denote the opposite monoidal structure.
} That is, $a \otimes^{op} b = b \otimes a$. Geometrically, passing from $\cC$ to $\cC^{op}$ corresponds to an action of spatial inversion.

\subsection{Module Categories}
For a $\cC$-symmetric quantum field theory $\mathcal{T}$, the interaction of the symmetry with the boundary conditions of $\mathcal{T}$ is characterized by a module category $\cM$ over $\cC$. See for example \cite{Cordova:2024vsq, Bhardwaj:2017xup, Huang:2021zvu, Choi_2023}. The boundary conditions of $\mathcal{T}$ are the objects of $\cM$. We will use the terms boundary condition and object interchangeably. Boundary conditions of $\mathcal{T}$ admit interfaces which are boundary local operators. The collection of topological boundary operators between two boundary conditions $m$ and $n$ forms a vector space denoted $\Hom(m,n)$. These are the morphisms in $\mathcal{M}$. The composition of two morphisms is defined by operator product expansion.

In correlation functions, two boundary conditions $m,n$ can be summed. This defines a new boundary condition $m + n$.\footnote{This sum must have non-negative integer coefficients so that the new boundary conditions admits a Hilbert space when quantized.} If a boundary condition $m$ cannot be split into a sum, it is called simple. We assume that $\cM$ is semisimple, meaning that every boundary condition is isomorphic to a sum of simple defects. We allow for $\cM$ to contain an arbitrary number of simple boundary conditions. The boundary conditions $m,n,\dots$ are assumed simple unless otherwise specified.

A boundary condition $m$ can be acted by a topological defect $a\in\cC$, turning it into $a\otimes m$ which decomposes into a sum of boundary conditions :
\begin{equation}
    \includegraphics[width=4.3cm,valign=m]{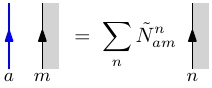}
\end{equation}
where we denote the empty theory in grey. Here by convention $a$ acts on $m$ from the left, and thus $\mathcal{M}$ is a left-module category.
We also consider the junction of one topological line $a$ and two boundary conditions $m,n$:
\begin{equation}
    \includegraphics[width=1.5cm,valign=m]{images/app_module_junction.pdf}.
\end{equation}
The junction hom-space is $\Hom(a\otimes m,n) \cong \mathbb{C}^{\tilde{N}_{am}^n}$. The quantum dimension $d_m$ of a boundary condition $m$ is defined as
\begin{equation}
    \includegraphics[width=3.2cm,valign=m]{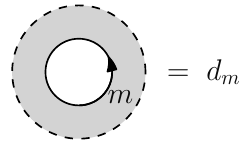}.
\end{equation}

Given two topological defects $a,b$ and a boundary condition $m$, there are two ways to act $a\otimes b$ on $m$: $(a\otimes b)\otimes m$ and $a\otimes (b\otimes m)$.
These defines two bases in $\Hom(a\otimes b\otimes m,n)$, and the two bases are related by a matrix $\tilde{F}_{a,b,m}^n$ called the module F-symbol:
\begin{equation}
    \includegraphics[width=6cm,valign=m]{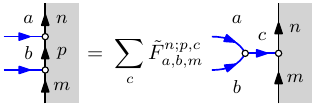}.
\end{equation}
Finally, in unitary QFTs there is a natural unitary structure on $\cM$. This is a collection of conjugate-linear maps $\dagger: \Hom(m,n) \to \Hom(n,m)$ for ever pair of objects in $\cM$. These maps are required to satisfy a positivity condition and be compatible with the unitary structure on $\cC$. See for example \cite{Jones_2017}.

When $\mathcal{T}$ is gapped, its infrared behavior is captured by a $\cC$-symmetric TQFT. The topological boundary conditions of this TQFT form a $\cC$-module category $\cM$ which fully characterizes the TQFT \cite{Huang:2021zvu}. The objects in $\cM$ label ground states of $\mathcal{T}$ and the simple objects are cluster decomposing vacua. When $\mathcal{T}$ is quantized on the real line, the clustering vacua label boundary conditions at infinity and so this module category captures the boundary conditions of the theory even above the deep infrared \cite{Cordova:2024vsq}. \newline

For any multi-fusion category $\cC$, $\cC$ itself is a module category, called its regular module. Physically, a regular module corresponds to the phase where $\cC$ is completely broken spontaneously.

\section{Weak Hopf Algebras}\label{app:WHA}
The notion of a weak Hopf algebra \cite{bohm1996coassociative,nill1998axioms,bohm1999weak} is a generalization of a group. See also \cite{nikshych2000finite,etingof_tensor_2015} for a comprehensive explanation. Here we collect basic definitions and related concepts.

\subsection{Axioms of a Weak Hopf Algebra}

A weak Hopf algebra is a vector space $H$ equipped with the following operations which satisfy a specific set of compatibility conditions:
\begin{itemize}
    \item a multiplication $\mu: H\otimes_\C H \to H$ and a unit $\eta :\mathbb{C} \to H$ with respect to $\mu$,
    \item a comultiplication $\Delta: H \to H\otimes_\C H$ and a counit $\epsilon: H \to \mathbb{C}$, and
    \item an antipode $S: H \to H$.
\end{itemize}
All of $\mu,\eta,\Delta,\epsilon,S$ are linear maps.
Comultiplication is a dual notion of multiplication;
the latter makes one element of $H$ out of two, the former creates (the tensor product of) two elements of $H$ out of one. In addition, we regard the unit $\eta$ to be a linear map $\eta:\mathbb{C}\to H$, so that the image $\eta(1)$, which we also write $1$, trivially multiplies under $\mu$.
Then, the counit is the dual notion of unit; it satisfies
\begin{equation*}
    (\epsilon\otimes \id) \circ \Delta = (\id\otimes \epsilon) \circ \Delta = \id.
\end{equation*}

For the tuple $(H,\mu,\eta,\Delta,\epsilon,S)$ to be a weak Hopf algebra, it has to satisfy the following conditions:
\begin{enumerate}
    \item The triple $(H, \mu, \eta)$ is an associative algebra with unit.
    \item The triple $(H, \Delta, \epsilon)$ is a coassociative coalgebra with counit. The coassociativity condition is
    \begin{equation*}
        (\id \otimes \Delta) \circ \Delta = (\Delta \otimes \id) \circ \Delta.
    \end{equation*}
    This is the dual of the associativity condition for multiplicataion.
    \item The comultiplication is an algebra morphism. That is,
    \begin{equation*}
        \Delta(ab) = \Delta(a)\Delta(b).
    \end{equation*}
    \item The unit satisfies
    \begin{equation*}
        (\Delta \otimes \id)\Delta(1) = (\Delta(1)\otimes 1)(1 \otimes \Delta(1)) = (1 \otimes \Delta(1))(\Delta(1)\otimes 1).
    \end{equation*}
    \item The counit satisfies
    \[\epsilon(abc) = \sum_{j_1,j_2}\epsilon(ab_{j_1})\epsilon(b_{j_2}c) = \sum_{j_1,j_2}\epsilon(ab_{j_2})\epsilon(b_{j_1}c)\]
    where $\Delta(b) = \sum\limits_{j_1,j_2} b_{j_1} \otimes b_{j_2}$.
    \item The antipode satisfies 
    \begin{align*}
        \mu(\id \otimes S)\Delta(h) &= (\epsilon \otimes \id)\Delta(1) (h\otimes 1),\\
        \mu(S \otimes \id)\Delta(h) &= (\id \otimes \epsilon) (1\otimes h)\Delta(1),\\
        \mu(\mu \otimes \id)(S \otimes \id \otimes S)(\Delta\otimes 1)\Delta(h) &= S(h).
    \end{align*}
\end{enumerate}

 A weak Hopf algebra is a Hopf algebra if it satisfies, instead of 4, 5 and 6 above, the following stronger conditions:
 \begin{enumerate}
      \item[4'] The unit is a coalgebra morphism. That is,
     \begin{equation*}
         \Delta(1) = 1 \otimes 1.
     \end{equation*}
     \item[5'] The counit is an algebra morphism. That is,
     \begin{equation*}
         \epsilon(ab) = \epsilon(a)\epsilon(b)
     \end{equation*}
     \item[6'] The aniopde satisfies
    \begin{equation*}
        \mu(\id \otimes S)\Delta(h) = \mu(S \otimes \id)\Delta(h) = 1\epsilon(h).
    \end{equation*}
 \end{enumerate}
In fact, demanding only one of 4', 5', and 6' is enough to make a weak Hopf algebra into a Hopf algebra \cite{bohm1999weak}.

\subsection{\texorpdfstring{$C^*$}{Cstar}-Weak Hopf Algebras}
In quantum field theory, we want the state space to be a Hilbert space. A $C^*$-algebra is an appropriate algebraic structure in this context \cite{gelfand1943imbedding}.
A $C^*$-algebra $A$ comes with a norm $||\bullet||$ and an antilinear involution $*$. The involution satisfies $(ab)^* = b^*a^*$ and $(a^*)^* = a$ and is also compatible with the norm: $||a^*a|| = ||a||^2$.

We would like to consider weak $C^*$-Hopf algebras \cite{bohm1999weak}, which have both the structure of a weak Hopf algebra and that of a $C^*$-algebra.
The only additional condition we need to impose is that the comultiplication is a $*$-algebra homomorphism, that is, $\Delta(h^*) = \sum_{j_1,j_2} h_{j_1}^* \otimes h_{j_2}^*$ when $\Delta(h) = \sum_{j_1,j_2} h_{j_1} \otimes h_{j_2}$. 
Then, the following conditions are automatic
\begin{equation}
    1^* = 1,\quad \epsilon(h^*) = \epsilon(h)^*,\quad S(h^*)^* = S^{-1}(h),
\end{equation}
where $*$ on a complex number means the complex conjugate.

\subsection{Modules and Representation Categories}

A left-$A$-module $M$ of an algebra $A$ is a vector space $M$ equipped with left-action of $A$. That is, a linear map $\rho: A \otimes M \to M$, $a\otimes m \mapsto a\cdot m$ satisfying $ ab \cdot m = a \cdot (b \cdot m)$ and $1\cdot m = m$. A right-module is defined similarly.

A linear map from a left-$A$-module $M_1$ to another left-$A$-module $M_2$ is called an $A$-module intertwiner if it commutes with the action of $A$.
We let $\Mod(A)$ denote the category of left-$A$-modules and $A$-module intertwiners. 

The comultiplication $\Delta$ in a weak Hopf algebra induces a monoidal structure on $\Mod(A)$ \cite{bohm2000weak}.\footnote{
    To define a monoidal structure we do not need the antipode $S$. The structure $(H,\mu,\eta,\Delta,\epsilon)$ satisfying the same axiom as a (weak) Hopf algebra but without the antipode is called a (weak) bialgebra.}
That is, given two modules $M_1$ and $M_2$, we can define the ``tensor product module'' $M_1\otimes M_2$.
The existence of tensor product is due to the coproduct;
the weak Hopf algebra $H$ can act on the tensor product over $\C$, $M_1 \otimes_\C M_2$, of two modules $M_1$ and $M_2$ through the coproduct $\Delta$: $ a \cdot (m_1\otimes m_2) := \sum a_{(1)}\cdot m_1 \otimes a_{(2)}\cdot m_2$, when $ \Delta a = \sum_{j_1,j_2} a_{j_1}\otimes a_{j_2}$. 
When $H$ is a weak Hopf algebra but not a Hopf algebra, $\Delta(1)$ does not acts on the naive tensor product $M_1 \otimes_\C M_2$ as the identity and therefore the naive tensor product is not compatible with the unit of $H$. To remedy this, the monoidal structure $\otimes$ on $\Mod(H)$ is modified to 
\begin{equation}
    M_1 \otimes M_2 \coloneqq  \Delta(1) \cdot (M_1 \otimes_\C M_2).
\end{equation}
On this space $\Delta(1)$ acts as the identity as $\Delta(1)$ is an idempotent in $H$.
For a Hopf algebra this modification is trivial as $\Delta(1) = 1\otimes 1$.
In the context of the main text, this projection is natural because the left vacuum of a state in $M_1$ must match the right vacuum of a state $M_2$ to consider a scattering problem between the two.

The antipode $S$ further induces a dual module $M^\vee$ of a module $M$, which is as a vector space isomorphic to $M$. The action of $H$ on $M^\vee$ is $a\cdot m^\vee := (S(a)\cdot m)^\vee$, where $m^\vee$ is an element $m$ in $M$ thought of as an element in $M^\vee$.

Together with the tensor product, the module category $\Mod(H)$ for a weak Hopf algebra $H$ forms a multi-fusion category \cite{nikshych2003invariants}. Indeed, converse is also true: a multi-fusion category is equivalent to the module category of a weak Hopf algebra \cite{hayashi1999canonical,ostrik2001module}. In particular, the strip algebra in the main text provides a way to create a weak Hopf algebra from a multi-fusion category.

When $H$ is a weak $C^*$-Hopf algebra, we can consider a $*$-representation of $H$, which is a $H$-module $V$ equipped with a non-degenerate hermitian inner product $(\bullet,\bullet)_V$ that satisfies $(u, h\cdot v) = (h^*\cdot u, v)$ for all $u,v \in M$. We let $\Rep(H)$ denote the category of $*$-representations of $H$ and $H$-module intertwiners.
$\Rep(H)$ for finite dimensional weak $C^*$-Hopf algebra $H$ is a unitary multi-fusion category \cite{Ciamprone:2021jrs}.

\bibliographystyle{JHEP}
\bibliography{references}

\end{document}